\PassOptionsToPackage{sort&compress}{natbib}
\documentclass[5p,times]{elsarticle}
\usepackage{enumitem}
\usepackage{kantlipsum}
\usepackage{stfloats}
\usepackage{xurl}
\usepackage{caption}
\usepackage{graphics}
\usepackage{verbatim}
\usepackage{pifont}
\usepackage{float}
\usepackage{amsmath}
\usepackage{cancel}
\usepackage{graphicx}
\usepackage{wrapfig}
\usepackage{float}
\usepackage{color} 
\usepackage[export]{adjustbox}
\usepackage{colortbl}
\usepackage{algpseudocode}
\usepackage{verbatim}
\usepackage{lipsum}
\usepackage{tikz}
\usepackage{enumitem}
\usepackage{enumitem,kantlipsum}
\usepackage{textcomp}
\usepackage{stfloats}
\usepackage{xurl}
\usepackage{caption}
\usepackage{graphics}
\usepackage{tabularx}
\usepackage{xcolor}
\usepackage{float}
\usepackage{forest}
\usepackage{microtype}
\usepackage{graphicx}
\usepackage{longtable}

\usepackage{bbding}
\usepackage{etoolbox}
\usepackage{graphicx}
\usepackage{float}
\usepackage{lscape}
\usepackage{ltxtable}
\usepackage{hhline}
\usepackage{caption}
\usepackage{subcaption}
\usepackage{algorithm} 
\usepackage{color} 
\usepackage{colortbl}
\usepackage{algpseudocode}
\usepackage{verbatim}
\usepackage{lipsum}
\usetikzlibrary{automata, arrows.meta, positioning}
\usetikzlibrary{decorations.pathreplacing}
\usepackage{amsthm}
\usepackage{booktabs}
\usepackage[flushleft]{threeparttable}
\usepackage{algorithm}
\usepackage{inputenc}
\usepackage{adjustbox}
\usepackage{url} 
\usepackage{multirow}
\usepackage{color,soul}
\usepackage{lipsum}
\usepackage{mathtools}
\usepackage{cuted}
\usepackage{xcolor}
\usepackage{microtype}
\usepackage{mdframed}
\usepackage{cancel}
\usepackage{wasysym}
\usepackage{pgfplotstable}
\usetikzlibrary{arrows.meta}
\usepackage{smartdiagram}
\usetikzlibrary{positioning, fit, calc, shapes, arrows,trees}
\usepackage[normalem]{ulem}
\definecolor{lemon}{rgb}{1.0, 1.0, 0.13}
\newcommand\low{\cellcolor{green!60}L}
\newcommand\med{\cellcolor{lemon!80}M}
\newcommand\high{\cellcolor{red!80}H}
\definecolor{columbiablue}{rgb}{0.61, 0.87, 1.0}

\tikzset{%
 thick arrow/.style={
 -{Triangle[angle=120:1pt 1]},
 line width=0.6cm, 
 draw=blue!20
 },
 arrow label/.style={
 text=black,
 align=center
 },
 set mark/.style={
 insert path={
 node [midway, arrow label, node contents=#1]
 }
 }
}
\newcommand\deleted{\bgroup\markoverwith{\textcolor{red}{\rule[0.5ex]{2pt}{0.4pt}}}\ULon}

\newcommand\doublecheck{\textcolor{black}{\checkmark\kern-0em\checkmark}}
\newcommand\semidoublecheck{\textcolor{black}{\checkmark\kern-0em\bcancel{\checkmark}}}

\usetikzlibrary{pgfplots.groupplots}

\definecolor{BulletsColor}{rgb}{0, 0, 0.9}
\newlist{myBullets}{itemize}{1}

\setlist[myBullets]{
 label={\textbullet},
 leftmargin=*,
 topsep=0ex,
 partopsep=0ex,
 parsep=0ex,
 itemsep=0ex,
}

\usepackage{cuted}
\usetikzlibrary{arrows.meta}
\usepackage{smartdiagram}
\usetikzlibrary{positioning, fit, calc, shapes, arrows,trees}

\definecolor{columbiablue}{rgb}{0.61, 0.87, 1.0}

\tikzset{%
 thick arrow/.style={
 -{Triangle[angle=120:1pt 1]},
 line width=0.8cm, 
 draw=blue!20
 },
 arrow label/.style={
 text=black,
 align=center
 },
 set mark/.style={
 insert path={
 node [midway, arrow label, node contents=#1]
 }
 }
}

\journal{Journal of Computer \& Security}

\begin{document}

\begin{frontmatter}
\title{
Evaluation Framework for Quantum Security Risk Assessment: A Comprehensive Strategy for Quantum-Safe Transition
}

\author[First]{Yaser Baseri}
\ead{yaser.baseri@umontreal.ca}
\author[Second]{Vikas Chouhan}
\ead{vikas.chouhan@unb.ca}
\author[Second]{Ali Ghorbani}
\ead{ghorbani@unb.ca}
\author[Third]{Aaron Chow}
\ead{aaron.chow@scotiabank.com}
\address[First]{Department of Computer Science and Operations Research, Universite de Montreal, Canada.}
\address[Second]{Canadian Institute for Cybersecurity (CIC), University of New Brunswick, Canada.}
\address[Third]{Scotiabank, Toronto, Canada.}

\begin{abstract}

The rise of large-scale quantum computing poses a significant threat to traditional cryptographic security measures. Quantum attacks, particularly targeting the mathematical foundations of current asymmetric cryptographic algorithms, render them ineffective. Even standard symmetric key cryptography is susceptible, albeit to a lesser extent, with potential security enhancements through longer keys or extended hash function outputs. Consequently, the cryptographic solutions currently employed to safeguard data will be inadequately secure and vulnerable to emerging quantum technology threats. In response to this impending quantum menace, organizations must chart a course towards quantum-safe environments, demanding robust business continuity plans and meticulous risk management throughout the migration process.  This study provides an in-depth exploration of the challenges associated with migrating from a non-quantum-safe cryptographic state to one resilient against quantum threats. We introduce a comprehensive security risk assessment framework that scrutinizes vulnerabilities across algorithmic, certificate, and protocol layers, covering the entire migration journey, including pre-migration, through-migration, and post-migration stages. Our methodology links identified vulnerabilities to the well-established STRIDE threat model, establishing precise criteria for evaluating their potential impact and likelihood throughout the migration process. Moving beyond theoretical analysis, we address vulnerabilities practically, especially within critical components like cryptographic algorithms, public key infrastructures, and network protocols. Our study not only identifies potential attacks and vulnerabilities at each layer and migration stage but also suggests possible countermeasures and alternatives to enhance system resilience, empowering organizations to construct a secure infrastructure for the quantum era. Through these efforts, we establish the foundation for enduring security in networked systems amid the challenges of the quantum era.

\end{abstract}


\end{frontmatter}

\section{Introduction}
 
\begin{figure*}[!ht]
 \begin{center} 
\resizebox{0.7\linewidth}{!}{
\begin{tikzpicture}[
 title/.style={minimum height=1cm,minimum width=4.9cm,font = {\large}},
 body/.style={draw,top color=white, bottom color=blue!20, rounded corners,minimum width=4.5cm,,minimum height=0.8cm,font = {\footnotesize}},
 typetag/.style={rectangle, draw=black!100, anchor=west}]
 \node (d0) [draw,top color=white, bottom color=blue!20, rounded corners,minimum height=1cm,minimum width=1.2\textwidth,font = {\large}] at (0,0) {NIST Crypto Standards
};
 
\node (d3) [title,below=of d0.center,xshift=0.15\textwidth] {Tools};
 \node (d31) [body,below=of d3.west, typetag, xshift=2mm,yshift=-1.5cm,minimum height=3.3cm] {\begin{minipage}[c]{4.00cm}\centering
RNG (800-90A/B/C)
\end{minipage}};
 \node (d32) [body,below=of d31.west, typetag,yshift=-2.55cm,minimum height=3.3cm] {\begin{minipage}[c]{4.00cm}\centering
KDF (800-108, 800-135)
\end{minipage}};
\node [top color=blue!40, bottom color=blue!40, rounded corners,minimum height=1cm,draw=black!100,fill opacity=0.1, fit={ (d3) (d31) (d32)}] {};
 
\node (d2) [title, left of=d3,xshift=-0.25\textwidth] {Symmetric Key Based
};
 \node (d21) [body,below=of d2.west, typetag, yshift=-0.3cm, xshift=2mm] {\begin{minipage}[c]{4.00cm}\centering
AES (FIPS 197)
\end{minipage}};
 \node (d22) [body,below=of d21.west, typetag] {\begin{minipage}[c]{4.00cm}\centering
TDEA (800-67)
\end{minipage}};
 \node (d23) [body,below=of d22.west, typetag] {\begin{minipage}[c]{4.00cm}\centering
Modes of Operations\\ (800 38A-38G)
\end{minipage}};
\node (d24) [body,below=of d23.west, typetag] {\begin{minipage}[c]{4.00cm}\centering
SHA-1/2 (FIPS 180) and SHA-3 (FIPS 202) 
\end{minipage}};
\node (d25) [body,below=of d24.west, typetag] {\begin{minipage}[c]{4.00cm}\centering
Randomized Hash (800-106)
\end{minipage}};
\node (d26) [body,below=of d25.west, typetag]{\begin{minipage}[c]{4.00cm}\centering
HMAC (FIPS 198)
\end{minipage}};
\node (d27) [body,below=of d26.west, typetag]{\begin{minipage}[c]{4.00cm}\centering
SHA-3 derived functions (800-185)
\end{minipage}};
\node [top color=blue!40, bottom color=blue!40, rounded corners,minimum height=1cm,draw=black!100,fill opacity=0.1, fit={ (d2) (d21) (d22) (d23) (d24) (d25) (d26) (d27)}] {};

\node (d1) [title,left of=d2,xshift=-0.25\textwidth] {Public Key Based
};
 \node (d11) [body,below=of d1.west, typetag, xshift=2mm, yshift=-1.5cm,minimum height=3.3cm] {\begin{minipage}[c]{4.00cm}\centering
Signature (FIPS 186)
\end{minipage}};
 \node (d12) [body,below=of d11.west, typetag,yshift=-2.55cm,minimum height=3.3cm] {
\begin{minipage}[c]{4.00cm}\centering
Key Establishment (800-56A/B/C)
\end{minipage}};

\node [top color=blue!40, bottom color=blue!40, rounded corners,minimum height=1cm,draw=black!100,fill opacity=0.1, fit={ (d1) (d11) (d12)}] {}; 
 
\node (d4) [title, right of=d3,xshift=0.25\textwidth] {Guideline};
 \node (d41) [body,below=of d4.west, typetag, yshift=-0.6cm, xshift=2mm,minimum height=1.55cm] {\begin{minipage}[c]{4.00cm}\centering
Hash Usage/Security (800-107)
\end{minipage}};
 \node (d42) [body,below=of d41.west, typetag, yshift=-0.8cm,minimum height=1.55cm] {\begin{minipage}[c]{4.00cm}\centering
Transition (800-121A)
\end{minipage}};
 \node (d43) [body,below=of d42.west, typetag, yshift=-0.75cm,minimum height=1.55cm] {\begin{minipage}[c]{4.00cm}\centering
Key Generation (800-133)
\end{minipage}};
\node (d44) [body,below=of d43.west, typetag, yshift=-0.8cm,minimum height=1.55cm] {\begin{minipage}[c]{4.00cm}\centering
Key Management (800-75)
\end{minipage}};
\node [top color=blue!40, bottom color=blue!40, rounded corners,minimum height=1cm,draw=black!100,fill opacity=0.1, fit={ (d4) (d41) (d42) (d43) (d44)}] {};

 \end{tikzpicture}
}
\caption{NIST Crypto Standards}
\label{fig:Temp-tree-standards}
\end{center}
\end{figure*}

{Quantum Computing (QC) represents a paradigm shift in computational capabilities, rooted in the principles of quantum mechanics. Unlike classical bits, which exist in a binary state of either 0 or 1, quantum bits (or qubits) can exist in superposition, meaning they can simultaneously represent both 0 and 1. This superposition, along with quantum entanglement and interference, allows quantum computers to process information in parallel, solving certain complex problems exponentially faster than classical computers ever could}~\cite{10064036,gill2022quantum}{. While this revolution promises breakthroughs in fields like materials science, artificial intelligence, and complex optimization, it also introduces significant risks to the cryptographic systems that underpin modern data security.}

{QC poses a significant threat to cryptographic algorithms. With the development of large-scale quantum computers in the near future, cryptographic standards widely used today, such as those established by the National Institute of Standards and Technology (NIST) as shown in Figure}~\ref{fig:Temp-tree-standards}{, will become vulnerable to new classes of attacks. More specifically, QC will primarily impact public key cryptography. Standard public key cryptographic algorithms, such as RSA (based on factoring), Diffie-Hellman, and Elliptic Curve Cryptography (based on the discrete logarithm problem), rely on the hardness of these mathematical problems. However, these algorithms will be susceptible to quantum attacks via Shor's algorithm}~\cite{shor1994algorithms}{. Shor's algorithm allows quantum computers to efficiently compute private keys from public keys, compromising the security of these cryptographic systems. Consequently, standard public key cryptographic algorithms must be replaced with quantum-resistant alternatives to maintain security.}
{Symmetric key cryptography, while more resilient, is not immune to quantum attacks. Grover's algorithm}~\cite{grover1996fast} {enables\break quantum computers to search through unsorted databases in $\mathcal{O}(\sqrt{N})$ time, effectively halving the security strength of symmetric key algorithms. For example, AES-128, which currently provides 128-bit security, would offer only 64-bit security in a quantum context. To preserve equivalent security levels, it is recommended to double the key size, such as upgrading from AES-128 to AES-256. Similarly, hash functions like SHA-3, essential for ensuring data integrity and digital signatures, will also face weakened security. The Brassard-Hoyer-Tapp (BHT) algorithm}~\cite{brassard1997quantum} {reduces the effective security strength of hash functions to one-third of their output size. For example, SHA3-256, which typically provides 128-bit security, would be reduced to approximately 85-bit security against quantum attacks.}

{As the development of quantum computers accelerates, organizations must proactively prepare their cryptographic infrastructure for the quantum era. Transitioning to quantum-safe cryptography involves more than merely adopting new algorithms; it necessitates a holistic migration strategy that addresses hardware, software, and cryptographic standards. A hybrid migration strategy, which deploys quantum-resistant algorithms alongside classical cryptographic methods, ensures that systems remain secure during the migration phase and that critical infrastructure is safeguarded against evolving quantum threats}~\cite{kwon2024compact, bindel2019hybrid, ghinea2022hybrid}.

{To facilitate a seamless transition, organizations must assess the specific risks posed by quantum computing to their cryptographic systems. A robust risk management strategy is essential for identifying vulnerabilities, prioritizing threats, and applying effective mitigation measures. According to NIST}~\cite{blank2011guide}{, an effective risk management framework consists of: (1) framing risks within the context of organizational objectives, (2) assessing risks by identifying and analyzing their potential impacts, (3) responding to risks by selecting and implementing mitigation strategies, and (4) continuously monitoring risks to ensure the effectiveness of controls. Without a structured approach to risk assessment and management, organizations may face significant security gaps during the transition to quantum-safe cryptography.}

\subsection{Motivation}

{The rise of QC presents an unprecedented threat to classical cryptographic systems. This imminent quantum threat compels organizations to migrate to quantum-safe cryptographic states. However, the process of migrating to quantum-safe systems introduces security risks that require careful evaluation and management. Previous research has examined aspects of the quantum threat landscape}~\cite{mosca2023, mosca2021quantum} {and proposed solutions. Some studies have explored areas such as financial risks}~\cite{deodoro2021quantum, covers2020financial} {and communication security}~\cite{sheng2021research, zheng2020quantum}{. However, these efforts often focus on narrow aspects of quantum migration, leaving broader areas like comprehensive risk management underexplored.

A gap exists in research addressing the full spectrum of security threats associated with quantum-safe migration. Some works have provided partial insights into threat assessment}~\cite{ThreatAssessment} {and migration strategies}~\cite{CYBER2016, CYBER2020}{, but a comprehensive risk assessment framework is still lacking. This study addresses this gap by introducing a holistic framework for assessing security risks across all stages of the quantum migration process: pre-migration, through-migration, and post-migration. The framework leverages the well-established STRIDE threat model}~\cite{van2021descriptive} {to analyze quantum-specific threats and provides criteria for evaluating the likelihood and impact of each risk.

In addition to theoretical analysis, our research delves into practical vulnerabilities, especially within critical systems like Public Key Infrastructure (PKI) and network protocols. By focusing on these practical concerns, we aim to provide organizations with the tools necessary to manage quantum-related risks effectively and ensure a secure transition to quantum-safe cryptography.}

\subsection{Contributions}

{This paper introduces a comprehensive framework to evaluate the risks associated with migrating to quantum-safe cryptographic systems. Our key contributions are:}
\begin{itemize}
    \item \textbf{In-depth Analysis of Quantum Threats:} {We provide a detailed analysis of security threats posed by quantum computing at each stage of the migration process (pre-migration, through-migration, post-migration), focusing on algorithmic, certificate, and protocol-level vulnerabilities. This includes identifying vulnerabilities that quantum attackers may exploit.}
    \item \textbf{Threat Modeling with STRIDE:} {We employ the STRIDE threat model to systematically map quantum-specific\break threats throughout the migration process, emphasizing the coexistence of classical and quantum-safe cryptographic systems during the transitional phase.}
    \item \textbf{Development of a Risk Assessment Framework:} {Our framework offers a structured approach to assessing\break quantum-specific risks across different migration stages and organizational levels, with custom criteria for evaluating the likelihood and impact of each risk.}
    \item \textbf{Practical Guidance for System Resilience:} {We propose practical countermeasures to address vulnerabilities at the algorithmic, protocol, and infrastructure levels, enabling organizations to strengthen their systems against the quantum threat.}
\end{itemize}

\subsection{Organization}
The remainder of the paper is organized as follows: Section~\ref{sec:related} reviews related work on quantum threat risk assessment. Sections~\ref{sec:ALA}, \ref{sec:PKI}, and \ref{sec:PLA} introduce our framework for threat analysis and security risk evaluation at the algorithmic, certificate, and protocol levels. Section~\ref{sec:emperical} {presents empirical validation and case studies.} Finally, Section~\ref{conclude} concludes the paper.

\section{Related Works}\label{sec:related}

\begin{table*}[!htbp]
\centering
\footnotesize
\caption{{Comparative Review of Our Research with Existing Related Works}}
\label{tab:comparison}
\resizebox{\linewidth}{!}{%
\begin{tabular}{|p{0.12\linewidth}|p{0.18\linewidth}|p{0.14\linewidth}|p{0.12\linewidth}|p{0.12\linewidth}|p{0.12\linewidth}|p{0.14\linewidth}|p{0.14\linewidth}|}
\hline
\textbf{Research} & \textbf{Main Contribution} & \textbf{Threat Analysis} & \textbf{Risk Assessment} & \textbf{Migration Approach} & \textbf{Hybrid Transition Strategy} & \textbf{Analysis Level} & \textbf{Mitigation Strategy} \\ \hline
ETSI (2017) \cite{ThreatAssessment} & Quantum-safe threat assessment & Simplified threat analysis &\begin{center} -- \end{center}  &\begin{center} -- \end{center}  &\begin{center} -- \end{center}  & Algorithmic, certificate, protocol levels  (general discussion) &\begin{center} -- \end{center} \\ \hline
Ma et al. (2021) \cite{ma2021caraf} & Crypto-agility risk assessment framework & Implicit (through crypto-agility) & Qualitative assessment & Generic guidelines for migration &\begin{center} -- \end{center}  & Algorithmic level & Algorithmic level mitigation \\ \hline
Sheng et al. (2021) \cite{sheng2021research} & Security risk assessment of quantum private communication systems & Communication threat analysis & Qualitative assessment &\begin{center} -- \end{center}  &\begin{center} -- \end{center}  & Communication system level & Quantum secure communication via QKD \\ \hline
White et al. (2022) \cite{IBM2022}&Migration guidance to quantum-Safe cryptography for IBM Z platform
&Focuses on algorithmic threats&\begin{center} -- \end{center}&Generic migration steps for IBM Z	& Hybrid approach for key exchange &Algorithmic level	&Algorithmic level mitigation \\ \hline 
Mosca \& Piani (2023) \cite{mosca2023} & Estimation of quantum threat timelines & Analysis of quantum threat timelines &\begin{center} -- \end{center}  & Generic approach to quantum safety &\begin{center} -- \end{center}  & Algorithmic level&\begin{center} -- \end{center}  \\ \hline

Hasan et al. (2024) \cite{hasan2024framework} & Framework for migrating to post-quantum cryptography & Discussion on migration challenges &\begin{center} -- \end{center}  & 
Migration strategy based on dependency analysis&\begin{center} -- \end{center}  & Algorithmic \& Dependency Analysis& Generic mitigation strategy based on  dependency analysis for some use cases   \\ \hline
Scholten et al. (2024) \cite{scholten2024assessing} & Benefits and risks assessment of quantum computers & Discussion on quantum computer risks &\begin{center} -- \end{center}  & Generic guidelines &\begin{center} -- \end{center}  & Algorithmic level 
& Algorithmic level mitigation  though QRNG, PQC and QKD\\ \hline
\textbf{Our Research} &  Comprehensive risk assessment framework and migration guidelines & Comprehensive threat analysis through algorithms, certificates, and protocols modeled in STRIDE threat model & Qualitative assessment & Detailed migration strategy & Comprehensive approach through algorithms, certificates, and protocols & In-depth analysis of algorithms, certificates, and protocols & Multi-level mitigation strategy including algorithmic, certificate, and protocol levels \\ \hline
\end{tabular}%
}
\end{table*}

{In this section, we explore existing research that encompasses quantum security risk assessment and migration, categorizing these works into distinct areas of focus. We also highlight the significance of our work in relation to these studies.}

\subsection{Quantum Threat Timelines}

 {The investigation of Quantum Threat Timelines has been a substantial field of inquiry. Mosca and Piani (2023)}~\cite{mosca2023} {and Mosca (2020)}~\cite{mosca2021quantum} {have delved into estimating the timeline for quantum threats. They've tapped into insights from quantum computing experts, illuminating the evolving landscape of quantum threats. Additionally, the ETSI report}~\cite{ThreatAssessment} {has synthesized findings from a threat assessment conducted according to ETSI guidelines for various usage scenarios. It envisions the eventual deployment of quantum computers and their influence on cryptographic systems. These works provide valuable context for grasping the urgency of quantum security risk assessment and migration.}

\subsection{Quantum Threat Assessment}

Quantum Threat Assessment encompasses research that delves into the cybersecurity challenges posed by quantum computing. Althobaiti and Dohler (2020)~\cite{althobaiti2020cybersecurity} discuss the vulnerabilities of current IoT security solutions to quantum attacks, highlighting the necessity for advanced cryptographic techniques. 
Additionally, Mosca's study (2018)~\cite{mosca2018cybersecurity} investigates organizations' readiness for quantum computers and focuses on risk assessment based on security shelf life, migration time, and the time left before quantum computers break security. These assessments are pivotal in understanding the specific threats that quantum computing poses to various domains. The GSMA conducted a study on the Post-Quantum Telco Network Impact Assessment in 2023~\cite{gsma2023telco}, providing insights into the impact of post-quantum cryptography on telecommunication networks, shedding light on the challenges and considerations in ensuring the security of such networks in the era of quantum computing.

\subsection{Quantum Computing and Cybersecurity}

{Research exploring the broader implications of quantum computing in cybersecurity falls under the category of ``Quantum Computing and Cybersecurity''. Bains et al. (2023)}~\cite{bains2023quantum}{ provide an in-depth analysis of risks and solutions associated with quantum computing in cybersecurity. They examine perspectives from cybersecurity professionals globally, offering insights into the evolving discourse on quantum intelligence and its implications for cybersecurity strategies. Additionally, Faruk et al. (2022)}~\cite{faruk2022review} {conduct a systematic survey of quantum cybersecurity, discussing quantum technology as both a threat and a solution. Their work provides a comprehensive overview of current trends, serving as a foundational resource for further research in the field. These studies shed light on the broader implications of quantum computing in the cybersecurity landscape.}



\subsection{Transitioning to Quantum-Safe Cryptography}
{Ma et al.} (2021)~\cite{ma2021caraf} {introduce CARAF, a crypto agility risk assessment framework that evaluates the risks resulting from the lack of crypto agility. This framework is particularly relevant in the context of transitioning to quantum-safe cryptographic systems, as it helps organizations determine appropriate mitigation strategies based on their risk tolerance. White et al. (2022) }\cite{IBM2022} {present migration guidance specifically tailored for the IBM Z platform. Their work focuses on providing practical steps for transitioning cryptographic systems on IBM Z hardware to quantum-resistant alternatives. While their focus is limited to a specific platform, it highlights the ongoing efforts towards developing migration strategies for various computing environments. Hasan et al. (2024)}~\cite{hasan2024framework}  {propose a framework to assist organizations in migrating to quantum-resistant cryptographic systems. Their work introduces cryptographic dependency analysis  to pinpoint situations where current cryptosystems might not provide adequate security for the intended lifespan of protected information assets. They showcase the framework's effectiveness through case studies that utilize  dependency mechanism to prioritize crypto-systems for replacement.}
 {Another notable contribution in understanding the implications of quantum computing is the work by Scholten et al. (2024)}~\cite{scholten2024assessing}{, which assesses both the benefits and risks of quantum computers. This study offers insights into the potential uses and risks associated with quantum computing, providing valuable information for security experts and policy decision-makers.}

\subsection{Mitigation Strategies and Hybrid Approaches}
 Beyond identifying vulnerabilities, recent research has\break focused on strategies for transitioning from classical to post-quantum cryptography, with hybrid approaches emerging as a notable strategy. These approaches combine classical and post-quantum cryptographic methods to ensure business continuity and mitigate risks during the migration process~\cite{Löhr2020}{. Studies have evaluated the feasibility and practicality of these transitions, considering factors such as performance, compatibility, and security}~\cite{Wang2021}{. In this context, Giacon et al.}~\cite{giacon2018kem} {propose a method for constructing secure Key-Encapsulation Mechanisms (KEMs) by combining multiple KEMs, ensuring CCA-security as long as at least one component KEM remains secure. This approach leverages cryptographic hash functions and block ciphers, enhancing the resilience of cryptographic systems against potential vulnerabilities. Bindel et al.}~\cite{bindel2017transitioning} {investigate hybrid digital signature schemes to ensure both unforgeability and non-separability, while addressing challenges related to backward compatibility and managing larger certificates. Furthermore, Bindel et al.}~\cite{bindel2019hybrid} {focus on hybrid key encapsulation mechanisms (KEMs). They propose several combiner functions to integrate classical and post-quantum mechanisms, making the resulting KEM resistant to quantum attacks. Additionally, they introduce refined security notions for KEMs, ensuring a strong theoretical foundation for secure hybrid cryptography. Collectively, these efforts provide a theoretical foundation for secure hybrid key exchanges and hybrid digital signatures, significantly aiding the transition to quantum-resistant cryptographic systems.}

\subsection{Security Analysis of Quantum Communication}

{Security Analysis of Quantum Communication focuses on secure data transmission in quantum communication systems. Sheng et al. (2021)}~\cite{sheng2021research} {conduct a security risk assessment of a quantum private communication system, enhancing its security protection capabilities. Furthermore, Zheng et al. (2020)}~\cite{zheng2020quantum}{ propose a quantum risk assessment model based on two three-qubit GHZ states, contributing to the security of quantum communication networks.}

\subsection{Significance of Our Work}

Our research significantly contributes to the field of quantum security risk assessment and migration by addressing practical challenges in a comprehensive manner. While existing studies have primarily focused on estimating quantum threat timelines or assessing quantum threats across various domains, our work stands out for its holistic approach and practical guidance.
We offer a unique security risk assessment framework that covers the entire migration process, from pre-migration to post-migration stages. This framework aligns identified vulnerabilities with the well-established STRIDE threat model, providing actionable insights for mitigating quantum security risks.
In comparison to existing research, which often lacks comprehensive risk assessment frameworks and migration guidelines, our work provides a clear path for organizations to navigate the complexities of quantum security. 
Our framework goes beyond theoretical analysis by placing a strong emphasis on critical components like Public Key Infrastructure (PKI), network, and communication protocols, ensuring its practical relevance. { Table}  \ref{tab:comparison} {details how our approach offers a more in-depth analysis and utilizes a multi-level mitigation strategy compared to existing works.}
In summary, our work bridges the gap between theoretical analysis and real-world implementation, offering invaluable guidance to organizations facing the quantum threat. Our comprehensive approach sets it apart from existing works, making it a valuable resource for ensuring the resilience of modern communication in the era of quantum computing.

\section{Quantum-Safe Migration Risk Assessment Approach}\label{risk-analysis-approach}

\begin{figure*}[!ht]
 \begin{center} 
\resizebox{0.7\linewidth}{!}{
\begin{tikzpicture}[
 title/.style={minimum height=1cm,minimum width=4.2cm,font = {\large}},
 body/.style={draw,top color=white, bottom color=blue!20, rounded corners,minimum width=4.1cm,,minimum height=1cm,font = {\footnotesize}},
 typetag/.style={rectangle, draw=black!100, anchor=west}]
 \node (d0) [draw,top color=white, bottom color=blue!20, rounded corners,minimum height=1cm,minimum width=1.27\textwidth,font = {\large}] at (0,0) {NIST Cybersecurity Framework

};
 
\node (d3) [title,below=of d0.center] {Detect
};
 \node (d31) [body,below=of d3.west, typetag, xshift=2mm,yshift=-0.9cm,minimum height=2.1cm] {\begin{minipage}[c]{3.4cm}\centering
Anomalies and Events 
\end{minipage}};
 \node (d32) [body,below=of d31.west, typetag,yshift=-1.4cm,minimum height=2.1cm] {\begin{minipage}[c]{3.4cm}\centering
Security Continuous Monitoring
\end{minipage}};
 \node (d33) [body,below=of d32.west, typetag,yshift=-1.35cm,minimum height=2.1cm] {\begin{minipage}[c]{3.4cm}\centering
Detection Processes
\end{minipage}};
\node [top color=blue!40, bottom color=blue!40, rounded corners,minimum height=1cm,draw=black!100,fill opacity=0.1, fit={ (d3) (d31) (d32) (d33)}] {};
 
\node (d2) [title, left of=d3,xshift=-0.2\textwidth] {Protect};
 \node (d21) [body,below=of d2.west, typetag, yshift=-0.35cm, xshift=2mm] {\begin{minipage}[c]{3.4cm}\centering
Identity Management and Access Control
\end{minipage}};
 \node (d22) [body,below=1.17 of d21.west, typetag] {\begin{minipage}[c]{3.4cm}\centering
Awareness and Training
\end{minipage}};
 \node (d23) [body,below=1.17 of d22.west, typetag] {\begin{minipage}[c]{3.4cm}\centering
Data Security 
\end{minipage}};
\node (d24) [body,below=1.17 of d23.west, typetag] {\begin{minipage}[c]{3.4cm}\centering
Info. Protection Processes and Procedures
\end{minipage}};
\node (d25) [body,below=1.17 of d24.west, typetag] {\begin{minipage}[c]{3.4cm}\centering
Maintenance
\end{minipage}};
\node (d26) [body,below=1.17 of d25.west, typetag]{\begin{minipage}[c]{3.4cm}\centering
Protective Technology
\end{minipage}};
\node [top color=blue!40, bottom color=blue!40, rounded corners,minimum height=1cm,draw=black!100,fill opacity=0.1, fit={ (d2) (d21) (d22) (d23) (d24) (d25) (d26)}] {};

\node (d1) [title,left of=d2,xshift=-0.2\textwidth] {Identify
};
 \node (d11) [body,below=of d1.west, typetag, yshift=-0.35cm, xshift=2mm] {\begin{minipage}[c]{3.4cm}\centering
Asset Management
\end{minipage}};
 \node (d12) [body,below=1.17 of d11.west, typetag] {\begin{minipage}[c]{3.4cm}\centering
Business Environment
\end{minipage}};
 \node (d13) [body,below=1.17 of d12.west, typetag] {\begin{minipage}[c]{3.4cm}\centering
Governance
\end{minipage}};
\node (d14) [body,below=1.17 of d13.west, typetag] {\begin{minipage}[c]{3.4cm}\centering
Risk Assessment
\end{minipage}};
\node (d15) [body,below=1.17 of d14.west, typetag] {\begin{minipage}[c]{3.4cm}\centering
Risk Management Strategy
\end{minipage}};
\node (d16) [body,below=1.17 of d15.west, typetag]{\begin{minipage}[c]{3.4cm}\centering Supply Chain Risk Management
\end{minipage}};
\node [top color=blue!40, bottom color=blue!40, rounded corners,minimum height=1cm,draw=black!100,fill opacity=0.1, fit={ (d1) (d11) (d12) (d13) (d14) (d15) (d16)}] {}; 
 
\node (d4) [title, right of=d3,xshift=0.2\textwidth] {Respond
};
 \node (d41) [body,below=of d4.west, typetag, yshift=-0.6cm, xshift=2mm,minimum height=1.55cm] {\begin{minipage}[c]{3.4cm}\centering
Response Planning
\end{minipage}};
 \node (d42) [body,below=of d41.west, typetag, yshift=-0.8cm,minimum height=1.55cm] {\begin{minipage}[c]{3.4cm}\centering
Communications 
\end{minipage}};
 \node (d43) [body,below=of d42.west, typetag, yshift=-0.75cm,minimum height=1.55cm] {\begin{minipage}[c]{3.4cm}\centering
Analysis
\end{minipage}};
\node (d44) [body,below=of d43.west, typetag, yshift=-0.8cm,minimum height=1.55cm] {\begin{minipage}[c]{3.4cm}\centering
Mitigation Improvements 
\end{minipage}};
\node [top color=blue!40, bottom color=blue!40, rounded corners,minimum height=1cm,draw=black!100,fill opacity=0.1, fit={ (d4) (d41) (d42) (d43) (d44)}] {};

\node (d5) [title, right of=d4,xshift=0.2\textwidth] {Recover};
 \node (d51) [body,below=of d5.west, typetag, xshift=2mm,yshift=-0.9cm,minimum height=2.1cm] {\begin{minipage}[c]{3.4cm}\centering
Recovery Planning
\end{minipage}};
 \node (d52) [body,below=of d51.west, typetag,yshift=-1.4cm,minimum height=2.1cm] {\begin{minipage}[c]{3.4cm}\centering
Improvements 
\end{minipage}};
 \node (d53) [body,below=of d52.west, typetag,yshift=-1.35cm,minimum height=2.1cm] {\begin{minipage}[c]{3.4cm}\centering
Communications
\end{minipage}};
\node (d1to5)[top color=blue!40, bottom color=blue!40, rounded corners,minimum height=1cm,draw=black!100,fill opacity=0.1, fit={ (d5) (d51) (d52) (d53)}] {};








 \end{tikzpicture}
}
\caption{{NIST Cybersecurity Framework~\cite{cybersecurity2018framework}}}
\label{fig:risk-assesment}
\end{center}
\end{figure*}

\begin{figure*}[!hb]
 \begin{center} 
\resizebox{0.85\linewidth}{!}{
\begin{tikzpicture}[
 title/.style={minimum height=1cm,minimum width=2.5cm, font = {\scriptsize}},
 body/.style={draw,top color=white, bottom color=blue!20, rounded corners,minimum width=2.5cm,,minimum height=1.2cm,, font = {\scriptsize}},
 typetag/.style={rectangle, draw=black!100, anchor=west}]
 \node (d0) [body] at (0,0) {Threat Source
};
 \node (d01) [title,below=of d0,yshift=1cm] {\begin{minipage}[l]{2.5cm}
\scriptsize{{with
characteristics
 (e.g., capability, intent, and targeting for adversarial threats)}} 
\end{minipage}};
 \node (d1) [body,right=of d0, xshift=4.2mm] {Threat Event
};
 \node (d11) [title,below=of d1,yshift=1cm] {\begin{minipage}[1]{2.5cm}
\scriptsize{{with a sequence of actions, activities, or scenarios}} 
\end{minipage}};
 \node (d21) [body,right=of d1, xshift=5.5mm] {\begin{minipage}[c]{2.5cm}\centering
Vulnerability
\end{minipage}};
 \node (d22) [title,below=of d21,yshift=1.15cm] {\begin{minipage}[c]{2.5cm}\centering
\scriptsize{with severity in the context of} 
\end{minipage}};
 \node (d23) [body,below=of d22,yshift=1.15cm] {\begin{minipage}[c]{2.5cm}\centering
Predisposing Conditions
\end{minipage}};
\node (d24) [title,below=of d23,yshift=1.15cm] {\begin{minipage}[c]{2.5cm}\centering
\scriptsize{with pervasiveness}
\end{minipage}};
\node (d25) [body,below=of d24,yshift=1.15cm] {\begin{minipage}[c]{2.5cm}\centering
Security Controls (Planned/ Implemented)
\end{minipage}};
\node (d26) [title,below=of d25,yshift=1.15cm] {\begin{minipage}[c]{2.5cm}\centering
\scriptsize{with effectiveness}
\end{minipage}};
\node [top color=blue!40, bottom color=blue!40, rounded corners,minimum height=1cm,draw=black!100,fill opacity=0.1, fit={ (d21) (d22) (d23) (d24) (d25) (d26)}] {};

\node (d3) [body,right=of d21,xshift=5.4mm] {Adverse Impact};
 \node (d31) [title,below=of d3,yshift=1cm] {\begin{minipage}[1]{2.5cm}
\scriptsize{{with risk
as a combination of Impact and Likelihood}} 
\end{minipage}};
\node (d4) [body,right=of d3,xshift=4.2mm] {Organizational Risk};

 \node (d41) [title,below=of d4,yshift=1cm] {\begin{minipage}[1]{2.5cm}
\scriptsize{{to organizational operations, organizational assets, individuals, other organizations, and the nation.}} 
\end{minipage}};

 \draw [thick arrow]
 (1.27,0) -- (2.67,0)[set mark={\scriptsize{initiates}}];

 \draw [thick arrow]
 (5.19,0) -- (6.59,0)[set mark={\scriptsize{Exploits}}];
 \draw [thick arrow]
 (9.62,0) -- (11.02,0)[set mark={\scriptsize{Causes}}];
 \draw [thick arrow]
 (13.55,0) -- (14.95,0)[set mark={\scriptsize{Produces}}];

 \end{tikzpicture}}
\caption{Generic Risk Model to Conduct Risk Assessment According to NIST SP 800-30 guideline~\cite{blank2011guide}}
\label{fig:NIST-SP-800-30}
\end{center}
\vspace{-0.3cm}
\end{figure*} 

Risk management is an ongoing process that involves identifying, assessing, and responding to risk. The primary objective of risk management is to reduce risk to a manageable level for an organization. To effectively manage risk, organizations must understand the likelihood that an event will occur and the potential impacts that may result. With this information, organizations can determine an acceptable level of risk that aligns with their organizational objectives and express this as their risk tolerance. Likelihood refers to the probability of a vulnerability being exploited by attackers or other threat sources, and it typically indicates the probability of intent, capability, and targets based on a specific time frame. Impact level describes the severity of an attack that occurs when a threat source exploits a vulnerability. The risk score or level is determined by the likelihood of a threat event and the impact that would result if the event occurred. A risk score can provide insight into the probability and severity of an event that could compromise the organization.

Two critical factors in risk assessment are the assessment approach and the analysis approach. Generally, there are three assessment approaches: (1) quantitative, (2) qualitative, and (3) semi-quantitative. The quantitative approach is useful for cost-based analysis, but the numeric values may be difficult to interpret without additional context. This approach is often faster and less expensive in terms of time and implementation. In the qualitative approach, impact and likelihood are defined by levels (low, medium, high). However, the levels of risk may be too narrow, making it challenging to prioritize risks accurately within a given level. The third approach is a combination of the previous methods, called the semi-quantitative approach. This approach involves creating range bins of values for impact and likelihood while assigning bins to a specific level, as in the qualitative method. Using this approach, it is easier to compare two risks relatively at a certain level while creating a meaningful gap between others. The choice of approach depends on the application and expenses of the domain and organizations.

{Organizations commonly conduct risk assessments to prioritize threats by determining the likelihood and impact of exploiting vulnerabilities. In this study, we utilize the risk assessment methodology provided by the National Institute of Standards and Technology (NIST) to evaluate the risk associated with quantum-safe migration. To conduct a thorough risk assessment, we follow the guidelines specified in NIST SP 800-30}~\cite{blank2011guide}{, a key resource recommended by the NIST Cybersecurity Framework} ~\cite{cybersecurity2018framework}{. The framework aims to assist organizations in managing cybersecurity risks and maintaining the reliable functioning of critical infrastructure. The core of the framework consists of five functions that provide a high-level structure for managing cybersecurity risk: (1) Identify risks, assets, and vulnerabilities; (2) Protect critical systems through safeguards; (3) Detect cybersecurity events in a timely manner; (4) Respond to incidents to mitigate damage; and (5) Recover systems efficiently after an incident   (see Figure} \ref{fig:risk-assesment}).

{According to the guideline specified in NIST SP 800-30 (see Figure} \ref{fig:NIST-SP-800-30}){, conducting a risk assessment involves five tasks: (1) identifying threat sources and events that could cause security issues, (2) identifying vulnerabilities that may result from those threat sources and events, (3) determining the likelihood of their occurrence, (4) determining the magnitude of impact for each vulnerability, and (5) assessing risk values for the identified threats. In the following section, we describe how we are carrying out these tasks specifically to assess the risks associated with quantum-safe migration.}

\begin{table}[!h]
\caption{Established Criteria for Likelihood Levels}
\small
\resizebox{\linewidth}{!}{%
\begin{tabular}{|l|l|p{0.95\linewidth}|}
\hline
\multirow{51}{*}{Likelihood} &\multirow{14}{*}{\cellcolor{red!80}High} & 
\hspace{-0.45cm}
\begin{minipage}[t]{1.05\linewidth}
		 \begin{itemize}
		 \vspace{0.1cm}
		 \item Serious security flaws in the system, services or underlying infrastructure~\cite{jelacic2020security}. 
		 \item Known exploit exists and can be launched from the Internet, semi-trusted or untrusted networks~\cite{jelacic2020security,jelacic2017stride}.
		 \item The threat source is highly motivated and capable; no controls or countermeasures are in place to prevent or at least significantly delay the successful exercise of the vulnerability~\cite{jelacic2017stride,jelacic2020security}.
		 \item No loyalty workforce and no insider threat monitoring~\cite{jelacic2020security}.
		 \item Personnel without proper security training~\cite{jelacic2020security}.
		 \item Highly exposed to external systems~\cite{jelacic2017stride}.
		 \item Highly integrated classic and quantum-resistant systems, which results in quantum system exposure.
		 \\
		 \end{itemize}
\end{minipage} \\ \hhline{|~|-|-|} 
 &\multirow{19}{*}{\cellcolor{lemon!80}Medium } & 
\hspace{-0.45cm}\begin{minipage}[t]{1.05\linewidth}
		 \begin{itemize}
		 \vspace{0.1cm}
		 \item Limited security flaws in the system, services, or underlying infrastructure~\cite{jelacic2020security}. 
		 \item Known exploit exists have countermeasure and requires to be launched via the Internet; or known exploit exists, have no countermeasure and requires to be launched via physical access of a malicious user to the target system~\cite{jelacic2017stride,jelacic2020security}.
		 \item Limited threat-source motivation or limited threat-source capability; but controls or countermeasures are in place that may impede the successful exercise of the vulnerability~\cite{jelacic2017stride}.
		 \item Restricted loyal workforce and limited insider threat monitoring in place~\cite{jelacic2020security}.
		 \item Personnel with limited security training~\cite{jelacic2020security}.		 \item Medium exposed to external systems~\cite{jelacic2017stride}.
		 \item Custom classic and quantum-resistant systems segmentation, limited quantum system exposure.\\
		 \end{itemize}
		 \end{minipage} \\ \hhline{|~|-|-|} 
&\multirow{15}{*}{\cellcolor{green!60}Low} & 
\hspace{-0.45cm}\begin{minipage}[t]{1.05\linewidth}
		 \begin{itemize}
		 \vspace{0.1cm}
		 \item No known security flaws in the system, services, or underlying infrastructure~\cite{jelacic2020security}. 
		 \item No known exploit (or an exploit with countermeasure) exists, but a malicious user needs to have administrative or elevated privileges in the target system~\cite{jelacic2017stride,jelacic2020security}. 
		 \item No threat-source motivation, sufficient capabilities, and controls to prevent the vulnerability from being exercised are ineffective~\cite{jelacic2017stride}.		 
		 \item Loyal workforce, advanced insider threat monitoring~\cite{jelacic2020security}.
		 \item Personnel with well security training and knowledgeable about the latest threats~\cite{jelacic2020security}.
		 \item Slightly exposed to external systems~\cite{jelacic2017stride}.
		 \item Excellent classic and quantum-resistant systems segmentation, no quantum system exposure.\\
		 \end{itemize}
		 \end{minipage} \\ \hhline{|~|-|-|} \hline
\end{tabular}}
\label{table:likelihood}
\end{table}



\textit{Task 1. Identify Threat Sources and Events:}
{The first step in conducting a risk assessment is to identify threats posed by quantum attackers. Quantum threats, as discussed earlier, can trigger a sequence of actions, activities, or scenarios referred to as threat events. These events can be described in general terms (e.g., phishing, distributed denial-of-service), with more specific tactics, techniques, and procedures, or in highly detailed terms (e.g., specific information systems, technologies, organizations, roles, or locations)}~\cite{blank2011guide}{. This task involves analyzing attack vectors that compromise safety and security at various stages of migration. The analysis should also identify vulnerabilities at multiple levels, including (a) algorithmic, (b) certificate, and (c) protocol.}

\begin{table}[!h]
\caption{Established Criteria for Impact Levels}
\small
\resizebox{\linewidth}{!}{%
\begin{tabular}{|l|l|p{0.95\linewidth}|}
\hline
\multirow{49}{*}{Impact} &\multirow{14}{*}{\cellcolor{red!80}High} &

\hspace{-0.45cm}\begin{minipage}[t]{1.05\linewidth}
		 \begin{itemize}
		 \vspace{0.1cm}
		 \item Threat that might impact the loss of human lives or serious injuries~\cite{jelacic2017stride,jelacic2020security}.
		 \item Threat that might impact saviour infrastructure damage~\cite{jelacic2017stride,jelacic2020security}.
		 \item Threat that might impact the loss of consumers' personal data~\cite{jelacic2017stride}.
		 \item Threat that might impact significant financial damage to assets~\cite{jelacic2017stride}.
		 \item Threat that might impact the functionality of the whole system~\cite{jelacic2017stride}. 
		 \item Threat that might cause the system inoperative or unavailable~\cite{jelacic2017stride}. 
		 \item Threat that might cause prolonged critical service malfunctions~\cite{jelacic2020security}.\\
		 \end{itemize}
		 \end{minipage} \\ \hhline{|~|-|-|} 
		 
		 &\multirow{15}{*}{\cellcolor{lemon!80}Medium} &

\hspace{-0.45cm}\begin{minipage}[t]{1.05\linewidth}
		 \begin{itemize}
		 \vspace{0.1cm}
		 
		 \item Threats that might cause failure in real-time operation of the process control system~\cite{jelacic2017stride}.
		 \item Threat that might impact limited infrastructure damage~\cite{jelacic2017stride,jelacic2020security}.
		 \item Threats that might cause unwanted functionality performing~\cite{jelacic2017stride}.
		 \item Threats that might significantly impact the satisfaction of clients, expose customers' personal data/secrets or damage the company's reputation~\cite{jelacic2017stride}.
		 \item Threats that might disclose, violet integrity or availability of logs or any records of the actions that occur in the system~\cite{jelacic2017stride}.
		 \item Threats that might cause fines and penalties by regulatory bodies and government agencies~\cite{jelacic2017stride}.
		 \item Threats that might cause prolonged non-critical service malfunction~\cite{jelacic2020security}.\\
		 \end{itemize}
		 \end{minipage} \\ \hhline{|~|-|-|} 
 		 &\multirow{7}{*}{\cellcolor{green!60}Low} &
 		 \hspace{-0.45cm}\begin{minipage}[t]{1.05\linewidth}
		 \begin{itemize}
		 \vspace{0.1cm}
 		 \item Threats that might cause delay, limited unavailability, or failure of non-critical services~\cite{jelacic2017stride,jelacic2020security}.
 \item Threats that might cause revealing of (a) non-critical information or (b) the information with non-direct financial impact or adverse impact on company image~\cite{jelacic2017stride,jelacic2020security}.\\
		 \end{itemize}
		 \end{minipage} \\ \hline
\end{tabular}%
}
\label{table:impact}
\end{table}
\textit{Task 2. Identify Vulnerabilities and Predisposing Conditions:}
In each stage of migration, there exist several vulnerabilities that attackers exploit to compromise a system. These vulnerabilities are presented at various levels, including (a) algorithmic, (b) certificate, and (c) protocol, which are described in the next sections. By analyzing vulnerabilities, we can identify the source of the threat event and attempt to implement mitigations to prevent future attacks.


\textit{Task 3. Determine Likelihood of Occurrence:}  {In this task, we employ a qualitative approach to evaluate the likelihood of vulnerabilities being exploited by the quantum attacker before, through, or after migration.  We establish a set of evaluation criteria and categorize the likelihood into three levels: \textit{Low (L)}, \textit{Medium (M)}, and \textit{High (H)}. The criteria used for this assessment are detailed in Table} ~\ref{table:likelihood}{, adapted from the criteria presented in}~\cite{jelacic2017stride, jelacic2020security, 10247152}.



\textit{Task 4. Determine Magnitude of Impact:} {The impact level quantifies the expected harm resulting from unauthorized disclosure, modification, or destruction of information, or loss of information system availability due to quantum threats}~\cite{nist2010800}{. To evaluate this impact on cryptosystems, we've established criteria mirroring the potential harm these threats pose. Similar to likelihood assessment, we use a qualitative approach with three impact levels: \textit{Low (L)}, \textit{Medium (M)}, and \textit{High (H)}. The detailed criteria for impact assessment are presented in Table}~\ref{table:impact}, adapted from \cite{jelacic2017stride, jelacic2020security,10247152} {to measure impact severity.}

\textit{Task 5. Assess Risk:} {The final objective is to determine the risk associated with each threat event. We define risk as the product of likelihood and impact, commonly visualized in a risk matrix. This study utilizes three risk levels: high, medium, and low. A high-risk level signifies multiple harmful consequences or catastrophic outcomes, a medium-risk level indicates severe results, and a low-risk level denotes a limited or negligible adverse effect. The risk score is calculated by multiplying the likelihood and impact levels. Figure}~\ref{fig:risk-matrix} {depicts the risk matrix, demonstrating how the risk level is derived from likelihood and impact assessments.}

\begin{figure}[!htbp]
\renewcommand{\arraystretch}{1.8}
 \scriptsize
{
{%
\subfloat{}{\begin{math}
 \qquad \raisebox{-1\normalbaselineskip}{Likelihood$\ \left\{\rule{0pt}{3\normalbaselineskip}\right.$}
\hspace{-0.15cm}
\begin{tabular}{p{0.1cm}}\\
\multicolumn{1}{l}{Low} \\ 
\multicolumn{1}{l}{Medium} \\
\multicolumn{1}{l}{High}
\end{tabular}
\end{math}
}
\hspace{-0.3cm}
\begin{math}
 \overbrace{\
\begin{tabular}{p{1.5cm}p{1.5cm}p{1.5cm}}
Low & Medium & High \\ \hline
\multicolumn{1}{|l|}{\cellcolor{green!60}Low} & \multicolumn{1}{l|}{\cellcolor{green!60}Low} & \multicolumn{1}{l|}{\cellcolor{lemon!80}Medium} \\ \hline
\multicolumn{1}{|l|}{\cellcolor{green!60}Low} & \multicolumn{1}{l|}{\cellcolor{lemon!80}Medium} & \multicolumn{1}{l|}{\cellcolor{red!80}High} \\ \hline
\multicolumn{1}{|l|}{\cellcolor{lemon!80}Medium} & \multicolumn{1}{l|}{\cellcolor{red!80}High} & \multicolumn{1}{l|}{\cellcolor{red!80}High} \\ \hline
\end{tabular}
}^{\mbox{Impact}}
\end{math}
 }}
 \caption{Qualitative Risk Assessment based on Likelihood and Impact Levels}
 \label{fig:risk-matrix}
\end{figure}

\subsection{ STRIDE: A Threat Model To Assess Risk}

We have developed a risk assessment process and established criteria for evaluating threat events that result from exploiting vulnerabilities caused by quantum threats on cryptosystems. This section introduces the STRIDE threat model, a well-known model that can help identify security vulnerabilities and mitigate risks~\cite{van2021descriptive,jelacic2017stride}. It is important to note that we will use both the established criteria and STRIDE mapping to evaluate vulnerabilities obtained from the literature and assess the risk of vulnerabilities throughout the entire migration process.

To protect against cyber attacks, researchers have created various threat model frameworks to identify, assess, and prioritize potential threats to an organization's assets, such as information, technology, or physical infrastructure~\cite{shostack2014threat}. Threat model frameworks are also crucial for the risk assessment process. While threat model frameworks identify threat events and vulnerabilities, risk assessment techniques rank risks related to those events and assist security teams in securing their systems. There are several widely recognized threat model frameworks, including STRIDE, DREAD, PASTA, TRIKE, and VAST~\cite{shevchenko2018threat}.  STRIDE stands out as one of the most mature and widely adopted threat modeling frameworks, particularly focusing on system design. This framework utilizes Design Data Flow Diagrams (DFD) to delineate the system's entities, events, and boundaries~\cite{sion2018solution}. The subsequent step involves identifying threats based on established threat names using the acronym STRIDE, representing Spoofing, Tampering, Repudiation, Information Disclosure, Denial of Service (DoS), and Elevation of Privilege (EoP). STRIDE has found application in modeling threats across diverse domains, including energy systems~\cite{zografopoulos2021cyber}, automotive systems~\cite{karahasanovic2017adapting}, and industrial control systems~\cite{jelacic2017stride,jelacic2020security}. In this study, we leverage the STRIDE threat model to map vulnerabilities within the STRIDE domain, offering researchers a more comprehensive understanding of each vulnerability.\\

{In the following sections, we present our framework for evaluating security risks in each migration stage across three distinct levels: algorithmic, certificate, and protocol. Recognizing the cascading nature of vulnerabilities from algorithm to certificate to protocol, our analysis prioritizes starting with the algorithmic level, followed by the certificate and protocol levels. We delve into the security threats introduced by quantum computing at various stages of migration. Subsequently, we conduct a thorough examination of potential attack vectors that jeopardize safety and security at each level and stage. This analysis encompasses identifying diverse vulnerabilities exploitable by quantum attackers, mapped using the STRIDE threat model. Finally, we assess the risk of migration for each level at each migration stage by applying custom criteria to the identified aspects and conducting detailed analyses.}



\section{Quantum Migration Threat Analysis and Risk Assessment for Algorithmic Level}\label{sec:ALA}
In order to analyze and assess the risks involved in quantum-safe migration, we start our evaluation with algorithmic-level analysis. We classify the algorithmic level into three broad stages: pre-migration, through-migration, and post-migration. For each stage of migration, we identify different vulnerabilities that a quantum attacker could exploit, evaluate QC threats using the STRIDE threat model, and assess the overall risk of vulnerabilities throughout the entire migration process.
We evaluate the algorithmic level into three broad stages: (1) pre-migration, (2) through-migration, and (3) post-migration. We identify the possible QC threats based on the STRIDE threat model and assess the risk in all three stages.

\subsection{Pre-Migration Algorithmic Level Analysis and Risk Assessment}
\begin{figure*}[!t]
\centering
\begin{minipage}{.48\textwidth}
  \centering
  \includegraphics[trim=0.2cm 0.1cm 0.2cm 0.1cm, clip=true, width=\linewidth, height=0.6\linewidth,frame]{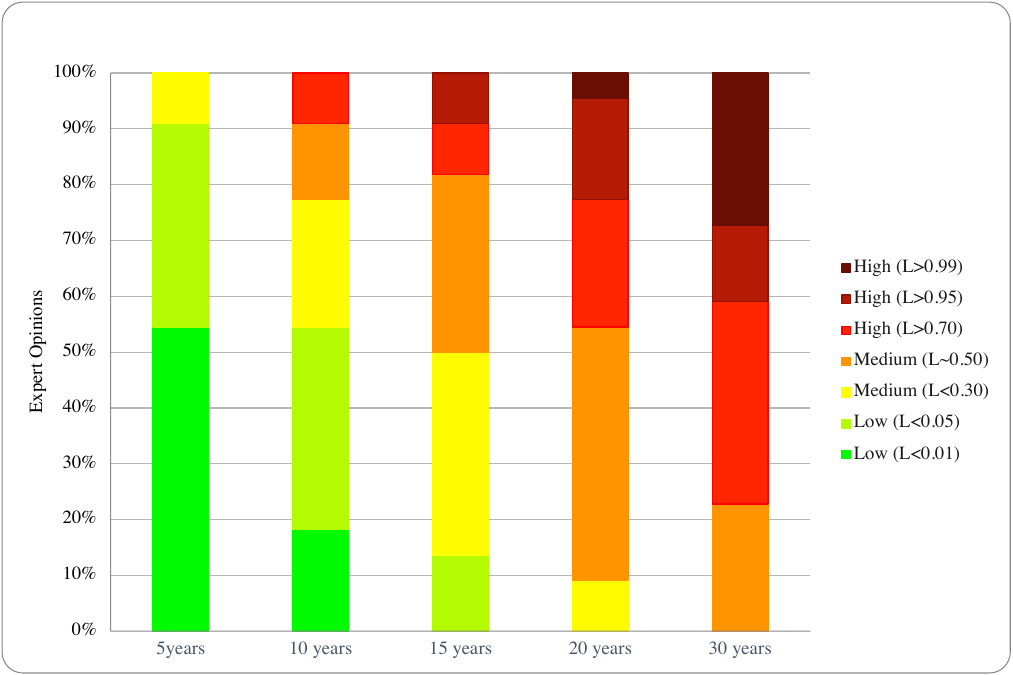}
  \captionof{figure}{Cumulative Expert Opinions Related to Quantum Threat to Classic Cryptography}
  \label{fig:chart1}
\end{minipage}%
\hfill
\begin{minipage}{.48\textwidth}
  \centering
\includegraphics[trim=0.2cm 0.1cm 0.1cm 0.6cm, clip=true, width=\linewidth, height=0.6\linewidth,frame]{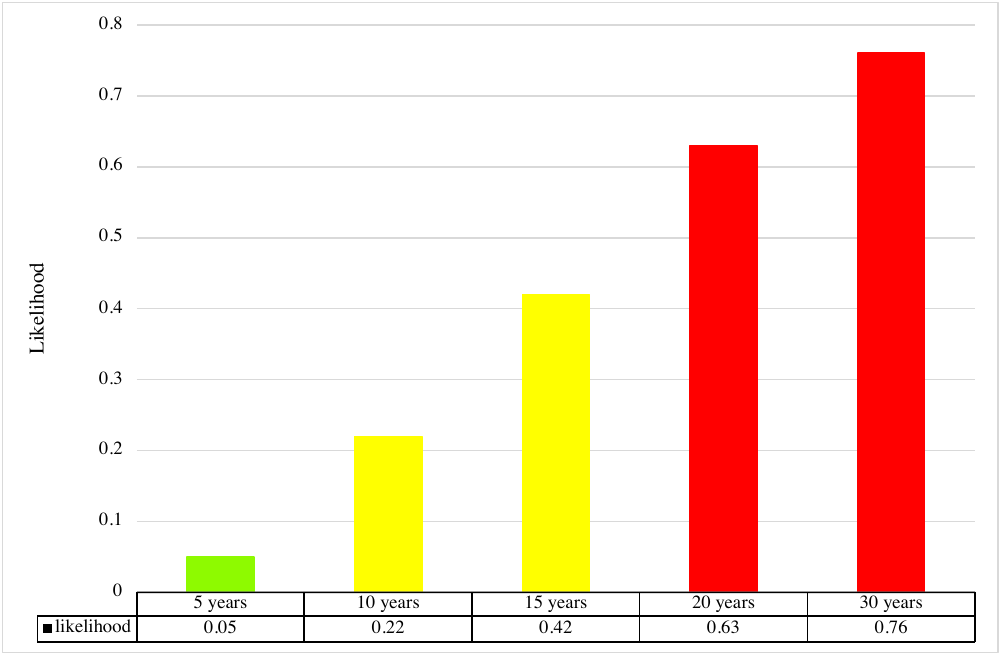}
  \captionof{figure}{Expected Likelihood of Quantum Threat for Classic Cryptography Within 30 Years}
    \label{fig:chart2}
\end{minipage}
\end{figure*}

The standard cryptographic algorithms currently used to provide security in various applications and communications are classical algorithms, which can be either asymmetric or symmetric. These algorithms can be broken or weakened by a quantum attacker equipped with quantum computers with sufficient resources. Specifically, the hard problems upon which the asymmetric cryptography commonly used today for secure communication relies will no longer be considered hard anymore. By using existing quantum algorithms such as Shor's algorithm, a quantum attacker has the potential to break the security of currently used asymmetric cryptographic algorithms.
Similarly, the level of security provided by symmetric cryptographic systems will also be affected. By using quantum algorithms such as Grover's algorithm and Brassard-Hoyer-Tapp (BHT) algorithm, a quantum attacker can weaken the security of symmetric cryptographic algorithms and communication mechanisms. Therefore, this section investigates the strengths and vulnerabilities in classical cryptographic algorithms before migration to potential quantum-safe cryptographic algorithms.

\subsubsection{Determine the Expected Likelihood of Quantum Threat} To understand the risks associated with quantum migration, it is imperative to
predict the emergence of quantum computers and the resultant risks to classical cryptosystems. Our analysis examines the timeline for quantum computers to appear within the next 5 to 30 years. This analysis is built on a cumulative likelihood of significant quantum threats to classical cryptosystems. Figure~\ref{fig:chart1} summarizes this evolution, incorporating
insights from multiple quantum experts regarding the quantum threat timeline~\cite{mosca20222021}. The “quantum threat'' is defined as
the probability of breaking RSA-2048 within 24 hours using a quantum machine. These assessments can be extended to
evaluate the likelihood of breaking other cryptographic algorithms based on their quantum security level.



For each period defined in Figure~\ref{fig:chart1} (i.e., 5, 10, 15, 20, and 30 years), the bar represents different fractions of expert opinions shown by different colors. Each color represents a fraction of expert opinions that agreed upon the same likelihood interval. For example, for the period of 5 years, 54.54\% of experts believe that quantum computers threaten classic cryptosystems with a likelihood less than 0.01, 36.36\% agreed on the likelihood of approximately 0.05, and 9.09\% of them polled to the likelihood less than 0.30 (see the legend of Figure~\ref{fig:chart1}).

To evaluate the \textit{``expected likelihood of the quantum threat for classical cryptosystems''} over a period (i.e., 5, 10, 15, 20, and 30 years), we accumulate different intervals for \textit{``likelihood of the quantum threat for classic cryptosystems''}, which are predicted by different experts participated in the poll. In our approach, for each period $period_j$ (e.g., 5 years), we calculate the expected likelihood of prediction (i.e., $E_{period_j}[likelihood]$) by multiplying all possible \textit{``agreed-upon likelihoods of predictions''} (i.e., $likelihood_{period_j} (\omega_i)$) for that period by the probability of those predictions (i.e., $Pr_{period_j} (\omega_i)$) and then summing them up.
\begin{align*}
 E_{period_j}[likelihood]=\sum_{\omega_i\subseteq [0,1]} likelihood_{period_j} (\omega_i) \times Pr_{period_j} (\omega_i)
\end{align*}

where $\omega_i$s are subsets of $[0,1]$ such that the union of all of them will be equal to $[0,1]$ (i.e., $\bigcup_{i=1}^{n} {\omega_i}=[0,1]$), and $Pr_{period_j} (\omega_i)$ for each period ${period_j}$ is evaluated via the fraction of expert opinions agreed upon prediction $\omega_i$ for that period, compared to the total number of predictions for that period.
By calculating the expected likelihood for each period, we would be able to predict the most possible likelihood for each period. In this way, the calculated expected likelihood of the quantum threat within 5, 10, 15, 20, and 30 years will be 0.05, 0.22, 0.42, 0.63, and 0.76, respectively.

To analyze the likelihood of the quantum threat to classic cryptosystems in a qualitative manner, we categorize them into three different levels: low, medium, and high (shown by different colors).
Based on the description mentioned above, we considered different qualitative levels for the likelihood of having quantum computers threaten classic cryptosystems. As shown in Figure~\ref{fig:chart2}, the expected likelihood of a quantum threat to classic cryptography within the period of less than 10 years is low, within the period of 15 years is medium, and within the period of 20 years or beyond is high. For our evaluation, we consider the medium qualitative level within 15 years (see Figure~\ref{fig:chart2}) for the likelihood of quantum computers threatening classic cryptosystems. This assumption can be easily changed for other periods.

\begin{figure}[!htbp]
 \centering
  \includegraphics[trim=0.2cm 0.1cm 0.2cm 0.1cm, clip=true, width=\linewidth, height=0.6\linewidth,frame]{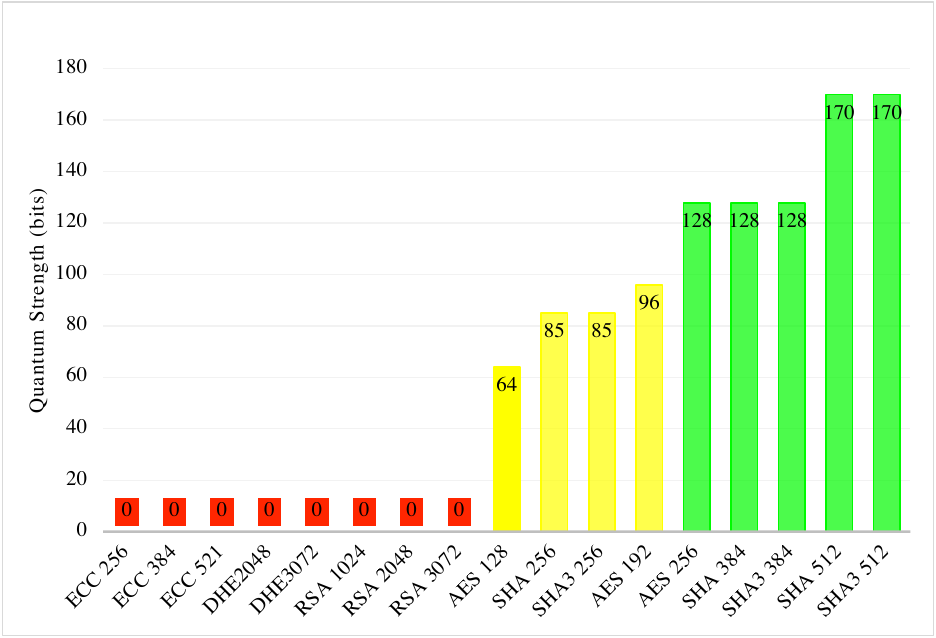}
  \caption{Expected Impact of Quantum Threat for Classic Cryptography}
  \label{fig:classic-impact}
\end{figure}


\subsubsection{Determine the Magnitude of Impact for Quantum\break Threats on Pre-Migration Algorithms} For pre-migration algorithmic level risk assessment, it's crucial to evaluate the impact of quantum threats on classic cryptographic algorithms. {Quantum security strength provides a measure of how well these algorithms can withstand attacks from quantum computers}~\cite{NISTpostquantum}{. This measure categorizes security resilience based on computational resources needed to break the algorithm, helping to assess vulnerabilities accurately.}

We determine the impact by considering the quantum security strength of each classic algorithm, as shown in Figure~\ref{fig:classic-impact}. High impact is indicated when the quantum strength of an algorithm is less than 64 bits, while low impact is when it's 128 bits or more. Medium impact falls between these ranges. This assessment considers both the likelihood and potential impact of quantum threats, as detailed in  Table~\ref{tab:Pre-Migration-Alg}.

\subsubsection{Evaluate the Risk of Quantum Threats on Pre-Migration Algorithms}
Table~\ref{tab:Pre-Migration-Alg} presents a summary of our findings related to algorithmic level analysis and risk assessment before migrating to a quantum-safe cryptographic state. 
 {It outlines both the classical and quantum security strengths of the algorithms, their vulnerabilities, and the quantum threats, categorized using the STRIDE model. The assessment evaluates the likelihood (L), impact (I), and overall risk (R) posed by quantum adversaries. Additionally, it identifies potential quantum-resistant alternatives.} This algorithmic-level analysis can assist in identifying the security strength of currently employed cryptographic algorithms and discovering mitigation measures in existing classical algorithms more quickly.

\begin{table*}[!hbpt]
\caption{Pre-Migration Algorithmic Level Analysis and Risk Assessment}
\small
\renewcommand{\arraystretch}{1.2}
\label{tab:Pre-Migration-Alg}
\resizebox{\textwidth}{!}{%
\begin{tabular}
{|l|l|l|l|l|l|p{0.22\linewidth}|p{0.5\linewidth}|l|l|l|p{0.23\linewidth}|}
\hline

 
\multirow{2}{*}{\textbf{Crypto Type}} & \multirow{2}{*}{\textbf{Algorithms}} & \multirow{2}{*}{\textbf{Variants}} & \multirow{2}{*}{\textbf{Key Length (bits)}} & \multicolumn{2}{l|}{\textbf{Strengths (bits)}}&\multirow{2}{*}{\textbf{Vulnerabilities}} &\multirow{2}{*}{\textbf{Quantum  Threats (STRIDE)}}& \multirow{2}{*}{\textbf{L}} & \multirow{2}{*}{\textbf{I}} & \multirow{2}{*}{\textbf{R}} & \multirow{2}{*}{\textbf{Possible QC-resistant Solutions}} \\ \cline{5-6}
  &  & & & \multicolumn{1}{l|}{\textbf{Classic}} & \textbf{Quantum} & & & & & & \\ \hline
\multirow{9}{*}{Asymetric} & \multirow{3}{*}{ECC~\cite{turner2010use}} & ECC 256 & 256 & \multicolumn{1}{l|}{128} & 0 & \multirow{8}{*}{{\begin{minipage}{\linewidth}
Broken by Shor's Algorithm~\cite{shor1994algorithms}.
\end{minipage}}} & &\med &\high&\high& \multirow{8}{*}{{\begin{minipage}{\linewidth}Algorithms presented in Table~\ref{tab:Post-Migration-Alg}.\end{minipage}}}\\\cline{3-6}\cline{9-11}
 & & ECC 384 & 384 & \multicolumn{1}{l|}{256} & 0 & & \multirow{6}{*}{{\begin{minipage}{\linewidth}
{For digital signature:}
\begin{myBullets}
\item {Spoofing: Shor's Algorithm allows forging of digital signatures.}
\item {Tampering: Integrity checks can be bypassed due to signature forgery.}
\item {Repudiation: Valid signatures can be forged, denying the origin of the message.}
\end{myBullets}
{For KEM/ENC:}
\begin{myBullets}
\item {Information Disclosure: KEM/ENC algorithms can be broken, revealing encrypted data.}
\end{myBullets}
\end{minipage}}}&\med& \high &\high& \\\cline{3-6}\cline{9-11}
 & & ECC 521 & 521 & \multicolumn{1}{l|}{256} & 0 & & &\med & \high&\high& \\ \cline{2-6}\cline{9-11}
 & \multirow{2}{*}{FFDHE~\cite{gillmor2016negotiated}} & DHE2048 & 2048 & \multicolumn{1}{l|}{112} & 0 & &&\med & \high&\high &\\\cline{3-6}\cline{9-11}
 & & DHE3072 & 3072 & \multicolumn{1}{l|}{128} & 0 & & &\med& \high& \high&\\ \cline{2-6}\cline{9-11}
 & \multirow{3}{*}{RSA~\cite{moriarty2016pkcs}} & RSA 1024 & 1024 & \multicolumn{1}{l|}{80} & 0 & 
& & \med& \high& \high & \\\cline{3-6}\cline{9-11}
 & & RSA 2048 & 2048 & \multicolumn{1}{l|}{112} & 0 & & &\med& \high &\high& \\\cline{3-6}\cline{9-11}
 & & RSA 3072 & 3072 & \multicolumn{1}{l|}{128} & 0 & & &\med& \high &\high& \\ \hline
\multirow{4}{*}{Symmetric} & \multirow{3}{*}{AES~\cite{schaad2003use}} & AES 128 & 128 & \multicolumn{1}{l|}{128} & 64 & \multirow{3}{*}{{\begin{minipage}{\linewidth}
Weakened by Grover's Algorithm~\cite{grover1996fast}.
\end{minipage}}}
& \multirow{3}{*}{{\begin{minipage}{\linewidth}
\begin{myBullets}
\vspace{0.2cm}
\item {Information Disclosure: Grover's algorithm reduces the effective key length, making brute-force attacks feasible.}
\end{myBullets}
\end{minipage}}}&\med& \med & \med & \multirow{3}{*}{{\begin{minipage}{\linewidth}Larger key sizes are needed.\end{minipage}}} \\
\cline{3-6}\cline{9-11}
 & & AES 192 & 192 & \multicolumn{1}{l|}{192} & 96 & & &\med & \med &\med & \\\cline{3-6}\cline{9-11}
 & & AES 256 & 256 & \multicolumn{1}{l|}{256} & 128 & & &\med& \low&\low& \\ 
 \cline{2-12}
 & \multirow{3}{*}{SHA2~\cite{eastlake2011us}} & SHA 256 & - & \multicolumn{1}{l|}{128} & {85} & \multirow{6}{*}{{\begin{minipage}{\linewidth}
Weakened by Brassard et al.'s Algorithm~\cite{brassard1997quantum}.
\end{minipage}}} & \multirow{6}{*}{{\begin{minipage}{\linewidth}
\begin{myBullets}
\item {Spoofing: Fake hash values can be created.}
\item {Tampering: Data integrity can be compromised by finding collisions.}
\end{myBullets}
\end{minipage}}}&\med& \med &\med & \multirow{6}{*}{{\begin{minipage}{\linewidth}Larger hash values are needed.\end{minipage}}} \\\cline{3-6}\cline{9-11} 
 & & SHA 384 & - & \multicolumn{1}{l|}{192} & 128 & & &\med & \low&\low& \\\cline{3-6}\cline{9-11}
 & & SHA 512 & - & \multicolumn{1}{l|}{256} & 170 & & & \med& \low&\low& \\ \cline{2-6}\cline{9-11}
 & \multirow{3}{*}{SHA3~\cite{eastlake2011us}} & SHA3 256 & - & \multicolumn{1}{l|}{128} & {85} && &\med& \med &\med & \\\cline{3-6}\cline{9-11} 
 & & SHA3 384 & - & \multicolumn{1}{l|}{192} & 128 & & &\med & \low&\low & \\\cline{3-6}\cline{9-11}
 & & SHA3 512 & - & \multicolumn{1}{l|}{256} & 170 & & &\med& \low&\low & \\ \hline
\end{tabular}%
}
\end{table*}

 \subsection{Through-Migration Algorithmic Level Analysis and Risk Assessment}

\begin{table*}[!htbp]
\caption{Through-Migration Algorithmic Level Analysis}
\scriptsize
\label{tab:Through-Migration-Alg}
\resizebox{\textwidth}{!}{%

\begin{tabular}{|p{0.1\linewidth}|p{0.1\linewidth}|p{0.15\linewidth}|p{0.1\linewidth}|p{0.3\linewidth}|p{0.3\linewidth}|p{0.3\linewidth}|}
\hline
\multirow{2}{*}{\textbf{Migration Strategy}} & \multirow{2}{*}{\textbf{Crypto Types}} & \multirow{2}{*}{\begin{minipage}{\linewidth}
\textbf{Combination Approaches}$^*$
\end{minipage}}& \multirow{2}{*}{\textbf{Mechanisms}} & 
\multirow{2}{*}{ \textbf{Pros}} & \multirow{2}{*}{\textbf{Cons}} & \multirow{2}{*}{\textbf{Quantum Threats (STRIDE)}} \\ 
 & & & & & & \\\hline
\multirow{90}{*}{Hybrid} & \multirow{60}{*}{\begin{minipage}{\linewidth}
    {KEM/ENC \cite{barker2017recommendation,steblia2020hybrid,giacon2018kem,giron2022post}} 
\end{minipage}}& {\begin{minipage}{\linewidth}
Concatenation
\end{minipage}} & Concatenation~\cite{barker2017recommendation}& 
{\begin{minipage}{\linewidth}
\begin{myBullets}
\vspace{0.1cm}
\item Supporting lightweight operations, simple logic, and easy implementation
\\
\end{myBullets}
\end{minipage}}
&
{\begin{minipage}{\linewidth}
\begin{myBullets}
\vspace{0.1cm}
\item Enabling the hybrid approach, a PQ key must be included in the code; as a result, the FIPS 140 validation code may need to be changed.
\item Only concatenating shared secret key's components (by the approach) does not protect its integrity,
\item Providing security proofs only for classical
adversaries~\cite{giron2022post}.
\\
\end{myBullets}
\end{minipage}}
& 
{\begin{minipage}{\linewidth}
\begin{myBullets}
\vspace{0.1cm}
\item Tampering (concatenation not supporting integrity),
\item Information Disclosure (when none of the algorithms used in the hybrid approach are secure against information disclosure).
\\
\end{myBullets}
\end{minipage}}\\ \cline{3-7} 
 & & \multirow{6}{*}{\begin{minipage}{\linewidth}
KDF
\end{minipage}} & Concat-KDF~\cite{campagna2020security,whyte2017quantum,kiefer2018hybrid,schanck2015circuit}
 & 
 {\begin{minipage}{\linewidth}
\begin{myBullets}
\vspace{0.1cm}
\item The approach combines all the outputs of key exchange through a single KDF. With the potentially longer keys, keeping the number of KDF applications the same~\cite{steblia2020hybrid}.
\item Reducing the effectiveness of brute-force attempts.
\\
\end{myBullets}
\end{minipage}}
 & {\begin{minipage}{\linewidth}
\begin{myBullets}
\vspace{0.1cm}
\item Enabling the hybrid approach, a PQ key must be included in the code; as a result, the FIPS 140 validation code may need to be changed.
\item In the random oracle model, Concat-KDF is IND-CPA secure as long as at least one KEM is OW-CPA secure. 
\item In the standard model, Concat-KDF is IND-CPA secure as long as at least one KEM is IND-CPA secure and the KDF is a weakly secure key derivation function for the appropriate source of key material~\cite{campagna2020security}.
\\
\end{myBullets}
\end{minipage}} & 
{\begin{minipage}{\linewidth}
\begin{myBullets}
\vspace{0.1cm}
\item Tampering (if the security of KDF is broken in a security model),
\item Information Disclosure (when none of the signature algorithms used in the hybrid approach are secure against information disclosure).
\\
\end{myBullets}
\end{minipage}} \\ \cline{4-7} & & & Cascade-KDF~\cite{campagna2020security}
 & 
{\begin{minipage}{\linewidth}
\begin{myBullets}
\vspace{0.1cm}
\item The approach produces a shared secret using a cascade that accepts a single secret in each iteration and combines all the outputs through the iteration of a KDF. With the potentially longer keys, it keeps the number of KDF applications the same~\cite{steblia2020hybrid}.
\item Reducing the effectiveness of brute-force attempts.\\
\end{myBullets}
\end{minipage}}
&
 {\begin{minipage}{\linewidth}
\begin{myBullets}
\vspace{0.1cm}
\item Enabling the hybrid approach, a PQ key must be included in the code; as a result, the FIPS 140 validation code may need to be changed.
\item Cascade-KDF is secure only in the random oracle model~\cite{campagna2020security},
\item Cascade-KDF is secure in the standard model by assuming KDF is a weakly secure key derivation function for the appropriate source of key material~\cite{campagna2020security}.
\\
\end{myBullets}
\end{minipage}}
& 
{\begin{minipage}{\linewidth}
\begin{myBullets}
\vspace{0.1cm}
\item Tampering (if the security of KDF is broken in a security model),
\item Information Disclosure (when none of the algorithms used in the hybrid approach are secure against information disclosure).
\\
\end{myBullets}
\end{minipage}} \\ \cline{3-7}
 & & \multirow{12}{*}{\begin{minipage}{\linewidth}
PRF
\end{minipage}} & Dual-PRF~\cite{bindel2019hybrid,dowling2020many} &
{\begin{minipage}{\linewidth}
\begin{myBullets}
\vspace{0.1cm}
 \item Keeping IND-CCA security if at least one of the two KEMs is IND-CCA secure and PRF components of Dual-PRF are IND-CCA secure~\cite{bindel2019hybrid}, 
\item Providing security proofs for classical, partial, and fully quantum adversaries.\\
\end{myBullets}
\end{minipage}}
& 
{\begin{minipage}{\linewidth}
\begin{myBullets}
\vspace{0.1cm}
\item Extra preprocessing for the first key~\cite{bindel2019hybrid}.\\
\end{myBullets}
\end{minipage}}
& 
{\begin{minipage}{\linewidth}
\begin{myBullets}
\vspace{0.1cm}
\item Information Disclosure (when none of the algorithms used in the hybrid approach are secure against information disclosure).
\\
\end{myBullets}
\end{minipage}} \\ \cline{4-7} 
 & & & Nested-Dual-PRF~\cite{bindel2019hybrid} & 

{\begin{minipage}{\linewidth}
\begin{myBullets}
\vspace{0.1cm}
\item Keeping IND-CCA security if at least one of the two KEMs is IND-CCA secure and PRF components of Nested-Dual-PRF are IND-CCA secure~\cite{bindel2019hybrid},
\item Providing security proofs for classical, partial, and fully quantum adversaries.\\
\end{myBullets}
\end{minipage}}
&
{\begin{minipage}{\linewidth}
\begin{myBullets}
\vspace{0.1cm}
\item Even more extra preprocessing is required for the first key~\cite{bindel2019hybrid}.
\\
\end{myBullets}
\end{minipage}}
& 
{\begin{minipage}{\linewidth}
\begin{myBullets}
\vspace{0.1cm}
\item Information Disclosure (when none of the algorithms used in the hybrid approach are secure against information disclosure).
\\
\end{myBullets}
\end{minipage}} \\ \cline{4-7} 
 & & & Split-key-PRF~\cite{giacon2018kem} & 
{\begin{minipage}{\linewidth}
\begin{myBullets}
\vspace{0.1cm}
\item Keeping IND-CCA security if the core function of the parallel combiner is split-key pseudorandom.~\cite{giacon2018kem},
\item Security of the model is offsetting by its cost in terms of efficiency when employed in a parallel combiner~\cite{giacon2018kem}.
\\
\end{myBullets}
\end{minipage}} 
 & 
 {\begin{minipage}{\linewidth}
\begin{myBullets}
\vspace{0.1cm}
\item Providing security proofs only for classical
adversaries~\cite{giron2022post}.
\\
\end{myBullets}
\end{minipage}}
& 
 {\begin{minipage}{\linewidth}
\begin{myBullets}
\vspace{0.1cm}
\item Information Disclosure (when none of the algorithms used in the hybrid approach are secure against information disclosure).
\\
\end{myBullets}
\end{minipage}} \\ \cline{3-7}
 & & \multirow{14}{*}{\begin{minipage}{\linewidth}
XOR Combination
\end{minipage}} 
& XOR~\cite{giacon2018kem,ghosh2015post,brendel2019breakdown}
 & 
 {\begin{minipage}{\linewidth}
\begin{myBullets}
\vspace{0.1cm}
\item Supporting lightweight operations, simple logic, and easy implementation.\\
\end{myBullets}
\end{minipage}}& {\begin{minipage}{\linewidth}\begin{myBullets}
\vspace{0.1cm}
\item Since XOR is reversible, the attacker can compromise one of the subkeys. If an attacker finds out any one of the subkeys and knows the corresponding salt, then they can recover the master key.
\item Vulnerability against related-key attacks on the cipher.
\item Preserving only IND-CPA security for~\cite{giacon2018kem}, one-way authenticated key exchanges
 (1W-AKE) for~\cite{ghosh2015post},
and breakdown-Resilient AKE security for~\cite{brendel2019breakdown},
\item Providing security proofs only for classical adversaries~\cite{ghosh2015post,giacon2018kem}.\\
\end{myBullets}
\end{minipage}}
& 
{\begin{minipage}{\linewidth}
\begin{myBullets}
\vspace{0.1cm}
\item Tampering (XOR not providing integrity),
\item Information Disclosure (when none of the algorithms used in the hybrid approach are secure against information disclosure).
\\
\end{myBullets}
\end{minipage}} \\ \cline{4-7} 
 & & & XOR-then-MAC~\cite{bindel2019hybrid} &

{\begin{minipage}{\linewidth}
\begin{myBullets}
\vspace{0.1cm}
\item Preventing the adversary from mix-and-match attacks by computing a message authentication code over the ciphertexts and attaching it to the encapsulation~\cite{bindel2019hybrid},
\item Keeping IND-CCA security~\cite{bindel2019hybrid},
\item Providing security proofs for classical, partial, and fully quantum adversaries,
\item Protecting the ciphertext from modification~\cite{bindel2019hybrid},
\item MAC suffices to use one-time MACs with multiple verification queries~\cite{bindel2019hybrid},
\item Relying solely on the security of one of the two combined KEMs and the (one-time) existential unforgeability of the MAC scheme.~\cite{bindel2019hybrid}.\\
\end{myBullets}
\end{minipage}}
 & {\begin{minipage}{\linewidth}
\begin{myBullets}
\vspace{0.1cm}
\item Depending on the MAC selected as part of the combination mechanism and its security, the adversary can intentionally modify the message content, calculate a new checksum, and eventually replace the original checksum with the new value.\\
\end{myBullets}
\end{minipage}}
 & 
 {\begin{minipage}{\linewidth}
\begin{myBullets}
\vspace{0.1cm}
\item Tampering (if the security of MAC is broken),

\item Information Disclosure (when none of the algorithms used in the hybrid approach are secure against information disclosure).
\\
\end{myBullets}
\end{minipage}}

 \\ \cline{4-7} 
 & & & XOR-then-PRF~\cite{giacon2018kem} & 

{\begin{minipage}{\linewidth}
\begin{myBullets}
\vspace{0.1cm}
\item Simply replacing the XOR for providing integrity protection on the ciphertexts~\cite{giacon2018kem}.
\\
\end{myBullets}
\end{minipage}} 
 & 
{\begin{minipage}{\linewidth}
\begin{myBullets}
\vspace{0.1cm}
\item Exploiting security issues of PRF, i.e., improper implementation or back-doors, to modify the message content intentionally.
\item Vulnerability against related-key attacks on the cipher~\cite{giacon2018kem}.
\item XOR-then-PRF combiner does not retaining CCA security~\cite{giacon2018kem},
\item Providing security proofs only for classical
adversaries~\cite{giron2022post}.
\\
\end{myBullets}
\end{minipage}} 
& 
{\begin{minipage}{\linewidth}
\begin{myBullets}
\vspace{0.1cm}
\item Tampering (XOR-then-PRF not supporting integrity due to vulnerability against related-key attacks),
\item Information Disclosure (when none of the algorithms used in the hybrid approach are secure against information disclosure).
\\
\end{myBullets}
\end{minipage}}
 \\ 
\cline{2-7} 
 & \multirow{20}{*}{
 \begin{minipage}{\linewidth} {Signature \cite{bindel2017transitioning,ghinea2022hybrid}}\end{minipage}} & Concatenation & Concatenation~\cite{bindel2017transitioning} &
{\begin{minipage}{\linewidth}
\begin{myBullets}
\vspace{0.1cm}
\item Supporting lightweight operations, simple logic, and easy implementation,
\item Retaining unforgeability when both signature algorithms are unforgeable~\cite{bindel2017transitioning}.
\\
\end{myBullets}
\end{minipage}} 
& 
{\begin{minipage}{\linewidth}
\begin{myBullets}
\vspace{0.1cm}
\item Not supporting non-separability property for both signature algorithms~\cite{bindel2017transitioning}.
\\
\end{myBullets}
\end{minipage}} 
& 
{\begin{minipage}{\linewidth}
\begin{myBullets}
\vspace{0.1cm}
	\item Any kind of attacks threatening both signature algorithms used in the hybrid approach can be applied in the combined signature too,
 \item Spoofing, Tampering, Repudiation (when none of the algorithms used in the hybrid approach are secure against Spoofing, Tampering, Repudiation respectively).
\\
\end{myBullets}
\end{minipage}}
 \\ \cline{3-7} 
 & & 
 \multirow{12}{*}{\begin{minipage}{\linewidth}
Nesting
\end{minipage}} & Weak Nesting~\cite{bindel2017transitioning} &
{\begin{minipage}{\linewidth}
\begin{myBullets}
\vspace{0.1cm}
\item Preserving unforgeability when the first signature algorithm is unforgeable~\cite{bindel2017transitioning},
\item Supporting non-separability property for the second signature algorithm~\cite{bindel2017transitioning}.
\\
\end{myBullets}
\end{minipage}} 
& 
{\begin{minipage}{\linewidth}
\begin{myBullets}
\vspace{0.1cm}
\item Unforgeability of weak nesting crucially depending on the unforgeability of first signature schemes instead of on both signature schemes as for the other proposed combiners~\cite{bindel2017transitioning}.
\\
\end{myBullets}
\end{minipage}}
& 
{\begin{minipage}{\linewidth}
\begin{myBullets}
\vspace{0.1cm}
	\item Any kind of attacks threatening the first signature algorithm used in the hybrid approach can be applied in the combined signature too,
 \item Spoofing, Tampering, Repudiation (when the first signature algorithm used in the hybrid approach is not secure against Spoofing, Tampering, and Repudiation, respectively.
\\
\end{myBullets}
\end{minipage}}
 \\ \cline{4-7} 
 & & & Strong Nesting~\cite{bindel2017transitioning,ghinea2022hybrid} & 
{\begin{minipage}{\linewidth}
\begin{myBullets}
\vspace{0.1cm}
\item Retaining unforgeability when both signature algorithms are unforgeable~\cite{bindel2017transitioning},
\item Preserving non-separability property for the second signature algorithm~\cite{bindel2017transitioning,ghinea2022hybrid}.
\\
\end{myBullets}
\end{minipage}} 
& 
{\begin{minipage}{\linewidth}
\begin{myBullets}
\vspace{0.1cm}
\item There is still a possible caution with Strong-Nesting~\cite{ghinea2022hybrid}. It leaks legitimate signatures for one of its underlying schemes from the hybrid counterpart, as was described in work~\cite{bindel2017transitioning}.
\\
\end{myBullets}
\end{minipage}} 
& 
{\begin{minipage}{\linewidth}
\begin{myBullets}
\vspace{0.1cm}
	\item Any kind of attacks threatening both signature algorithms used in the hybrid approach can be applied in the combined signature too,
 \item Spoofing, Tampering, Repudiation (when none of the signature algorithms used in the hybrid approach are secure against Spoofing, Tampering, Repudiation respectively).
\\
\end{myBullets}
\end{minipage}}
\\ \cline{4-7} 
 & & & Dual Nesting~\cite{bindel2017transitioning} & 
{\begin{minipage}{\linewidth}
\begin{myBullets}
\vspace{0.1cm}
\item Preserving the unforgeability of each message under its corresponding signature scheme~\cite{bindel2017transitioning},
\item Retaining unforgeability of both messages when the outer signature scheme is unforgeable~\cite{bindel2017transitioning}.
\\
\end{myBullets}
\end{minipage}} 
& 
{\begin{minipage}{\linewidth}
\begin{myBullets}
\vspace{0.1cm}
\item Dual-message combiner is not designed for providing the unforgeability of both messages under either signature scheme~\cite{bindel2017transitioning}.
\\
\end{myBullets}
\end{minipage}} 
&
{\begin{minipage}{\linewidth}
\begin{myBullets}
\vspace{0.1cm}
	\item Any kind of attacks threatening both signature algorithms used in the hybrid approach can be applied in the combined signature too,
 \item Spoofing, Tampering, Repudiation (when none of the signature algorithms used in the hybrid approach are secure against Spoofing, Tampering, Repudiation respectively).
\\
\end{myBullets}
\end{minipage}}
\\ \hline
\end{tabular}%
 }\vspace{3pt}
\footnotesize{$^*$The approaches that combine the output of classic and post-quantum key exchanges to construct hybrid ones.}
\end{table*}

 
In this section, we investigate different approaches to combining multiple independent algorithms and providing a hybrid strategy for migrating from a non-quantum-safe cryptographic state to a quantum-safe cryptographic state. Such a strategy enables the use of both classical and quantum-safe components in a cryptographic system to achieve a balance between security and efficiency.

There are different approaches to providing a hybrid strategy. These approaches utilize different combiners to combine multiple Key Encapsulation/Encryption (KEM/ENC) or signature mechanisms, allowing organizations to prepare themselves for the quantum-safe era and providing a smooth transition from a pre-migration non-quantum-safe state to a post-migration quantum-safe state (i.e., classical to post-quantum cryptographic algorithms) that supports business continuity within the migration time. It also enables organizations to be crypto-agile. Crypto-agility refers to the ability of a cryptographic system to adapt to new cryptographic algorithms as they become available~\cite{8755433}. This is because the security of quantum-safe components is not yet fully understood and may be susceptible to attacks as QC technology advances. Therefore, a hybrid strategy enables organizations to switch to new cryptographic algorithms that offer improved security as they become available without replacing the entire system.

\subsubsection{Hybrid KEM/ENC Strategy} Within the Hybrid Key Exchange Mechanism/Encryption (KEM/ENC) strategy, various independent KEM/ENC algorithms are amalgamated through KEM/ENC combiners, creating a hybrid KEM/ENC algorithm that attains security levels equivalent to the most robust constituent. A KEM/ENC combiner serves as a framework specifying how distinct KEM/ENC algorithms can be combined, often incorporating additional cryptographic primitives. The resulting hybrid algorithm ensures security on par with the strongest individual algorithm, preventing the compromise of incorrect bits. This strategic approach is crucial for fortifying KEM/ENC algorithms against the imminent threat of quantum computers. It significantly enhances security, especially in the context of potential quantum threats. Notably, specific combiners, including Concatenation~\cite{barker2017recommendation}, Concat-KDF~\cite{campagna2020security,whyte2017quantum,kiefer2018hybrid,schanck2015circuit}, Cascade-KDF~\cite{campagna2020security}, Dual-PRF~\cite{bindel2019hybrid,dowling2020many}, Nested-Dual-PRF~\cite{bindel2019hybrid}, Split-key-PRF~\cite{giacon2018kem}, XOR~\cite{giacon2018kem,ghosh2015post,brendel2019breakdown}, XOR-then-MAC~\cite{bindel2019hybrid}, and XOR-then-PRF~\cite{giacon2018kem}, present unique advantages and limitations.
However, the selection of a specific combiner requires careful consideration of its associated trade-offs. For instance, Concatenation offers simplicity, supporting lightweight operations and easy implementation. Nevertheless, it introduces vulnerabilities such as the potential compromise of the shared secret key's integrity and provides security proofs only against classical adversaries~\cite{giron2022post}. Concat-KDF, on the other hand, combines key exchange outputs through a single Key Derivation Function (KDF), reducing the effectiveness of brute-force attempts. However, its implementation may necessitate adjustments to FIPS 140 validation code, and security proofs are limited to classical adversaries~\cite{campagna2020security}. Each combiner comes with its set of pros, cons, and potential quantum threats, detailed comprehensively in Table~\ref{tab:Through-Migration-Alg}. Due to space limitations, a thorough exploration of the technical intricacies of each solution is provided within the table.

\subsubsection{Hybrid Signature} The hybrid signature strategy involves combining multiple independent signatures to ensure the unforgeability of the resulting signature under a chosen message attack (EUF-CMA) scenario. EUF-CMA implies that an adversary can interact with a signing oracle to obtain signatures on any desired messages but remains unable to produce a forged signature on a new message. Additionally, non-separability is deemed an essential property for hybrid signatures, preventing an adversary from disassembling the hybrid signature into valid signatures from individual component signature schemes~\cite{bindel2017transitioning}.
Non-separability serves the purpose of thwarting potential attackers from manipulating a hybrid signature into something that a verifier might accept as originating from a single-scheme signature, thereby distorting the original intention of the signer. Various combiners, including Concatenation~\cite{bindel2017transitioning}, Weak Nesting~\cite{bindel2017transitioning}, Strong Nesting~\cite{bindel2017transitioning,ghinea2022hybrid}, and Dual Nesting~\cite{bindel2017transitioning}, each present distinct advantages and drawbacks. For detailed insights into the different signature combiners, their respective strengths, weaknesses, and associated security threats, please refer to Table~\ref{tab:Through-Migration-Alg}. The table comprehensively discusses concatenation and three types of nested signatures, acknowledging the intricate technical nuances of each approach.

In-depth technical analyses of each hybrid strategy, covering specific approaches, security properties, potential vulnerabilities, and quantum threats within the STRIDE model, are provided in Table~\ref{tab:Through-Migration-Alg}. This table serves as a comprehensive reference, offering organizations a detailed exploration of the technical aspects of through-migration hybrid strategies at the algorithmic level. It encompasses different combination approaches, mechanisms, pros, cons, and quantum threats associated with each strategy.
\begin{figure}[!htbp]
\renewcommand{\arraystretch}{1.2}
 \footnotesize
{

{%
\subfloat{}{\begin{math}
 \qquad \raisebox{-0.5\normalbaselineskip}{Risk\ (Premitive \ 1) $\ \left\{\rule{0pt}{2\normalbaselineskip}\right.$}
\hspace{-0.15cm}
\begin{tabular}{p{0.1cm}}\\
\multicolumn{1}{l}{L} \\ 
\multicolumn{1}{l}{M} \\
\multicolumn{1}{l}{H}
\end{tabular}
\end{math}
}
\hspace{-0.3cm}
\begin{math}
 \overbrace{\
\begin{tabular}{p{1cm}p{1cm}p{1cm}}
L & M & H \\ \hline
\multicolumn{1}{|l|}{\low} & \multicolumn{1}{l|}{\low} & \multicolumn{1}{l|}{\low} \\ \hline
\multicolumn{1}{|l|}{\low} & \multicolumn{1}{l|}{\med} & \multicolumn{1}{l|}{\med} \\ \hline
\multicolumn{1}{|l|}{\low} & \multicolumn{1}{l|}{\med} & \multicolumn{1}{l|}{\high} \\ \hline
\end{tabular}
}^{\mbox{Risk\ (Premitive \ 2)}}
\end{math}
}}
 \caption{ Through-Migration  Risk Assessment (Based on the
Risk Levels of the Primitives Involved in the Combination)}
 \label{figure:hybridrisk}
\end{figure}
\subsubsection{Through-Migration Algorithmic Level Risk Assessment} According to the definition of the hybrid approach, a hybrid mechanism combines multiple independent primitives (i.e., algorithms for key encapsulation (KEM), encryption (ENC), or signature) in parallel. The combined algorithm (hybrid) is considered secure as long as at least one of the combined primitives remains secure. In fact, the security of the combination is determined by the strongest primitive in the combination. Consequently, the level of risk considered for the combined hybrid algorithm is defined as the minimum level of risk caused by any of the two algorithms involved in the combination (see Figure~\ref{figure:hybridrisk}). Note that weak nesting does not support unforgeability as expected from the hybrid approach. In fact, the unforgeability level of weak nesting is as much as 
unforgeability level of the first algorithm in the combination. Hence unlike other combiners in which the level of the risk considered for this combination a the minimum level of risk among the two algorithms, the risk for weak nesting is equivalent to the level of the risk considered for the first algorithm in the combination.

 \subsection{Post-Migration Algorithmic Level Analysis and Risk Assessment}\label{post-algo}
 
\begin{figure*}[!htbp]
\vspace{-0.5cm}
\begin{center} 
\resizebox{0.9\linewidth}{!}{
\begin{tikzpicture}[line width=0.035cm]
\node (d0) [draw,top color=white, bottom color=blue!20, rounded corners,minimum height=1cm,minimum width=4cm] at (0,0) {Attacks};
\node (d1) [draw,top color=white, bottom color=blue!20, rounded corners,minimum height=1cm,minimum width=4cm, right of=d0,xshift=11cm,,yshift=-2.5cm] {Math. Analysis Attacks};

\node (d2) [draw,top color=white, bottom color=blue!20, rounded corners,minimum height=1cm,minimum width=4cm, left of=d0,xshift=-11cm,yshift=-2.5cm] {Side-channel Attacks};

\node (d21) [draw,top color=white, bottom color=blue!20, rounded corners,minimum height=1cm,minimum width=4cm,yshift=-1.5cm, below of=d2, xshift=-9.5cm] {Active Attacks};
\node (d211) [draw,top color=white, bottom color=blue!20, rounded corners,minimum height=1cm,minimum width=4cm,yshift=-1.5cm, below of=d21] {Fault Attacks};

\node (d22) [draw,top color=white, bottom color=blue!20, rounded corners,minimum height=1cm,minimum width=4cm,yshift=-1.5cm, below of=d2,
 xshift=9.4cm] {Passive Attacks};
 \node (d223) [draw,top color=white, bottom color=blue!20, rounded corners,minimum height=1cm,minimum width=4cm,yshift=-1.5cm, below of=d22,
 xshift=3.1cm] {Electromagnetic Attacks};
 \node (d222) [draw,top color=white, bottom color=blue!20, rounded corners,minimum height=1cm,minimum width=4cm,right of=d223,
 xshift=3.5cm] {Template Attacks};
 \node (d221) [draw,top color=white, bottom color=blue!20, rounded corners,minimum height=1cm,minimum width=4cm,right of=d222,
 xshift=3.5cm] {Cold-Boot Attacks};
\node (d224) [draw,top color=white, bottom color=blue!20, rounded corners,minimum height=1cm,minimum width=4cm,yshift=-1.5cm, below of=d22,
 xshift=-3.1cm] { Advanced (correlation/differential) Power Analysis};
\node (d225) [draw,top color=white, bottom color=blue!20, rounded corners,minimum height=1cm,minimum width=4cm, left of=d224,
 xshift=-5.5cm] {Simple Power Analysis Attacks};
\node (d226) [draw,top color=white, bottom color=blue!20, rounded corners,minimum height=1cm,minimum width=4cm, left of=d225,
 xshift=-4.1cm] {Timing Attacks};
 \node (d11) [draw,top color=white, bottom color=blue!20, rounded corners,minimum height=1cm,minimum width=4cm,yshift=-1.5cm, below of=d1] {Cryptanalysis Attacks};

 \draw[-latex] (d0) -| (d2);
 \draw[-latex] (d0) -| (d1);
 \draw[-latex] (d2) -| (d21);
 \draw[-latex] (d21) -- (d211);
 \draw[-latex] (d22) -| (d221); 
 \draw[-latex] (d22) -| (d222);
 \draw[-latex] (d22) -| (d223);
 \draw[-latex] (d22) -| (d224);
 \draw[-latex] (d22) -| (d225); 
 \draw[-latex] (d22) -| (d226); 
 \draw[-latex] (d2) -| (d22);
 \draw[-latex] (d1) -- (d11);
\end{tikzpicture}}

\caption{Taxonomy of Attacks for Post-Migration Algorithmic-Level Analysis (for NIST-Standardized and $4^{th}$-Round Candidates)}
\label{fig:EoP-tree}
\end{center}
\end{figure*} 
 
\begin{table*}[!htbp]
 \scriptsize
\caption{Post-Migration Algorithmic Level Analysis and Risk Assessment }
\label{tab:Post-Migration-Alg}
\resizebox{\textwidth}{!}{%
\begin{tabular}{|l|l|p{0.09\linewidth}|p{0.17\linewidth}|p{0.48\linewidth}|p{0.35\linewidth}|l|l|l|}
\hline
\multirow{2}{*}{\textbf{PQ Approaches}} & \multirow{2}{*}{\textbf{Crypto Types}} & \multirow{2}{*}{\textbf{Candidates}}  & \multirow{2}{*}{\textbf{Vulnerabilities and Attacks}} & \multirow{2}{*}{\textbf{Possible Countermeasures}} & \multirow{2}{*}{\textbf{Quantum Threats (STRIDE)}} & \multirow{2}{*}{\textbf{L}} & \multirow{2}{*}{\textbf{I}} & \multirow{2}{*}{\textbf{R}} \\ 
 & & & & & & & & \\\hline
\multirow{40}{*}{Lattice-based} & \multirow{15}{*}{KEM/ENC} & \multirow{15}{*}{Kyber~\cite{bos2018crystals}}  & Fault Attacks~\cite{ravi2019number,oder2016practical,ravi2020drop} & 
{\begin{minipage}{\linewidth}
\begin{myBullets}
\vspace{0.1cm}
\item Masking decryption process by splitting secret key~\cite{oder2016practical,ravi2020drop},
 \item Checking the secret and error components of the LWE instances for known trivial weaknesses~\cite{ravi2019number}.
\vspace{0.1cm}\end{myBullets}
\end{minipage}}
 & 
{\begin{minipage}{\linewidth}
\begin{myBullets}
\vspace{0.1cm}
\item Information Disclosure through message and key recovery~\cite{ravi2019number, oder2016practical, ravi2020drop}.
\vspace{0.1cm}\end{myBullets}
\end{minipage}}
 &\med &\med &\med \\ \cline{4-9} 
 & & & Simple Power Analysis~\cite{hamburg2021chosen} & 
 {\begin{minipage}{\linewidth}
\begin{myBullets}
\vspace{0.1cm}
 \item (a) Masking of input~\cite{hamburg2021chosen}; (b) Randomizing the order of executed operations within an NTT computation or by inserting random dummy operations inside the NTT~\cite{hamburg2021chosen}.
\vspace{0.1cm}\end{myBullets}
\end{minipage}}
 & 
 {\begin{minipage}{\linewidth}
\begin{myBullets}
\vspace{0.1cm}
 \item Information disclosure via key recovery~\cite{hamburg2021chosen}.
\vspace{0.1cm}\end{myBullets}
\end{minipage}}
 &\med & \med&\med \\ \cline{4-9} 
 & & &  Advanced Power Analysis~\cite{pessl2019more,kamucheka2021power, dubrova2022breaking} & {\begin{minipage}{\linewidth}
\begin{myBullets}
\vspace{0.1cm}
 \item Masking the Number Theoretic Transform (NTT), which is an integral part of efficient implementations of many lattice-based schemes~\cite{pessl2019more}.
 \item No countermeasures for the attack mentioned in~\cite{dubrova2022breaking}.
\vspace{0.1cm}\end{myBullets}
\end{minipage}}
 & 
 {\begin{minipage}{\linewidth}
\begin{myBullets}
\vspace{0.1cm}
 \item Information Disclosure through the recovery of the transmitted symmetric key~\cite{pessl2019more}.
\vspace{0.1cm}\end{myBullets}
\end{minipage}}
 &\high & \med&\high\\ \cline{4-9}  
 & & &  Electromagnetic Attacks~\cite{ravi2020generic,xu2021magnifying,ravi2020drop} & 
 {\begin{minipage}{\linewidth}
 \begin{myBullets}
\vspace{0.1cm}
    \item Masking ECC procedures, including decryption/decapsulation~\cite{ravi2020generic,oder2016practical}, protecting FO transform in CCA~\cite{ravi2020generic}, and securing the secret key~\cite{xu2021magnifying}.
    \item Discarding ciphertexts with special structure or low entropy~\cite{xu2021magnifying}.
    \item Splitting the secret into random shares and randomizing entire decryption/decapsulation~\cite{xu2021magnifying}.
\vspace{0.1cm}\end{myBullets}
\end{minipage}}
 & 
 {\begin{minipage}{\linewidth}
\begin{myBullets}
\vspace{0.1cm}
\item Information Disclosure via full key extraction~\cite{xu2021magnifying,ravi2020generic} or secret message bit disclosure~\cite{ravi2020drop}.
\vspace{0.1cm}\end{myBullets}
\end{minipage}}

 &\low &\med & \low\\ \cline{4-9}  
 & & &  Template Attacks~\cite{ravi2021exploiting} & 
 {\begin{minipage}{\linewidth}
\begin{myBullets}
\vspace{0.1cm}
\item No countermeasures for the attack mentioned in~\cite{ravi2021exploiting}.
\vspace{0.1cm}\end{myBullets}
\end{minipage}}
 & 
 {\begin{minipage}{\linewidth}
\begin{myBullets}
\vspace{0.1cm}
\item Information Disclosure via message recovery~\cite{ravi2021exploiting}.
\vspace{0.1cm}\end{myBullets}
\end{minipage}}
&\med & \med&\med\\ \cline{4-9} 
 & & &  Cold-Boot Attacks~\cite{albrecht2018cold} & 
 {\begin{minipage}{\linewidth}
\begin{myBullets}
\vspace{0.1cm}
 \item Storing the secret in the time domain instead of the frequency domain~\cite{albrecht2018cold}.
\vspace{0.1cm}\end{myBullets}
\end{minipage}}
 & 
 {\begin{minipage}{\linewidth}
\begin{myBullets}
\vspace{0.1cm}
\item Information Disclosure through secret key recovery~\cite{albrecht2018cold}.
\vspace{0.1cm}\end{myBullets}
\end{minipage}}
 & \low & \med & \low \\ \cline{2-9}
 & \multirow{23}{*}{Signature} & \multirow{11}{*}{Dilithium~\cite{ducas2018crystals}}  & Fault Attacks~\cite{ravi2019number,bruinderink2018differential} & 
{\begin{minipage}{\linewidth}
\begin{myBullets}
\vspace{0.1cm}
\item Identifying weaknesses in LWE instances through secret and error component analysis~\cite{ravi2019number},
\item Applying generic countermeasures such as Double computation, Verification-after-sign, and Additional randomness~\cite{bruinderink2018differential}.
\vspace{0.1cm}\end{myBullets}
\end{minipage}}
 & 
{\begin{minipage}{\linewidth}
\begin{myBullets}
\vspace{0.1cm}
\item Spoofing and Tampering via key recovery~\cite{ravi2019number,bruinderink2018differential},
\item Tampering, Repudiation and Elevation of privilege through signature forgery~\cite{bruinderink2018differential},
\vspace{0.1cm}\end{myBullets}
\end{minipage}}
&\med & \med&\med \\ \cline{4-9}  
 & & &  Advanced Power Analysis~\cite{migliore2019masking,marzougui2022profiling} & 
 {\begin{minipage}{\linewidth}
\begin{myBullets}
\vspace{0.1cm}
  \item Masking with linear secret sharing~\cite{migliore2019masking},
  \item Boolean and arithmetic masking using variable splitting and sharing sensitive variable~\cite{marzougui2022profiling}.
\vspace{0.1cm}\end{myBullets}
\end{minipage}}
 & 
{\begin{minipage}{\linewidth}
\begin{myBullets}
\vspace{0.1cm}
\item Spoofing, Tampering, Repudiation, and Elevation of Privilege via forged signatures~\cite{marzougui2022profiling}.
\item Spoofing, Tampering, Repudiation, and Elevation of Privilege by disclosing secret variables~\cite{migliore2019masking}.\vspace{0.1cm}\end{myBullets}
\end{minipage}}
 &\med & \med&\med \\ \cline{4-9}  
 & & &  Electromagnetic Attacks~\cite{ravi2019exploiting,singh2022end} & 
{\begin{minipage}{\linewidth}
\begin{myBullets}
\vspace{0.1cm}
\item Re-ordering of operations within the signing procedure and embedding the vulnerable addition operation deep enough inside the signing procedure~\cite{ravi2019exploiting},
\item Bit-slicing design for NTT, the most critical sub-block, to provide a spatial intra-instruction redundancy~\cite{singh2022end}.
\vspace{0.1cm}\end{myBullets}
\end{minipage}}
 & 
{\begin{minipage}{\linewidth}
\begin{myBullets}
\vspace{0.1cm}
\item Spoofing and Tampering via disclosing some info. about secret key~\cite{singh2022end}.
\item Tampering, Repudiation, and elevation of Privilege through forged signatures on messages~\cite{ravi2019exploiting}.
 \vspace{0.1cm}\end{myBullets}
\end{minipage}}
 &\low & \med&\low\\ \cline{4-9} 

 & & &  Template Attacks~\cite{cryptoeprint:2023/050} & 
 {\begin{minipage}{\linewidth}
\begin{myBullets}
\vspace{0.1cm}
\item Shuffling and Secret sharing~\cite{cryptoeprint:2023/050}.
\vspace{0.1cm}\end{myBullets}
\end{minipage}}
 & 
 {\begin{minipage}{\linewidth}
\begin{myBullets}
\vspace{0.1cm}
\item Spoofing and  Tampering  via revealing information
on the signer’s secret key~\cite{cryptoeprint:2023/050}.
\item Repudiation and Elevation of Privilege through signature forgery using the revealed signer's secret key~\cite{cryptoeprint:2023/050}.
\vspace{0.1cm}\end{myBullets}
\end{minipage}}
&\med & \med&\med\\ \cline{3-9} 

 & & \multirow{12}{*}{Falcon~\cite{fouque2018falcon}}  & Fault Attacks~\cite{mccarthy2019bearz} & 
{\begin{minipage}{\linewidth}
\begin{myBullets}
\vspace{0.1cm}
 \item (a) Double computation of signature~\cite{mccarthy2019bearz}; (b)  Immediate verification after signing~\cite{mccarthy2019bearz}; (c) Zero checking of the sampled vector~\cite{mccarthy2019bearz}.\vspace{0.1cm}\end{myBullets}
\end{minipage}}
 & 
{\begin{minipage}{\linewidth}
\begin{myBullets}
\vspace{0.1cm}
 \item Spoofing and Tampering by retrieving the private-key~\cite{mccarthy2019bearz},
 \item Repudiation and Elevation of Privilege via  signature forgery  through retrieved   private key~\cite{mccarthy2019bearz}.
\vspace{0.1cm}\end{myBullets}
\end{minipage}}
&\med & \med&\med \\ \cline{4-9}  
 & & &  Timing Attacks~\cite{mccarthy2019bearz} & 
 {\begin{minipage}{\linewidth}
\begin{myBullets}
\vspace{0.1cm}
 \item (a) Blind-Vector algorithm extended the use of the Fisher-Yates shuffling procedure to enhance random shuffles for side-channel protection~\cite{mccarthy2019bearz}; (b)  Sample discard performing extra cache read from random addresses to distort statistics~\cite{mccarthy2019bearz}.
\vspace{0.1cm}\end{myBullets}
\end{minipage}}
 & 
 {\begin{minipage}{\linewidth}
\begin{myBullets}
\vspace{0.1cm}
\item Spoofing and Tampering via private key retrieval~\cite{mccarthy2019bearz},
\item Repudiation and Privilege Elevation through signature forgery using the retrieved private key~\cite{mccarthy2019bearz}.\vspace{0.1cm}\end{myBullets}
\end{minipage}}
&\med & \med&\med \\ \cline{4-9}  
 & & &  Simple Power Analysis~\cite{guerreau2022hidden} & 
 {\begin{minipage}{\linewidth}
\begin{myBullets}
\vspace{0.1cm}
\item Effectively reduce the Hamming weight gap as discussed in Guerreau's work~\cite{guerreau2022hidden}.
\vspace{0.1cm}\end{myBullets}
\end{minipage}}
 & 
 {\begin{minipage}{\linewidth}
\begin{myBullets}
\vspace{0.1cm}
\item Exploiting Spoofing, Tampering, Repudiation, and Privilege Elevation via secret key recovery~\cite{guerreau2022hidden}.
\vspace{0.1cm}\end{myBullets}
\end{minipage}}
&\med & \med&\med\\ \cline{4-9} 
 & & & Electromagnetic Attacks~\cite{karabulut2021falcon} & {\begin{minipage}{\linewidth}
\begin{myBullets}
\vspace{0.1cm}
 \item (a) Hiding by making power consumption constant,~\cite{karabulut2021falcon}; (b)  Masking using randomizing the intermediates values~\cite{karabulut2021falcon}.
\vspace{0.1cm}\end{myBullets}
\end{minipage}}
 & 
 {\begin{minipage}{\linewidth}
\begin{myBullets}
\vspace{0.1cm}
\item Spoofing and Tampering through secret key extraction~\cite{karabulut2021falcon},
\item Tampering, Repudiation, and Elevation of Privilege via forged signatures on arbitrary messages~\cite{karabulut2021falcon}.
\vspace{0.1cm}\end{myBullets}
\end{minipage}}
 & \low & \med& \low \\ \hline
\multirow{50}{*}{Code-based} & \multirow{50}{*}{KEM/ENC} & \multirow{28}{*}{McEliece~\cite{bernstein2017classic}
} & Cryptanalysis Attacks~\cite{bernstein2008attacking} & {\begin{minipage}{\linewidth}
\begin{myBullets}
\vspace{0.1cm}
 \item (a) Increasing binary code length~\cite{bernstein2008attacking}; (b) Using decoding list to increase weight $w$~\cite{bernstein2008attacking}.
\vspace{0.1cm}\end{myBullets}
\end{minipage}} & 
{\begin{minipage}{\linewidth}
\begin{myBullets}
\vspace{0.1cm}
 \item Information Disclosure through decoding the ciphertext~\cite{bernstein2008attacking}.
\vspace{0.1cm}\end{myBullets}
\end{minipage}}
&\med & \med&\med \\ \cline{4-9} 
 & & &  Fault Attacks~\cite{cayrel2020message,kreuzer2020fault} & {\begin{minipage}{\linewidth}
\begin{myBullets}
\vspace{0.1cm}
 \item No countermeasures for the attack mentioned in~\cite{cayrel2020message},
 \item (a) Checking the weight of the output of the decryption map 
 to discover fault injection~\cite{kreuzer2020fault}; (b) Re-encrypting the output of the decryption map and 
 comparing it with the ciphertext to discover fault injection~\cite{kreuzer2020fault}.
 \vspace{0.1cm}\end{myBullets}
\end{minipage}} & 
{\begin{minipage}{\linewidth}
\begin{myBullets}
\vspace{0.1cm}
 \item Information Disclosure through real-time message recovery~\cite{cayrel2020message} or by computing an alternative valid secret key~\cite{kreuzer2020fault}.
\vspace{0.1cm}\end{myBullets}
\end{minipage}}
&\med & \med&\med \\ \cline{4-9}  
 & & & Timing Attacks~\cite{strenzke2008side} & 
{\begin{minipage}{\linewidth}
\begin{myBullets}
\vspace{0.1cm}
 \item Artificially raising the degree of error locator polynomial (defined Goppa Codes) for any degree lower than a threshold~\cite{strenzke2008side}.
\vspace{0.1cm}\end{myBullets}
\end{minipage}}
 & 
{\begin{minipage}{\linewidth}
\begin{myBullets}
\vspace{0.1cm}
\item Information Disclosure via partial or complete revelation of a secret~\cite{strenzke2008side}.
\vspace{0.1cm}\end{myBullets}
\end{minipage}}
&\med & \med&\med \\ \cline{4-9}  
 & & & Simple Power Analysis~\cite{petrvalsky2015countermeasure,guo2022key} & 
{\begin{minipage}{\linewidth}
\begin{myBullets}
\vspace{0.1cm}
 \item Eliminating special power traces patterns, branch statements, and data dependency to ensure that power consumption and execution time are constant~\cite{petrvalsky2015countermeasure}
 \item No countermeasures for the attack mentioned in~\cite{guo2022key}.
\vspace{0.1cm}\end{myBullets}
\end{minipage}} & 
{\begin{minipage}{\linewidth}
\begin{myBullets}
\vspace{0.1cm}
\item Information Disclosure through complete secret permutation matrix recovery~\cite{petrvalsky2015countermeasure} or full secret key recovery~\cite{guo2022key}.
\vspace{0.1cm}\end{myBullets}
\end{minipage}}
&\med & \med&\med \\ \cline{4-9}  
 & & & Advanced Power Analysis~\cite{jedlicka2022secure} & 
{\begin{minipage}{\linewidth}
\begin{myBullets}
\vspace{0.1cm}
\item Parallelization and shuffling~\cite{jedlicka2022secure,chen2015horizontal}.
 \item Masking the cryptosystem by adding Goppa codewords to a ciphertext during the permutation process~\cite{jedlicka2022secure,petrvalsky2016differential}.
\vspace{0.1cm}\end{myBullets}
\end{minipage}}
 & 
{\begin{minipage}{\linewidth}
\begin{myBullets}
\vspace{0.1cm}
\item Information Disclosure through private key exposure via permutation matrix~\cite{petrvalsky2016differential}.
\item Information Disclosure through secret key recovery from a limited number of decryptions~\cite{chen2015horizontal}.
\vspace{0.1cm}\end{myBullets}
\end{minipage}}
&\med & \med&\med \\ \cline{4-9}  
 & & & Electromagnetic Attacks~\cite{lahr2020side} & 
{\begin{minipage}{\linewidth}
\begin{myBullets}
\vspace{0.1cm}
 \item No countermeasures for the attack mentioned in~\cite{lahr2020side}.
\vspace{0.1cm}\end{myBullets}
\end{minipage}}
 & 
 {\begin{minipage}{\linewidth}
\begin{myBullets}
\vspace{0.1cm}
\item Information Disclosure via successful plaintext recovery~\cite{lahr2020side}.\vspace{0.1cm}\end{myBullets}
\end{minipage}}
&\med & \med&\med \\ \cline{4-9}  
 & & & Template Attacks~\cite{strenzke2008side} & 
{\begin{minipage}{\linewidth}
\begin{myBullets}
\vspace{0.1cm}
 \item Eliminating the memory access dependency on the content of lookup-table resulting in constant running time, no jumps depending on secret input, and only access memory addresses depending on public input~\cite{strenzke2008side}).
\vspace{0.1cm}\end{myBullets}
\end{minipage}}
 & 
 {\begin{minipage}{\linewidth}
\begin{myBullets}
\vspace{0.1cm}
 \item Information Disclosure through  revealing the permutation that is part of the secret key~\cite{strenzke2008side}.
\vspace{0.1cm}\end{myBullets}
\end{minipage}}
 & \low& \med& \low\\ \cline{4-9} 
 & & &  Cold-Boot Attacks~\cite{polanco2019cold} & 
 {\begin{minipage}{\linewidth}
\begin{myBullets}
\vspace{0.1cm}
 \item No countermeasures for the attack mentioned in~\cite{polanco2019cold}.
\vspace{0.1cm}\end{myBullets}
\end{minipage}}
 & 
 {\begin{minipage}{\linewidth}
\begin{myBullets}
\vspace{0.1cm}
 \item Information Disclosure through private key recovery~\cite{polanco2019cold}.
\vspace{0.1cm}\end{myBullets}
\end{minipage}}
 &\med &\med &\med \\ \cline{3-9}
 & & \multirow{6}{*}{BIKE~\cite{aragon2017bike}}  & Fault Attacks~\cite{cayrel2020message,xagawa2021fault} & {\begin{minipage}{\linewidth}
\begin{myBullets}
\vspace{0.1cm}
\item (a) Default failing which initiates a variable with the fail result, and if a condition is satisfied, then the variable will be overwritten by the sensitive data~\cite{xagawa2021fault}; (b) Assembly-level instruction duplicating~\cite{xagawa2021fault}; (c) Random delaying~\cite{xagawa2021fault}.
 \item No countermeasures for the attack mentioned in~\cite{cayrel2020message}.
\vspace{0.1cm}\end{myBullets}
\end{minipage}}
 & 
{\begin{minipage}{\linewidth}
\begin{myBullets}
\vspace{0.1cm}
 \item Information Disclosure through real-time message recovery~\cite{cayrel2020message,xagawa2021fault}).
\vspace{0.1cm}\end{myBullets}
\end{minipage}}
 &\med & \med&\med \\ \cline{4-9}  
 & & &  Timing Attacks~\cite{guo2020key,guo2022don} & {\begin{minipage}{\linewidth}
\begin{myBullets}
\vspace{0.1cm}
 \item (a) Increasing the number of bytes that are generated initially to double the previous amount and removing additional random data generation call~\cite{guo2022don}; (b) Constant-time random number generation~\cite{guo2022don}.
\item No countermeasures for the attack mentioned in~\cite{guo2020key}.
\vspace{0.1cm}\end{myBullets}
\end{minipage}}
 & 
{\begin{minipage}{\linewidth}
\begin{myBullets}
\vspace{0.1cm}
 \item Information Disclosure via secret key recovery~\cite{guo2020key,guo2022don}.
\vspace{0.1cm}\end{myBullets}
\end{minipage}}
&\med & \med&\med\\ \cline{3-9}
 & & \multirow{14}{*}{HQC~\cite{melchor2018hamming}} & Fault Attacks~\cite{cayrel2020message,xagawa2021fault} & {\begin{minipage}{\linewidth}
\begin{myBullets}
\vspace{0.1cm}
\item (a) Default failing which initiates a variable with the fail result, and if a condition is satisfied, then the variable will be overwritten by the sensitive data~\cite{xagawa2021fault}; (b) Assembly-level instruction duplicating~\cite{xagawa2021fault}; (c) Random delaying~\cite{xagawa2021fault}.
 \item No countermeasures for the attack mentioned in~\cite{cayrel2020message}.
\vspace{0.1cm}\end{myBullets}
\end{minipage}}
 & 
{\begin{minipage}{\linewidth}
\begin{myBullets}
\vspace{0.1cm}
 \item Information Disclosure through  real-time message recovery~\cite{cayrel2020message,xagawa2021fault}.
\vspace{0.1cm}\end{myBullets}
\end{minipage}}
&\med & \med&\med \\ \cline{4-9}  
 & & & Timing Attacks~\cite{guo2020key,guo2022don,wafo2020practicable} & {\begin{minipage}{\linewidth}
\begin{myBullets}
\vspace{0.1cm}
\item (a) Double the initial byte generation and eliminate extra random data calls~\cite{guo2022don}; (b) Implement constant-time random number generation~\cite{guo2022don}.
\item Develop a constant-time decoding algorithm to perform (a) implementation, (b) field arithmetic, (c) syndromes computation and roots computation, and (d) error locator polynomial computation in constant time~\cite{wafo2020practicable},
\item No countermeasures for the attack mentioned in~\cite{guo2020key}.
\vspace{0.1cm}\end{myBullets}
\end{minipage}}
 & 
{\begin{minipage}{\linewidth}
\begin{myBullets}
\vspace{0.1cm}
 \item Information Disclosure via secret key recovery~\cite{guo2020key,guo2022don,wafo2020practicable}.
\vspace{0.1cm}\end{myBullets}
\end{minipage}}
&\med & \med&\med\\ \cline{4-9} 
 & & & Simple Power Analysis~\cite{schamberger2020power} & 
{\begin{minipage}{\linewidth}
\begin{myBullets}
\vspace{0.1cm}
\item No countermeasures for the attack mentioned in~\cite{schamberger2020power}.
\vspace{0.1cm}\end{myBullets}
\end{minipage}} & 
{\begin{minipage}{\linewidth}
\begin{myBullets}
\vspace{0.1cm}
 \item Information Disclosure through the retrieval of a significant portion of the secret key~\cite{schamberger2020power}.
\vspace{0.1cm}\end{myBullets}
\end{minipage}} &\high &\med & \high \\\cline{4-9}  
 & & &  Electromagnetic Attacks~\cite{goy2022new} & 
{\begin{minipage}{\linewidth}
\begin{myBullets}
\vspace{0.1cm}
\item Masking using linear secret sharing and dividing the knowledge in $n$ shares~\cite{goy2022new}.
\vspace{0.1cm}\end{myBullets}
\end{minipage}} & 
{\begin{minipage}{\linewidth}
\begin{myBullets}
\vspace{0.1cm}
 \item Information Disclosure via secret key recovery~\cite{goy2022new}).
\vspace{0.1cm}\end{myBullets}
\end{minipage}} & \low& \med&\low \\ \hline
\multirow{3}{*}{Hash-based} & \multirow{3}{*}{Signature} & \multirow{3}{*}{SPHINCS+~\cite{bernstein2019sphincs+}}  & Fault Attacks~\cite{castelnovi2018grafting,genet2018practical} & 
{\begin{minipage}{\linewidth}
\begin{myBullets}
\vspace{0.1cm}
    \item (a) Redundant computation of signatures~\cite{castelnovi2018grafting}; (b) Computing few-time signatures (FTS) using public values~\cite{castelnovi2018grafting}; (c) Linking hyper-tree layers to detect computation faults leading to invalid signatures~\cite{castelnovi2018grafting},
    \item (a) Fault detection through recomputation of sub-trees with swapped nodes and enhanced hash function~\cite{genet2018practical}; (b) Storing one-time signatures for reuse~\cite{genet2018practical}; (c) Mismatch detection by recomputing vulnerable instructions on different hardware modules~\cite{genet2018practical}.
\vspace{0.1cm}\end{myBullets}
\end{minipage}}
 & 
\begin{minipage}{\linewidth}
\begin{myBullets}
\vspace{0.1cm}

\item Spoofing and Tampering via exploiting vulnerabilities like key recovery or universal signature forgery~\cite{castelnovi2018grafting}.
\item Tampering and Repudiation through message signatures forgery using techniques such as voltage glitch injection and collecting faulty signatures~\cite{genet2018practical}.
\vspace{0.1cm}\end{myBullets}
\end{minipage}
&\med & \med&\med \\ \cline{4-9}  
 & & & Advanced Power Analysis~\cite{kannwischer2018differential} & 
{\begin{minipage}{\linewidth}
\begin{myBullets}
\vspace{0.1cm}
 \item Hiding the order of the Mix procedures~\cite{kannwischer2018differential}.
\vspace{0.1cm}\end{myBullets}
\end{minipage}}
 & 
{\begin{minipage}{\linewidth}
\begin{myBullets}
\vspace{0.1cm}
 \item Spoofing and tampering through secret key recovery~\cite{kannwischer2018differential}.
 \item Tampering, Repudiation, and Privilege Elevation by. creating signatures for arbitrary messages~\cite{kannwischer2018differential}.
\vspace{0.1cm}\end{myBullets}
\end{minipage}}
 &\med & \med&\med \\ \hline
\multirow{18}{*}{Isogeny-based} & \multirow{18}{*}{KEM/ENC} & \multirow{18}{*}{SIKE~\cite{jao2011towards}} & Cryptanalysis Attacks~\cite{castryck2022efficient} & 
 {\begin{minipage}{\linewidth}
\begin{myBullets}
\vspace{0.1cm}
\item No countermeasures for the attack mentioned in~\cite{castryck2022efficient}.
\vspace{0.1cm}\end{myBullets}
\end{minipage}}
 & 
{\begin{minipage}{\linewidth}
\begin{myBullets}
\vspace{0.1cm}
 \item Information Disclosure through  secret key recovery~\cite{castryck2022efficient}).
\vspace{0.1cm}\end{myBullets}
\end{minipage}}
 &\high & \med& \high\\ \cline{4-9} 
 & &  & Fault Attacks\cite{xagawa2021fault,tasso2021resistance} & {\begin{minipage}{\linewidth}
\begin{myBullets}
\vspace{0.1cm}
\item (a) Default failure handling involves initializing a variable with a fail result; if a specific condition is met, the variable is overwritten with sensitive data to prevent fault attacks~\cite{xagawa2021fault}; (b) Assembly-level instruction duplication~\cite{xagawa2021fault}; (c) Random delaying~\cite{xagawa2021fault}.
  \item (a) Successive pushing of curves through small-degree isogenies for kernel generator computation~\cite{tasso2021resistance}; (b) Probability-based recovery of correct elliptic curve coefficients during public key generation~\cite{tasso2021resistance}; (c) Updated implementation includes verification at the end of public key generation~\cite{tasso2021resistance}.
\vspace{0.1cm}\end{myBullets}
\end{minipage}}
 & 
{\begin{minipage}{\linewidth}
\begin{myBullets}
\vspace{0.1cm}
 \item Information Disclosure via secret key recovery~\cite{xagawa2021fault,tasso2021resistance}.
\vspace{0.1cm}\end{myBullets}
\end{minipage}}
&\med & \med&\med \\ \cline{4-9}  
 & & & Advanced Power Analysis~\cite{de2022sike} & 
{\begin{minipage}{\linewidth}
\begin{myBullets}
\vspace{0.1cm}
 \item Checking curve $E_A$ is a supersingular curve which uses points of order $3^{e_3}$ generating $E_A[3^{e_3}]$~\cite{de2022sike}.
\vspace{0.1cm}\end{myBullets}
\end{minipage}}
 & {\begin{minipage}{\linewidth}
\begin{myBullets}
\vspace{0.1cm}
 \item Information Disclosure through full key recovery~\cite{de2022sike}.
\vspace{0.1cm}\end{myBullets}
\end{minipage}} 
&\med & \med&\med\\ \cline{4-9}  
 & & & Electromagnetic Attacks~\cite{de2022sike} & 
{\begin{minipage}{\linewidth}
\begin{myBullets}
\vspace{0.1cm}
 \item Checking curve $E_A$ is a supersingular curve which uses points of order $3^{e_3}$ generating $E_A[3^{e_3}]$~\cite{de2022sike}.
\vspace{0.1cm}\end{myBullets}
\end{minipage}}
 & {\begin{minipage}{\linewidth}
\begin{myBullets}
\vspace{0.1cm}
 \item Information Disclosure by full key recovery~\cite{de2022sike}.
\vspace{0.1cm}\end{myBullets}
\end{minipage}} &\low &\med & \low \\ \cline{4-9}  
 & & &  Cold-Boot Attacks~\cite{villanueva2020cold} & 
 {\begin{minipage}{\linewidth}
\begin{myBullets}
\vspace{0.1cm}
\item No countermeasures for the attack mentioned in~\cite{villanueva2020cold}.
\vspace{0.1cm}\end{myBullets}
\end{minipage}}
 & 
{\begin{minipage}{\linewidth}
\begin{myBullets}
\vspace{0.1cm}
\item Information Disclosure through secret key recovery~\cite{villanueva2020cold}.
\vspace{0.1cm}\end{myBullets}
\end{minipage}}
 &\med &\med &\med \\ \hline
\end{tabular}%
}\vspace{3pt}
\tiny{$^*$ We perform risk evaluation with the presumption of considering the countermeasures mentioned in the table.}
\end{table*}

As previously discussed, the advent of QC will bring about significant changes in the landscape of cryptographic algorithm attacks. While the impact of QC on symmetric cryptographic algorithms is less pronounced (see Figure~\ref{fig:classic-impact}), as they can be effectively secured through longer keys or extended hash function outputs, QC poses a severe threat to widely-used public key cryptographic algorithms. Consequently, existing public key cryptographic algorithms and standards need to be replaced.

To protect against the security threat of QC on widely-used public key cryptographic algorithms and migrate to an environment in a quantum-safe cryptographic state, the development of quantum-safe cryptographic algorithms is essential. NIST has launched a program to standardize such algorithms, recognizing the vulnerability of current cryptographic methods to quantum computers. This program includes a competition for post-quantum cryptographic algorithms, focusing on securing Key Exchange (KEM), Encryption (ENC), and Signature algorithms against QC threats.

Various post-quantum cryptographic algorithms have been proposed, falling into categories such as code-based~\cite{bernstein2017classic, melchor2018hamming, aragon2017bike}, hash-based~\cite{bernstein2019sphincs+}, lattice-based~\cite{bos2018crystals, ducas2018crystals, fouque2018falcon}, and isogeny-based~\cite{jao2011towards} cryptographic algorithms. NIST has taken proactive steps to address QC threats by soliciting proposals for post-quantum public-key exchange and digital signature algorithms. In 2022, NIST selected four quantum-safe (post-quantum) cryptographic algorithms and approved four additional candidates for its $4^{th}$ round, as detailed in Table~\ref{tab:Post-Migration-Alg}~\cite{NIST_2022July}. These candidates are recommended for adoption to ensure quantum-safe cryptography. Notably, NIST is seeking feedback on the initial public drafts of three Federal Information Processing Standards (FIPS): FIPS 203 for Module-Lattice-Based Key-Encapsulation Mechanism \cite{fips203}, FIPS 204 for Module-Lattice-Based Digital Signature \cite{fips204}, and FIPS 205 for Stateless Hash-Based Digital Signature \cite{fips205}.

NIST Post-Quantum Cryptography (NIST PQC) is a public competition that aims to develop new cryptographic standards that can withstand attacks from quantum computers. However, even if a cryptographic algorithm is post-quantum secure, it may still be vulnerable to other types of attacks, such as side-channel and cryptanalysis attacks. A side-channel attack is a type of attack that exploits information leaked during the execution of a cryptographic algorithm, such as power consumption, electromagnetic radiation, or timing information (see Figure~\ref{fig:EoP-tree}). By analyzing this information, an attacker may be able to extract secret information such as a private key. Cryptanalysis attacks, on the other hand, are attacks that aim to break the encryption or signature schemes of a cryptographic algorithm. These attacks typically involve analyzing the structure and properties of the algorithm to find weaknesses that can be exploited to recover the secret information.

Several successful side-channel and cryptanalysis attacks have been reported on  NIST-standardized and $4^{th}$-round PQC candidates by now. It is important to note that the evaluation process is ongoing, and more attacks may be discovered in the future. This section explores the attacks, possible countermeasures, and threats related to each attack for post-quantum cryptographic algorithms considered by NIST~\cite{NIST_2022July} as quantum-safe cryptographic algorithms. 
A quantum attacker may continue to attempt to crack PQ cryptography (e.g., recovery of secrets/plaintexts and forging the signature) by exploiting weaknesses (e.g., side-channels) or through mathematical analysis. Figure~\ref{fig:EoP-tree} provides a taxonomy of the vulnerabilities a quantum attacker can exploit to break the security of post-quantum cryptographic algorithms considered for standardization in NIST round 4.
Thus, an attacker can execute side-channel or mathematical analysis-type attacks. Therefore, we evaluate the risks associated with each attack on candidates in Table~\ref{tab:Post-Migration-Alg}.

\subsubsection{Post-Migration Algorithmic Level Risk Assessment} In appraising the risks associated with each attack on NIST-standardized and $4^{th}$-round PQC candidates, we provide a qualitative risk assessment based on the evaluation criteria for likelihood (refer to Table~\ref{table:likelihood}) and impact (refer to Table~\ref{table:impact}). These criteria are grounded in presumptions derived from a meticulous consideration of potential countermeasures outlined in the table. The ultimate risk assessment integrates both likelihood and impact, as illustrated in Figure~\ref{fig:risk-matrix}.

For the likelihood evaluation, we scrutinize exploitability (via physical access, network, or the Internet), available countermeasures (listed in Table~\ref{tab:Post-Migration-Alg}), and the establishment criteria detailed in Table~\ref{table:likelihood}. Our analysis for the likelihood level is categorized as follows:

\begin{enumerate}[wide, font=\itshape, labelwidth=!, labelindent=0pt, label=\textit{Category} \arabic*.]
\item A known exploit exists and can be launched from the Internet or Network; likelihood is high if no countermeasures are available; otherwise, the likelihood is medium.
\item A known exploit exists and has no countermeasure; likelihood is medium if it requires physical access to launch an attack on the target system.
\item No known exploit (or an exploit with countermeasure) exists; likelihood is low if a malicious user needs administrative or elevated privileges in the target system to launch an attack.
\end{enumerate}

In the impact evaluation, we consider the establishment criteria outlined in Table~\ref{table:impact}. Based on the evaluation criteria, the threats caused by quantum attackers might significantly impact satisfaction, expose personal data/secrets, or damage organizations’ reputations; hence the impact level should be considered as medium.\\

Given limited space, a detailed technical analysis of each post-migration algorithm, encompassing vulnerabilities, potential attacks, countermeasures, and their quantum threats in the STRIDE model, is presented in Table~\ref{tab:Post-Migration-Alg}. The table also offers a comprehensive evaluation of probability and impact levels, addressing risks associated with diverse post-migration algorithms, making it a valuable reference for organizations seeking an in-depth understanding of post-migration hybrid strategies at the algorithmic level.

\section{Quantum Migration Threat Analysis and Risk Assessment for Certificate Level}\label{sec:PKI}

In order to analyze and assess the risks involved in quantum-safe migration across certificate levels, we evaluate the certificate level at each of the three migration stages: pre-migration, through-migration, and post-migration. For each stage, we identify different vulnerabilities that a quantum attacker could exploit, evaluate QC threats using the STRIDE threat model, and assess the overall risk of vulnerabilities throughout the entire migration process. 
\begin{table*}[!htbp]
\small
\caption{{Pre-Migration Certificate Level Analysis and Risk Assessment}}

\resizebox{\textwidth}{!}{%
\begin{tabular}{|l|l|p{0.3\linewidth}|p{0.3\linewidth}|p{0.21\linewidth}|p{0.25\linewidth}|l|l|l|p{0.23\linewidth}|}
\hline
\multirow{2}{*}{\textbf{Certificate Type}}&\multirow{2}{*}{\textbf{Version}} & \multirow{2}{*}{\textbf{Fields of Certificate}} & \multirow{2}{*}{\textbf{Purpose}} & \multirow{2}{*}{\textbf{Recommended Crypto Suite}} & \multirow{2}{*}{\textbf{Quantum Threats (STRIDE)}}& \multirow{2}{*}{\textbf{L}} & \multirow{2}{*}{\textbf{I}} & \multirow{2}{*}{\textbf{R}} & \multirow{2}{*}{\textbf{Possible QC-resistant Solutions}}\\
& & & & & & & & &\\\hline
\multirow{12}{*}{Classic (X.509)~\cite{cooper2008internet}} & v1&
{\begin{minipage}{\linewidth}
\begin{myBullets}
\vspace{0.2cm}
 \item Basic Fields:
 \begin{itemize}[label={\textbullet},
 topsep=0ex,
 partopsep=0ex,
 parsep=0ex,
 itemsep=0ex]
 \item Version number,
 \item Serial number,
 \item Signature Algorithm Identifier,
 \item Issuer name,
 \item Validity period,
 \item Subject Name,
 \item Subject’s public key information.\\
 \end{itemize}
\end{myBullets}
\end{minipage}}
 & 
{\begin{minipage}{\linewidth}
\begin{myBullets}
\vspace{0.2cm}
 \item Managing identity and security in computer networking and over the Internet which includes securing email, communications, and digital signatures.
 \item Supporting authentication, integrity, confidentiality, repudiation.\\
\end{myBullets}
\end{minipage}}
& \multirow{12}{*}{\begin{minipage}{\linewidth}
\begin{myBullets}
\vspace{0.2cm}
 \item Public-Key Crypto: RSA, ECDHE, ECDSA (broken by Shor's Algo.)
\item Symmetric
Crypto: AES, SHA2 (Weakened by Grover's Algo.)
\\
\end{myBullets}
\end{minipage}}
& \multirow{1}{*}{\begin{minipage}{\linewidth}
\begin{myBullets}
\vspace{-1pt}

\item {Spoofing: Quantum attacker can forge Issuer's signature to impersonate a trusted entity due to broken signature algorithms including RSA, ECDSA. }
\item {Tampering: Quantum attacker can alter certificate content due to broken SHA2 hash function. }
\item {Repudiation: Attacker can forge a certificate and use it to deny issuing it (difficult but possible). }
\item {Info. Disclosure: Quantum attacker can potentially recover private keys from public keys using Shor's Algorithm.}

\end{myBullets}
\end{minipage}} & \cellcolor{lemon!80} & \cellcolor{red!80}&\cellcolor{red!80} & 
\multirow{12}{*}{
{\begin{minipage}{\linewidth}
\begin{myBullets}
\vspace{0.2cm}
 \item Dual Certificate~\cite{bindel2017transitioning,vogt2021quantum,bindel2019x}.
 \item For X.509 (v1) and (v2), upgrade to X.509 (v3) and use extension fields to support both classic and post-quantum cryptography~\cite{bindel2017transitioning,vogt2021quantum,bindel2019x}.
 \item For X.509 (v3), use extension fields to support both classic and post-quantum cryptography~\cite{bindel2017transitioning,vogt2021quantum,bindel2019x}.
\end{myBullets}

\end{minipage}}}
 \\ \cline{2-4}
 & v2&
{\begin{minipage}{\linewidth}
\begin{myBullets}
\vspace{0.2cm}
 \item Basic Fields inherited from v1,
 \item Version 2 Additional Fields:
 \begin{itemize}[label={\textbullet},
 topsep=0ex,
 partopsep=0ex,
 parsep=0ex,
 itemsep=0ex]
 \item Issuer Unique ID,
 \item Subject Unique ID.\\
 \end{itemize}
\end{myBullets}
\end{minipage}}
 & 
{\begin{minipage}{\linewidth}
\begin{myBullets}
\vspace{0.2cm}
 \item To handle the possibility of reuse of subject and/or issuer names over time, in addition to all in addition to handle all the purpose of version 1~\cite{ibmx509cert},
 \item Supporting authentication, integrity, confidentiality, repudiation.\\
\end{myBullets}
\end{minipage}}
& 
& 
& \multirow{-5}{*}{
\med}& \multirow{-5}{*}{\high}&\multirow{-5}{*}{\high} &
 \\  \cline{2-4}
 &v3 & 
{\begin{minipage}{\linewidth}
\begin{myBullets}
\vspace{0.2cm}
 \item Basic Fields inherited from v1,
 \item Version 2 Additional Fields,
 \item Version 3 Additional Field:
 \begin{itemize}[label={\textbullet},
 topsep=0ex,
 partopsep=0ex,
 parsep=0ex,
 itemsep=0ex]
 \item Extensions.\\
 \end{itemize}
\end{myBullets}
\end{minipage}}
 & 
{\begin{minipage}{\linewidth}
\begin{myBullets}
\vspace{0.2cm}
 \item Includes the notion of extension, in addition to all in addition to supporting all purpose of version 2~\cite{ibmx509cert},
 \item Supporting authentication, integrity, confidentiality, repudiation.\\
\end{myBullets}
\end{minipage}}
&&   & \cellcolor{lemon!80}& \cellcolor{red!80}&\cellcolor{red!80} & 
\\ \hline
\end{tabular}%
}
\label{tab:Pre-migration-Cert}
\end{table*}

\subsection{Pre-Migration Certificate Level Analysis and Risk Assessment}

The current X.509 Public Key Infrastructure (PKI) standard and certificates~\cite{cooper2008internet} employ public key cryptosystems to manage identity and security across the Internet. The X.509 certificate encompasses procedures for certificate management, revocation, and supports authentication, integrity, confidentiality, and repudiation. Certificates, issued by Certificate Authorities (CAs) or subordinate entities, adhere to a hierarchical structure, ensuring a chain of trust. Each certificate comprises a specific structure, including a public key, a signature, and additional information about the issuing CA, revocation status, validity, and algorithm details (refer to Figure~\ref{fig:HybridCertMech.a}). When a CA issues a certificate, it attests to a user (relying party) that a specific public key is intricately linked to the identity and/or attributes of a particular entity~\cite{chokhani2003internet}.

Certificate susceptibilities and the reasons for each quantum threat mentioned in the table are based on the functionality support provided by the STRIDE aspects, as detailed in the purpose column. However, the emergence of quantum computing introduces a formidable threat to classical public keys and signatures employed in certificates. This vulnerability poses potential quantum risks to the traditional X.509 certificate. Our comprehensive analysis, detailed in Table~\ref{tab:Pre-migration-Cert}, meticulously explores all versions of the X.509 certificate. It examines QC threats, evaluates associated risks, and proposes QC-resistant solutions to fortify the existing classical X.509 certificate and Public Key Infrastructure (PKI) against quantum threats before transitioning to the quantum era. As depicted in the table, three versions of the classical X.509 certificate exist. The initial version of the classical certificate (v1) underwent updates to the second version (v2), introducing issuer and subject unique identity fields to handle the potential reuse of subject and/or issuer names over time. The latest version (v3) certificate was introduced to support customization by introducing an extension field, in addition to the second version (v2).
{All versions (v1, v2, and v3) are susceptible to spoofing, tampering, repudiation, and information disclosure attacks from quantum adversaries because the certificates support authentication, integrity, confidentiality, and repudiation.}

\subsubsection{Pre-Migration Certificate  Level Risk Assessment} 
{To analyze the risk, we need to evaluate the likelihood and impact of QC threats for classical certificates, similar to the pre-migration algorithmic level risk assessment. To evaluate the likelihood, we consider 15 years as the timeline of emerging quantum computers. Based on this consideration, we assign a medium qualitative level for the likelihood of threatening classical cryptosystems by QC (see Figure~}\ref{fig:chart2}{). These assumptions for likelihood can be easily changed to be aligned for another period. We determine the impact based on the quantum security strength of different classical algorithms used within the recommended crypto suite of the classical certificate (see Table}~\ref{tab:Pre-migration-Cert}{). Referring to the established criteria for impact levels in Table}~\ref{table:impact}{ and  analyzing the dependency of certificate quantum threat on the quantum security strength of different classical algorithms, the potential consequences of compromised classical certificates due to QC advancements classify the impact as high. The final risk assessment evaluation is based on likelihood and impact, shown in Table}~\ref{tab:Pre-migration-Cert}.

 \subsection{Through-Migration Certificate Level Analysis and Risk Assessment}
\begin{figure*}[!htpb]
 \begin{center} 
 \begin{subfigure}[t]{.22\linewidth}
 \centering
 \resizebox{\linewidth}{!}{
\begin{tikzpicture}[
 title/.style={minimum height=1cm,minimum width=4cm,font = {\large}},
 body/.style={draw, rounded corners,minimum width=4.5cm,minimum height=0.8cm,font = {\tiny}},
 typetag/.style={rectangle, draw=black!100, anchor=west}]
\node (d3) [title] at (0,0) {
\begin{minipage}{3cm}
\tiny
X509 Certificate
\begin{itemize}[label={\textbullet}, topsep=0ex, leftmargin=5ex, partopsep=0ex, parsep=0ex, itemsep=0.5ex]
 \item TBS Certificate
 \begin{itemize}[label={\textbullet}, topsep=0ex, leftmargin=5ex, partopsep=0ex, parsep=0ex, itemsep=0.5ex]
 \item Version Number,
 \item Serial Number,
 \item Signature Algo. Identifier,
 \item Issuer Name,
 \item Validity Period,
 \item Subject Name,
 \item Subject’s Public Key Info.
 \item Issuer Unique ID,
 \item Subject Unique ID.
 \item X509 (v3) Extensions.
 \end{itemize}
 \item Signature Algorithm
 \item Signature Value
\end{itemize}
\end{minipage}};
\node [top color=white!40, bottom color=blue!70, rounded corners,minimum height=4.8cm,draw=black!100,fill opacity=0.2, fit={ (d3)}] {};
 \end{tikzpicture}
}
 \caption{}\label{fig:HybridCertMech.a}
\end{subfigure}
\hfill
 \begin{subfigure}[t]{.22\linewidth}
 \centering
 \resizebox{\linewidth}{!}{
\begin{tikzpicture}[
 title/.style={minimum height=1cm,minimum width=4cm,font = {\large}},
 body/.style={draw, rounded corners,minimum width=4.5cm,,minimum height=0.8cm,font = {\tiny}},
 typetag/.style={rectangle, draw=back!100, anchor=west}]
\node (d2) [title] at (0,0) {
\begin{minipage}{4cm}
\tiny
X509 Certificate
\begin{itemize}[label={\textbullet}, topsep=0ex, leftmargin=5ex, partopsep=0ex, parsep=0ex, itemsep=0.5ex]
 \item TBS Certificate
 \begin{itemize}[label={\textbullet}, topsep=0ex, leftmargin=5ex, partopsep=0ex, parsep=0ex, itemsep=0.5ex]
 \item Version Number,
 \item $\cdots$
 \item $\cdots$ 
 \end{itemize}
 \item Signature Algorithm
 \item Signature Value
\end{itemize}
\end{minipage}};
\node (222222) [top color=white!40, bottom color=blue!70, rounded corners,minimum height=2.4cm,draw=black!100,fill opacity=0.2, fit={ (d2)}] {};

\node (d3) [title, below=of d2,yshift=0.2cm] {
\begin{minipage}{4cm}
\tiny
PQ X509 Certificate
\begin{itemize}[label={\textbullet}, topsep=0ex, leftmargin=5ex, partopsep=0ex, parsep=0ex, itemsep=0.5ex]
 \item TBS Certificate
 \begin{itemize}[label={\textbullet}, topsep=0ex, leftmargin=5ex, partopsep=0ex, parsep=0ex, itemsep=0.5ex]
 \item Version Number,
 \item $\cdots$
 \item $\cdots$ 
 \end{itemize}
 \item PQ Signature Algorithm
 \item PQ Signature Value
\end{itemize}
\end{minipage}};
\node (d33333)[top color=white!40, bottom color=blue!70, rounded corners,minimum height=2.4cm,draw=black!100,fill opacity=0.2, fit={ (d3)}] {};
 \end{tikzpicture}
}
 \caption{}\label{fig:HybridCertMech.b}
\end{subfigure}
 \hfill
 \begin{subfigure}[t]{.22\linewidth}
 \centering
 \resizebox{\linewidth}{!}{
\begin{tikzpicture}[
 title/.style={minimum height=1cm,minimum width=4cm,font = {\large}},
 body/.style={draw, rounded corners,minimum width=4.5cm,minimum height=0.8cm,font = {\tiny}},
 typetag/.style={rectangle, draw=black!100, anchor=west}]
\node (d3) [title] at (0,0) {
\begin{minipage}{4cm}
\tiny
X509 Certificate
\begin{itemize}[label={\textbullet}, topsep=0ex, leftmargin=5ex, partopsep=0ex, parsep=0ex, itemsep=0.5ex]
 \item TBS Certificate
 \begin{itemize}[label={\textbullet}, topsep=0ex, leftmargin=5ex, partopsep=0ex, parsep=0ex, itemsep=0.5ex]
 \item Version Number,
 \item Serial Number,
 \item Signature Algo. Identifier,
 \item Issuer Name,
 \item Validity Period,
 \item Subject Name,
 \item Subject’s Public Key Info.
 \item Issuer Unique ID,
 \item Subject Unique ID.
 \item X509 (v3) Extensions: 
 \begin{itemize}[label={\textbullet}, topsep=0ex, leftmargin=5ex, partopsep=0ex, parsep=0ex, itemsep=0.5ex]
 \item PQ X509 Certificate
 \begin{itemize}[label={\textbullet}, topsep=0ex, leftmargin=5ex, partopsep=0ex, parsep=0ex, itemsep=0.5ex]
 \item TBS Certificate,
 \item $\cdots$
 \item $\cdots$
 \item Signature Algorithm
 \item Signature Value
 \end{itemize}
 \end{itemize}
 \end{itemize}
 \item Signature Algorithm
 \item Signature Value
\end{itemize}
\end{minipage}};
 \node (d31) [rounded corners,minimum height=4.5cm,draw=red!80,fill opacity=0.2, xshift=0mm,yshift=-1.1cm,dashed,rounded corners,minimum width=4.4cm, minimum height=1.7cm]{};
\node [top color=white!40, bottom color=blue!70, rounded corners,minimum height=4.5cm,draw=black!100,fill opacity=0.2, fit={ (d3) (d31)}] {};
 \node (d4) [draw, below=of d3, rounded corners, xshift=0.25cm,yshift=3.03cm,rounded corners,minimum width=2.8cm, minimum height=1.4cm]{};
 \end{tikzpicture}
}
 \caption{}\label{fig:HybridCertMech.c}
\end{subfigure}
\hfill
 \begin{subfigure}[t]{.22\linewidth}
 \centering
 \resizebox{\linewidth}{!}{
\begin{tikzpicture}[
 title/.style={minimum height=1cm,minimum width=4cm,font = {\large}},
 body/.style={draw, rounded corners,minimum width=4.5cm,minimum height=0.8cm,font = {\tiny}},
 typetag/.style={rectangle, draw=black!100, anchor=west}]
\node (d3) [title] at (0,0) {
\begin{minipage}{3cm}
\tiny
X509 Certificate
\begin{itemize}[label={\textbullet}, topsep=0ex, leftmargin=5ex, partopsep=0ex, parsep=0ex, itemsep=0.5ex]
 \item TBS Certificate
 \begin{itemize}[label={\textbullet}, topsep=0ex, leftmargin=5ex, partopsep=0ex, parsep=0ex, itemsep=0.5ex]
 \item Version Number,
 \item Serial Number,
 \item Signature Algo. Identifier,
 \item Issuer Name,
 \item Validity Period,
 \item Subject Name,
 \item Subject’s Public Key Info.
 \item Issuer Unique ID,
 \item Subject Unique ID.
 \item X509 (v3) Extensions: 
 \begin{itemize}[label={\textbullet}, topsep=0ex, leftmargin=5ex, partopsep=0ex, parsep=0ex, itemsep=0.5ex]
 \item Alt. Public Key Info.
 \item Alt. Signature Algo.
 \item Alt. Signature Value 
 \end{itemize}
 \end{itemize}
 \item Signature Algorithm
 \item Signature Value
\end{itemize}
\end{minipage}};
 \node (d31) [rounded corners,minimum height=4.5cm,draw=red!80,fill opacity=0.2, xshift=0mm,yshift=-1.08cm,dashed,rounded corners,minimum width=4.5cm, minimum height=1cm]{};
\node [top color=white!40, bottom color=blue!70, rounded corners,minimum height=5.35cm,draw=black!100,fill opacity=0.2, fit={ (d3) (d31)}] {};
 \end{tikzpicture}
}
 \caption{}\label{fig:HybridCertMech.d}
\end{subfigure}
\caption{{ Mechanisms for Hybrid Certificate: (a) Original X.509 Certificate, (b) $1^{st}$ Mechanism: X.509 Dual Certificate, (c) $2^{nd}$ Mechanism: PQ Certificate Embedded in the Extension of Classic One, (d) $3^{rd}$ Mechanism: PQ Public Key and Signature Information Embedded in the Extension of Classic Certificate.}}
\label{fig:HybridCertMech}
\end{center}
\end{figure*}

\begin{table*}[!tbh]
\caption{{Through-Migration Certificate Level Analysis}}
\small
\label{tab:Through-Migration-Cert}
\resizebox{\textwidth}{!}{%
\begin{tabular}{|l|p{0.35\linewidth}|p{0.17\linewidth}|p{0.18\linewidth}|p{0.3\linewidth}|p{0.26\linewidth}|p{0.21\linewidth}|}
\hline
\multirow{2}{*}{\textbf{Migration Strategy}} & \multirow{2}{*}{\textbf{Best Practices for Chain of Trust}} & \multirow{2}{*}{\textbf{Approaches}} & \multirow{2}{*}{\textbf{Mechanisms}} & \multirow{2}{*}{\textbf{Pros.}} & \multirow{2}{*}{\textbf{Cons.}} & \multirow{2}{*}{\textbf{Quantum Threats (STRIDE)}} \\ 
& & & & & & \\ \hline

\multirow{20}{*}{Hybrid~\cite{truskovsky-lamps-pq-hybrid-x509-00,driscoll-pqt-hybrid-terminology-02}} & \multirow{3}{*}{\begin{minipage}{\linewidth}
\begin{myBullets}
\vspace{-20pt}
\item Initiate the migration strategy from the root CA, considering the extended validity periods of root CA certificates, ranging from 10 to 25 years~\cite{ImpendRootCA}. Commencing migration at the root level mitigates the challenges associated with complex and time-consuming redistribution of certificates.
\item Acknowledge the pivotal role of the root CA in establishing trust within the PKI. The root CA, being the core of trust, necessitates the implementation of the highest trust levels and well-established signature schemes to ensure the overall security of the public key infrastructure.
\item Recognize the comparatively shorter validity periods of intermediate CA and end-entity certificates, typically ranging between 1 to 10 years~\cite{ImpendRootCA}. Since authentication cannot be retroactively compromised, there is no imperative need for simultaneous migration of all components in the chain of trust for post-quantum cryptography. This phased approach aligns with the evolving cryptographic landscape and facilitates a smoother transition.\\
\vspace{45pt}
\end{myBullets}
\end{minipage}} &
Dual Certificate~\cite{bindel2017transitioning,vogt2021quantum,bindel2019x} & Two Separate Certificates & 
\begin{minipage}{\linewidth}
\begin{myBullets}
\vspace{0.1cm}
\item Only a few changes of standards and applications/devices for PQ one~\cite{vogt2021quantum},
\item Only moderate increase of certificate size for PQ one,
\item Smooth transition to quantum-safe certificates.
\\
\end{myBullets}
\end{minipage}
& 
\begin{minipage}{\linewidth}
\begin{myBullets}
\vspace{0.1cm}

\item Since the subject, issuer, and other metadata are repeated in both, dual certificates are slightly bigger than one,
\item PKI software needs to be changed to manage parallel hierarchies.
\\
\end{myBullets}
\end{minipage}
& 
\multirow{17}{*}{ \begin{minipage}{\linewidth}
\begin{myBullets}
\vspace{0.1cm}
 \item {Spoofing, Tampering, Repudiation, Info. Disclosure, DoS, and Elevation of Privileges (when none of the certificates used in the hybrid approach are secure against them.)}
  \end{myBullets}
 \vspace{0.1cm}
\end{minipage}} \\ \cline{3-6} 
& & \multirow{2}{*}{ \begin{minipage}{\linewidth}
\vspace{30pt}
Composite Certificate \cite{bindel2017transitioning,vogt2021quantum,bindel2019x}\end{minipage} } & Post-Quantum Cert. in Extension & 
 \begin{minipage}{\linewidth}
\begin{myBullets}
\vspace{0.1cm}
\item Combines security of pre- and post-quantum algorithms.
\\
\end{myBullets}
\end{minipage}
& 
 \begin{minipage}{\linewidth}
\begin{myBullets}
\vspace{0.1cm}
\item Since the subject, issuer, and other metadata are repeated in both original and in-extension certificates are slightly bigger than one,
 \item Abrupt migration for all applications at the same time,
 \item Needs changes of standards (e.g., RFC 5280~\cite{cooper2008internet}) for two signatures and two public keys in a certificate,
 \item Size of certificates increases the most.\\
\end{myBullets}
\end{minipage}
&
\\ \cline{4-6} 
 & & & \begin{minipage}{\linewidth} PQ Public key and Signature Information in the Extension \end{minipage} & 
 \begin{minipage}{\linewidth}
\begin{myBullets}
\vspace{0.1cm}
 \item Smooth transition to quantum-safe certificates,
 \item Combines security of pre- and post-quantum algorithms.\\
 \end{myBullets}
\end{minipage}
 & 
 \begin{minipage}{\linewidth}
\begin{myBullets}
\vspace{0.1cm}
 \item Needs changes of standards (e.g., RFC 5280~\cite{cooper2008internet}) to store and verify two signatures and two public keys in a certificate - Size of certificates increases.\\
 \end{myBullets}
\end{minipage} 
 & 

\\ \hline
\end{tabular}%
}
\end{table*}

Organizations navigating the transition to a quantum-safe state recognize the pivotal role of through-migration certificate levels. Central to this process are hybrid strategies, integrating classical and post-quantum solutions. This approach ensures backward compatibility and acts as a bridge between quantum-safe and non-quantum-safe cryptographic states. Seamlessly facilitating connectivity, the hybrid strategy guides organizations through the migration period until the comprehensive transition to a quantum-safe cryptographic state is achieved.

Within the certificate management framework of the hybrid strategy, multiple certificates (e.g., one classical and one post-quantum) are merged into a unified certificate, offering a flexible solution for servers to adapt to clients with diverse cryptography capabilities. The security of this hybrid approach is contingent on the resilience of at least one certificate within the combined set. As long as one certificate remains intact, the hybrid strategy maintains its security, providing a robust and adaptable solution for the evolving cryptographic landscape.

In adherence to RFC5280~\cite{RFC5280}, an X.509 certificate is restricted to one TBS certificate, containing a single subject public key and signed using only one CA signature (as illustrated in Figure~\ref{fig:HybridCertMech.a}). Crafting a hybrid certificate that integrates classical and post-quantum solutions poses a challenge due to this limitation. Various approaches are presented to overcome these challenges, preserving backward compatibility while concurrently supporting classical and post-quantum certificates.

The first approach, known as dual certificates~\cite{bindel2017transitioning,vogt2021quantum,bindel2019x}, involves generating two separate certificates-one designed for classical  algorithms and another for post-quantum algorithms, as depicted in Figure~\ref{fig:HybridCertMech.b}. The second approach, termed the composite certificate~\cite{bindel2017transitioning,vogt2021quantum,bindel2019x}, entails creating a hybrid certificate utilizing an extension mechanism. This extension field operates in two manners: (i) a post-quantum certificate is initially formulated and then set as an extension within the classical certificate, as illustrated in Figure~\ref{fig:HybridCertMech.c}. However, a primary drawback of this approach is the presence of information redundancy, leading to increased certificate sizes due to repeated information in both certificates. This extension mechanism necessitates alterations in standards (e.g., RFC 5280~\cite{RFC5280}) to manage and authenticate two signatures and two public keys within a certificate; (ii) only additional information, such as PQ public key and signature details, is embedded in the extension to mitigate redundancies of similar information in the certificate. This adaptation aims to ensure a seamless transition to quantum-safe certificates, as demonstrated in Figure~\ref{fig:HybridCertMech.d}. Nevertheless, this extension mechanism also requires modifications in standards (e.g., RFC 5280) to handle and validate two signatures and two public keys within a certificate.

For a detailed examination of post-migration certificate-level analysis and risk assessment, including the hybrid strategy, best practices, and different certificate-combining approaches, along with their respective advantages, disadvantages, and potential quantum threats within the STRIDE model, refer to Table~\ref{tab:Through-Migration-Cert}. Due to space limitations, the comprehensive technical details are elaborated within the table itself.

\subsubsection{Through-Migration Certificate Level Risk Assessment} The hybrid certificate approach merges classical and\break post-quantum cryptography, offering a secure and efficient\break method for facilitating communication between classical and quantum secure systems. By integrating classical and quantum secure certificates (see Figure~\ref{fig:HybridCertMech}), this approach ensures the combined certificate's security, provided at least one of the primitives within the combination remains secure~\cite{barker2017recommendation,vogt2021quantum}. Evaluating the security of this hybrid certificate involves scrutinizing the algorithms employed and the combiners utilized (e.g., concatenation and dual nesting). Consequently, the security of the hybrid certificate hinges upon the strength of the most secure certificate within the combination. Hence, the level of risk associated with the hybrid certificate is contingent upon the minimum risk posed by either of the two certificates involved in the combination (see Figure~\ref{figure:hybridrisk}).

 \subsection{Post-Migration Certificate Level Analysis and Risk Assessment}

\begin{table*}[!htbp]
\caption{Post-Migration Certificate Level Analysis and Risk Assessment}

\label{tab:Post-migration-Cert}
\resizebox{\textwidth}{!}{%
\small
\begin{tabular}{|p{0.27\linewidth}|p{0.13\linewidth}|p{0.5\linewidth}|p{0.4\linewidth}|p{0.2\linewidth}|p{0.22\linewidth}|l|l|l|}

\hline
\multirow{2}{*}{\begin{minipage}{\linewidth}\textbf{Possible Quantum-Secure Certificate} \end{minipage}} & \multirow{2}{*}{\textbf{Purpose}} &\multirow{2}{*}{\textbf{Challenges and Attacks}} & \multirow{2}{*}{\textbf{Possible Countermeasures}} & \multirow{2}{*}{\textbf{Recommended Crypto Suite} } & \multirow{2}{*}{\textbf{Quantum Threats (STRIDE)}} & \multirow{2}{*}{\textbf{L}} & \multirow{2}{*}{\textbf{I}} & \multirow{2}{*}{\textbf{R}} \\
 & & & & & & & & \\ \hline

 \begin{minipage}{\linewidth}{ \vspace{0.2cm}A potential quantum-resistant solution for X.509 certificates involves replacing traditional public-key cryptography with post-quantum (PQ) cryptographic algorithms. Additionally, to enhance the security of X.509 certificates, adopting symmetric cryptography with longer keys is recommended. This combination of post-quantum public-key algorithms and stronger symmetric encryption keys helps safeguard X.509 certificates against potential threats posed by quantum computers, which have the capability to break traditional encryption methods.}\vspace{0.2cm}\end{minipage}
& 
 \begin{minipage}{\linewidth}
\begin{myBullets}
\vspace{0.1cm}
 \item Implementing Qunatum-safe PKI and Identity systems.
\\
 \end{myBullets}
\end{minipage} 
& 

\begin{minipage}{\linewidth}
\begin{myBullets}
\vspace{0.2cm}
\item Increased Certificate Size:
\begin{itemize}
\item DoS attacks: Larger certificates may strain network resources and processing capabilities, potentially enabling DoS attacks.
\item Fragmentation and Retransmission: Certificates exceeding the maximum packet size may need fragmentation, leading to performance impact and retransmissions.
\item Network Congestion: Larger certificate sizes can contribute to network congestion, especially under high traffic volume.
\end{itemize}
\item Implementation Challenges:
\begin{itemize}
\item Buffer Overflow Vulnerabilities: Larger post-quantum certificates could trigger buffer overflow vulnerabilities in applications, potentially leading to code injection attacks~\cite{homoliak2014characteristics,madan2005stackoffence,certbuffoverfl}.
\item Limited Rollback: Transitioning to post-quantum algorithms may lack straightforward rollback options if unforeseen security issues arise.
\end{itemize}
\item Side-Channel Attacks: Post-quantum algorithms may be susceptible to side-channel attacks, potentially leaking sensitive information (See  Table~\ref{tab:Post-Migration-Alg}).
\end{myBullets}
\end{minipage}
& 
 \begin{minipage}{\linewidth}
\begin{myBullets}
\vspace{0.2cm}
 \item To prevent DoS: providing required bandwidth, building redundancy into infrastructure, deploying DDoS resilience hardware/software modules like firewalls, adopting DDoS Protection Appliance, configuring network hardware against DDoS attacks, etc~\cite{garcia2022deep,liu2018practical}.
 \item To avoid fragmentation and trigger retransmission: considering the extreme size of a typical message for the case when the quantum-safe certificate is used instead of a classic one~\cite{muller2020retrofitting,essay89509}.
 \item To control congestion: using congestion control mechanisms to prevent or remove congestion~\cite{jay2018internet,bohloulzadeh2020survey,jay2019deep}. 
 \item To immune against stack-based {buffer overflow} for certificate on servers by malicious clients: Canary, DEP (Data Execution Prevention, ASLR (Address Space Layout Randomization)~\cite{zhou2022final,nicula2019exploiting}. 
 \item To immune against no fall back: adopting crypto-agility and supporting multiple quantum-safe algorithms~\cite{ma2021caraf,ott2019identifying,ma2021crypto}.
 
 \item To immune against side-channel attacks, refer to the solutions mentioned in Table~\ref{tab:Post-Migration-Alg}.\\
 \end{myBullets}
\end{minipage} 
&
 \begin{minipage}[!b]{\linewidth}
\begin{myBullets}
 \item All the algorithms mentioned in Table~\ref{tab:Post-Migration-Alg}
 \end{myBullets}
 \end{minipage} 
 & \begin{minipage}{\linewidth}
\begin{myBullets}
\vspace{0.1cm}
 \item {Info. Disclosure: Potential leakage of sensitive information through side-channel attacks on PQ algorithms as mentioned in Table}~\ref{tab:Post-Migration-Alg},
 \item {DoS: Increase in certificate size due to transitioning to post-quantum cryptography, leading to the possibility of Denial of Service attacks. }\\
 \end{myBullets}
\end{minipage} & \med& \med& \med\\ \hline
\end{tabular}%
}\vspace{3pt}

\begin{tablenotes}
\scriptsize
 \item[a] \tiny{$^*$ We perform risk evaluation with the presumption of considering the countermeasures mentioned in the table.}
\end{tablenotes}
\end{table*}

As previously discussed, an X.509 certificate has a TBS certificate containing a subject public key, which is signed using a CA signature~\cite{RFC5280} (as shown in Figure~\ref{fig:HybridCertMech.a}). Since classic cryptographic algorithms are used in X.509 certificates (e.g., subject public key and CA signature), they are vulnerable to quantum attacks using Shor’s and Grover's algorithms. In this section, we analyze possible approaches to develop a quantum-resistant certificate, evaluate its applicable challenges and attacks, describe possible countermeasures, determine QC threats, and assess the risks associated with the threats.

The NIST post-quantum cryptographic algorithms, detailed in Table~\ref{tab:Post-Migration-Alg}, aim to create cryptographic standard algorithms resistant to quantum attacks. These algorithms could be used to form a quantum-safe X.509 certificate, presenting a viable option for developing a quantum-resistant certificate. While enabling the implementation of a quantum-safe public key infrastructure and identity system post-migration, this approach introduces challenges due to post-quantum algorithms, including certificate size increase, potential denial of service, fragmentation, retransmission issues, network congestion, and buffer overflow. Quantum attackers may exploit these vulnerabilities to inject malicious code or attempt data exfiltration via side-channels, as indicated in Table~\ref{tab:Post-Migration-Alg}.
Table~\ref{tab:Post-migration-Cert} thoroughly explores these challenges and potential attacks, offering a comprehensive discussion of countermeasures to immunize against them. These countermeasures are designed to mitigate various threats, including denial of service~\cite{garcia2022deep, liu2018practical}, fragmentation and retransmission~\cite{muller2020retrofitting, essay89509}, network congestion~\cite{jay2018internet, bohloulzadeh2020survey, jay2019deep}, and buffer overflow~\cite{zhou2022final, nicula2019exploiting}. By addressing these issues head-on, these countermeasures contribute to fortifying the resilience of quantum-resistant certificates.

\subsubsection{Post-Migration Certificate Level Risk Assessment} In our comprehensive risk analysis, we methodically evaluate the likelihood and impact of potential quantum threats. Reference points for the evaluation criteria related to likelihood and impact can be found in Table~\ref{table:likelihood} and Table~\ref{table:impact}, respectively. For the likelihood assessment, considering that all the attacks outlined in Table \ref{tab:Post-migration-Cert} can potentially be launched from the Internet or network and effective countermeasures are available, we designate the likelihood as medium. To assess the impact, we consider that a quantum attacker's threats may result in restricted damage to infrastructure, unwanted functionality, or disclosure of personal data or secrets. Based on the evaluation criteria outlined in Table~\ref{table:impact}, we conclude that the level of impact should be considered medium. The final risk assessment evaluation is based on likelihood and impact, shown in Table~\ref{tab:Post-migration-Cert}.

\section{Quantum Migration Threat Analysis and Risk Assessment for Protocol Level}\label{sec:PLA}

Security protocols are widely used throughout organizations to authenticate the origin and protect the confidentiality and integrity of the information that is communicated and stored. Current security protocols, such as SSL and TLS, which rely on public-key algorithms, are effective at preventing classical computer attacks on network communications. However, the emergence of a fault-tolerant quantum computer could threaten the security of these and other protocols by compromising the underlying mathematical challenges in mere hours or seconds~\cite{Deloite}.

 Several protocols are available today, but our investigation focuses on the analysis of widely used standard security protocols as outlined in the Canadian National Quantum-Readiness guidelines~\cite{CFDIR}. This analysis encompasses the identification of various vulnerabilities that could be exploited by a quantum attacker at each stage of the migration process for each protocol. Employing the STRIDE threat model, we systematically pinpoint Quantum Computing (QC) threats in each migration stage and assess the associated risks emanating throughout the entirety of the migration process.\\

\subsection{Pre-Migration Protocol Level Analysis and Risk
Assessment}

\begin{table*}[!htbp]
\small
\caption{Pre-Migration Protocol Level Analysis and Risk Assessment}
\resizebox{\textwidth}{!}{%
\begin{tabular}{|l|p{0.25\linewidth}|p{0.3\linewidth}|p{0.3\linewidth}|p{0.35\linewidth}|l|l|l|p{0.3\linewidth}|}
\hline
\multirow{2}{*}{\textbf{Protocols}} & \multirow{2}{*}{\textbf{Main Components}} & \multirow{2}{*}{\textbf{Purposes}} & \multirow{2}{*}{\textbf{Crypto Suites the Protocols Uses}} & \multirow{2}{*}{\textbf{Quantum Threats (STRIDE)}}& \multirow{2}{*}{\textbf{L}} & \multirow{2}{*}{\textbf{I}} & \multirow{2}{*}{\textbf{R}} & \multirow{2}{*}{\textbf{Possible QC-resistant Solutions}}\\
& & & & & & & &\\\hline
{SSH (v2)}~\cite{RFC4251,RFC4252,RFC4253,RFC4254,RFC4256,RFC4335,RFC4419,RFC5656,RFC8308}
&
{\begin{minipage}{\linewidth}
\begin{myBullets}
\vspace{0.2cm}
\item Transport Layer Protocol 
\item User Authentication Protocol 
\item Connection Protocol\\
\end{myBullets}
\end{minipage}} &
{\begin{minipage}{\linewidth}
\begin{myBullets}
\vspace{0.2cm}
 \item Providing secure remote login and other secure network services over an insecure network,
 \item Supporting authentication, integrity, and confidentiality.
 \\
\end{myBullets}
\end{minipage}}
 &
 {\begin{minipage}{\linewidth}
\begin{myBullets}
\vspace{0.2cm}
 \item Public-Key Crypto: EdDSA, ECDSA, RSA and DSA, ECDH and DH (broken by Shor's Algo.)
\item Symmetric
Crypto: AES, RC4, 3DES, DES, ChaCha20-Poly1305, SHA1/SHA2, MD5 (Weakened by Grover's Algo.)
\\
\end{myBullets}
\end{minipage}}
 & {\begin{minipage}{\linewidth}
\begin{myBullets}
\vspace{0.2cm}

 \item  {Spoofing: Vulnerabilities in cryptographic algorithms used for key exchange and user authentication, susceptible to Shor's algorithm, can be exploited.}
 \item  {Tampering: Cryptographic primitives like HMAC-SHA1/SHA2, ensuring integrity protection, can be broken, allowing unauthorized data modification.}
 \item  {Info. Disclosure: Weaknesses in encryption algorithms like AES in SSH can lead to the disclosure of sensitive information transmitted over SSH connections.}
 \\
\end{myBullets}
\end{minipage}} & \med&\high & \high &
{\begin{minipage}{\linewidth}
\begin{myBullets}
\vspace{0.2cm}
 \item OQS-OpenSSH~\cite{qs-openssh,oqs-ssh,oqs-liboqs},
 \item OQS-libssh~\cite{oqs-ssh,oqs-liboqs}.
 \\
\end{myBullets}
\end{minipage}}
\\ \hline
TLS (v1.3)\cite{RFC2246,RFC4346,RFC5246,RFC8446,RFC7525,RFC8447,RFC7627}& 
{\begin{minipage}{\linewidth}
\begin{myBullets}
\vspace{0.2cm}
\item Handshake Protocol,
\item Record Protocol,
\item Change Cipher Spec Protocol,
\item Alert Protocol.\\
\end{myBullets}
\end{minipage}}
 & 
{\begin{minipage}{\linewidth}
\begin{myBullets}
\vspace{0.2cm}
 \item Providing communications security~\cite{rescorla2018transport}.
 \item Supporting authentication, integrity, confidentiality.\\
\end{myBullets}
\end{minipage}}
& 
{\begin{minipage}{\linewidth}
\begin{myBullets}
\vspace{0.2cm}
 \item Public-Key Crypto: RSA, ECDH (broken by Shor's Algo.)
\item Symmetric
Crypto: AES, SHA2, ChaCha20-Poly1305 (Weakened by Grover's Algo.)
\\
\end{myBullets}
\end{minipage}}
& {\begin{minipage}{\linewidth}
\begin{myBullets}
\vspace{0.2cm} 
 \item  {Spoofing: Compromising key exchange mechanisms allows impersonation of legitimate servers or clients.}
 \item {Tampering: Breaking the integrity protection mechanisms like HMAC-SHA2, allowing them to modify data in transit.}
 \item  {Info. Disclosure: Exploiting encryption algorithm weaknesses leads to sensitive information disclosure in TLS connections.}\\
\end{myBullets}
\end{minipage}} & \med &\high&\high& 
{\begin{minipage}{\linewidth}
\begin{myBullets}
\vspace{0.2cm}
 \item OQS-OpenSSL~\cite{post-quantum-tls, open-quantum-safe-tls},
 \item KEMTLS (post-quantum version of TLS in which post-quantum KEMs are used instead of signatures for handshake authentication)~\cite{schwabe2020post}.\\
\end{myBullets}
\end{minipage}}
 \\ \hline
mTLS~\cite{RFC5246,RFC8446}& 
{\begin{minipage}{\linewidth}
\begin{myBullets}
\vspace{0.2cm}
\item Handshake Protocol,
\item Record Protocol,
\item Change Cipher Spec Protocol,
\item Alert Protocol.\\
\end{myBullets}
\end{minipage}}
 & 
{\begin{minipage}{\linewidth}
\begin{myBullets}
\vspace{0.2cm}
 \item Providing communications security.
 \item Supporting authentication, integrity, confidentiality.\\
\end{myBullets}
\end{minipage}}
& 
{\begin{minipage}{\linewidth}
\begin{myBullets}
\vspace{0.2cm}
 \item Public-Key Crypto: RSA, ECDH (broken by Shor's Algo.)
\item Symmetric
Crypto: AES, SHA2, ChaCha20-Poly1305 (Weakened by Grover's Algo.)
\\
\end{myBullets}
\end{minipage}}
& {\begin{minipage}{\linewidth}
\begin{myBullets}
\vspace{0.2cm}
 \item {Spoofing, Tampering, Info. Disclosure: Since mTLS builds on TLS, it inherits the same vulnerabilities.}\\
\end{myBullets}
\end{minipage}} & \med &\high&\high& 
{\begin{minipage}{\linewidth}
\begin{myBullets}
\vspace{0.2cm}
 \item Mutual use of OQS-OpenSSL~\cite{post-quantum-tls, open-quantum-safe-tls} by both parties,
 \item Mutual use of KEMTLS~\cite{schwabe2020post} by both parties.\\
\end{myBullets}
\end{minipage}}
 \\ \hline

sFTP~\cite{secsh-filexfer,draft-ietf-secsh-filexfer-13} & {\begin{minipage}{\linewidth}
\begin{myBullets}
\vspace{0.2cm}
\item SFTP does not have distinct sub-protocols; however it operates as a subsystem of the SSH protocol, utilizing the secure communication channel established by SSH for file transfer.\\
\end{myBullets}
\end{minipage}} &{\begin{minipage}{\linewidth}
\begin{myBullets}
\vspace{0.2cm}
 \item Providing secure access, transfer, and management of files over any reliable data stream via the Secure Shell (sSH).

 \item Supporting client/server authentication, integrity and confidentiality.\\
\end{myBullets}
\end{minipage}} & 
{\begin{minipage}{\linewidth}
\begin{myBullets}
\vspace{0.2cm}
 \item Public-Key Crypto: RSA, DSS, DH (broken by Shor's Algo.)
\item Symmetric
Crypto: 3DES, blowfish, twofish, serpent,IDEA, CAST, AES, HMAC- (MD5,SHA1) (Weakened by Grover's Algo.)
\\
\end{myBullets}
\end{minipage}}
& {\begin{minipage}{\linewidth}
\begin{myBullets}
\vspace{0.2cm}
 \item {Spoofing, Tampering, Info. Disclosure: sFTP utilizes SSH for secure communication, making it susceptible to the same quantum attacks as SSH described earlier.} 
 \\
\end{myBullets}
\end{minipage}}& \med & \high& \high& 
 {\begin{minipage}{\linewidth}
\begin{myBullets}
\vspace{0.2cm}
 \item sFTP in which SSH is replaced by OQS-OpenSSH~\cite{post-quantum-SSH} or OQS-libssh~\cite{oqs-ssh}.
\end{myBullets}
\end{minipage}}
\\ \hline
FTPS~\cite{RFC4217} & {\begin{minipage}{\linewidth}
\begin{myBullets}
\vspace{0.2cm}
    \item Explicit FTPS (FTPES), in which the client requests security on port 21 for SSL/TLS.
    \item Implicit FTPS, in which security is auto-initiated on the connection to the server on port 990, operating as SSL/TLS.\\
\end{myBullets}
\end{minipage}}
&{\begin{minipage}{\linewidth}
\begin{myBullets}
\vspace{0.2cm}
 \item Providing security support for File Transfer Protocol (FTP) via the use of Transport Layer Security (TLS).
 
 \item Support connection authentication, integrity and confidentiality.\\
\end{myBullets}
\end{minipage}}

&{\begin{minipage}{\linewidth}
\begin{myBullets}
\vspace{0.2cm}
 \item Public-Key Crypto: RSA, ECDH (broken by Shor's Algo.)
\item Symmetric
Crypto: AES, SHA2, ChaCha20-Poly1305 (Weakened by Grover's Algo.)
\\
\end{myBullets}
\end{minipage}}
& {\begin{minipage}{\linewidth}
\begin{myBullets}
\vspace{0.2cm}
  \item {Spoofing, Tampering, Info. Disclosure:
FTPS relies on TLS for security, inheriting the vulnerabilities mentioned for TLS.}\\
\end{myBullets}
\end{minipage}} & \med &\high & \high& 
{\begin{minipage}{\linewidth}
\begin{myBullets}
\vspace{0.2cm}
 \item FTPS which provides file transfer via OQS-OpenSSL~\cite{post-quantum-tls, open-quantum-safe-tls} or KEMTLS~\cite{schwabe2020post} instead of TLS.\\
\end{myBullets}
\end{minipage}}
 \\ \hline 

SAML (v2)~\cite{RFC7522} &
{\begin{minipage}{\linewidth}
\begin{myBullets}
\vspace{0.2cm}
\item SAML Authentication Request Protocol,
\item SAML Single Logout Protocol:.
\end{myBullets}
\end{minipage}}
& 
{\begin{minipage}{\linewidth}
\begin{myBullets}
\vspace{0.2cm}
 \item Providing authentication to multiple applications
 \item Supporting authentication and authorization 
\\
\end{myBullets}
\end{minipage}}
 & {\begin{minipage}{\linewidth}
\begin{myBullets}
\vspace{0.2cm}
 \item Public-Key Crypto: RSA, DSA (broken by Shor's Algo.)
\item Symmetric Crypto: SHA2 (Weakened by Grover's Algo.)
 \\
\end{myBullets}
\end{minipage}} &
 {\begin{minipage}{\linewidth}
\begin{myBullets}
\vspace{0.2cm}
 \item {Spoofing: Shor's algorithm can break RSA and DSA used in SAML, allowing  to forge assertions and impersonate legitimate users, leading to unauthorized access.}

\item {Elevation of Privilege:  Once a quantum attacker spoofs a SAML assertion and gains system access, they can exploit system vulnerabilities to escalate their privileges.}
 \\
\end{myBullets}
\end{minipage}}
 & \med&\high & \high & {\begin{minipage}{\linewidth}
\begin{myBullets}
\vspace{0.2cm}
 \item SAML in which RSA or DSA, as two recommended public-key algorithms in crypto suites, are replaced by a PQ one. 
 There is no limitation on the extreme size of typical public keys used for SAML; thus, any public-key PQ crypto algorithms can be used for PQ one. For symmetric crypto, longer keys should be used. The certificate used in SAML should be replaced by PQ one as mentioned in Section~\ref{sec:PKI}. 
 Optional use of TLS should be replaced with the optional use of possible PQ TLS solutions mentioned in Table~\ref{tab:Pre-migration-protocol}. \\
\end{myBullets}
\end{minipage}}\\ \hline
OAuth (v2)~\cite{RFC6749,RFC6750,RFC6819,RFC7662,RFC7636} & 
{\begin{minipage}{\linewidth}
\begin{myBullets}
\vspace{0.2cm}
 \item OAuth 2.0 itself doesn't have sub-protocols. \\
\end{myBullets}
\end{minipage}}
& 
{\begin{minipage}{\linewidth}
\begin{myBullets}
\vspace{0.2cm}
 \item Granting a website or application via assertion token to access resources hosted by other web apps on behalf of a user,
 \item Supporting authorization,
 \item Token protection via signature.
 \item Optional use of TLS to pass tokens securely.\\
\end{myBullets}
\end{minipage}}
& {\begin{minipage}{\linewidth}
\begin{myBullets}
\vspace{0.2cm}
 \item Public-Key Crypto: RSA, ECDHE (Use via TLS broken by Shor's Algo.)
\item Symmetric Crypto: AES, SHA2, ChaCha20-
Poly1305 (Used via TLS and weakened by Grover's Algo.), HMAC-SHA1 for Token signature (Weakened by Grover's Algo.)
\\
\end{myBullets}
\end{minipage}}&
 {\begin{minipage}{\linewidth}
\begin{myBullets}
\vspace{0.2cm}
 \item  {Elevation of Privilege: Vulnerabilities in the cryptographic mechanisms used for token protection, such as HMAC-SHA1, can lead to unauthorized access to resources.}
 \\
\end{myBullets}
\end{minipage}}
& \med& \med & \med & {\begin{minipage}{\linewidth}
\begin{myBullets}
\vspace{0.2cm}
\item OAuth (v2) in which tokens signature (i.e., HMAC-SHA1) is replaced by a PQ option. For token signature, longer keys should be used considering the maximum length of access tokens should be 2048 bytes.
 There is no cryptographic mechanism except the optional use of TLS considered for OAuth (v2) mentioned in RFC 6819~\cite{RFC6819,lodderstedt2013oauth} and RFC6749~\cite{RFC6749}. PQ TLS should be used instead of TLS as mentioned Table~\ref{tab:Post-migration-Cert}.\\

\end{myBullets}
\end{minipage}}\\ \hline
IKE (v2)~\cite{RFC7296} & {\begin{minipage}{\linewidth}
\begin{myBullets}
\vspace{0.2cm}
\item IKE\_SA Establishment
  \item Child SA Establishment
  \item Authentication
  \item Key Exchange
  \item Encrypted Payloads
  \item Integrity Protection 
  \vspace{0.2cm}
\end{myBullets}
\end{minipage}}& 
{\begin{minipage}{\linewidth}
\begin{myBullets}
\vspace{0.2cm}
 \item Providing Security Association (SA) setup, policy negotiation, and key management.
 \item Supporting key exchange, EAP authentication, integrity, and confidentiality. \\
\end{myBullets}
\end{minipage}}
& 
{\begin{minipage}{\linewidth}
\begin{myBullets}
\vspace{0.2cm}
 \item Public-Key Crypto: DH (broken by Shor's Algo.)
\item Symmetric
Crypto: AES, HMAC-SHA1/SHA2, 3DES, MD5, ChaCha20-Poly1305 (Weakened by Grover's Algo.)
\\
\end{myBullets}
\end{minipage}}
& {\begin{minipage}{\linewidth}
\begin{myBullets}
\vspace{0.2cm}
\item {Spoofing: Vulnerabilities in IKE's DH cryptography, prone to Shor's algorithm, allow impersonation, enabling potential traffic interception or manipulation during negotiations.}
\item {Tampering: Weakened hashing functions (HMAC-SHA1/SHA2) in IKE, susceptible to Grover's algorithm, allow undetected modification of data packets during exchanges.}
\item {Info. Disclosure: Symmetric key algorithms in IKE are susceptible to Grover's algorithm, enabling quantum attackers to decrypt sensitive information transmitted over IKE-secured connections, threatening confidentiality.}
 \\
\end{myBullets}
\end{minipage}} & \med& \high & \high
&{\begin{minipage}{\linewidth}
\begin{myBullets}
\vspace{0.2cm}
 \item PQ IKE mentioned in RFC 8784\cite{RFC8784}: ``Mixing Preshared Keys in the IKE (v2) for Post-quantum Security''~\cite{fluhrer2020mixing}. \\
\end{myBullets}
\end{minipage}} \\ \hline
IPsec~\cite{RFC4301} & 
{\begin{minipage}{\linewidth}
\begin{myBullets}
\vspace{0.2cm}
 \item IKE (Internet Key Exchange) 
 \item AH (Authentication Header)
 \item ESP (Encapsulating Security Protocol)\\
\end{myBullets}
\end{minipage}}
 & 
 {\begin{minipage}{\linewidth}
\begin{myBullets}
\vspace{0.2cm}
 \item Providing secure authenticated reliable communication.
 \item Supporting data origin authentication, connection-less integrity, confidentiality.\\
\end{myBullets}
\end{minipage}}
 & 
 {\begin{minipage}{\linewidth}
\begin{myBullets}
\vspace{0.2cm}
 \item Public-Key Crypto: DH, ECDH, RSA. ECDSA (broken by Shor's Algo.) 
\item Symmetric
Crypto: AES, HMAC-SHA1/SHA2, 3DES, MD5, ChaCha20-Poly1305 (Weakened by Grover's Algo.)
\\
\end{myBullets}
\end{minipage}}
 &
 {\begin{minipage}{\linewidth}
\begin{myBullets}
\vspace{0.2cm}
   \item {Spoofing: Quantum algorithms such as Shor's can compromise public-key cryptography (DH, ECDH, RSA) used in IKE, enabling attackers to impersonate legitimate devices through forged messages.}
  \item {Tampering: IPSec's AH protocol, utilizing hashing functions is vulnerable to Grover's Algorithm, enabling undetected data packet modification during transit, risking sensitive info. corruption.}
\item {Info. Disclosure: Quantum attackers can use Grover's algorithm to exploit weaknesses in IPSec's symmetric encryption and pre-shared keys (PSKs). This could lead to the decryption of sensitive data traveling within IPSec tunnels if the key lengths are not sufficiently long.}
 \\
\end{myBullets}
\end{minipage}}
 & \med& \high & \high & 
 {\begin{minipage}{\linewidth}
\begin{myBullets}
\vspace{0.15cm}
 \item OpenVPN~\cite{post-quantum-vpn},
 \item StrongSwan~\cite{strongSwan},
 \item WireGuard~\cite{hulsing2021post}.\\
\end{myBullets}
\end{minipage}}
 \\ \hline
Kerberos (v5)~\cite{neuman2005kerberos} &
{\begin{minipage}{\linewidth}
\begin{myBullets}
\vspace{0.2cm}
 \item AS (Authentication Service) Exchange
 \item TGS (Ticket Granting Service) Exchange
 \item CS (Client/Server) Exchange \\
\end{myBullets}
\end{minipage}}
& {\begin{minipage}{\linewidth}
\begin{myBullets}
\vspace{0.2cm}
 \item Providing a mechanism for authenticating access to systems over an untrusted network like the Internet.
 \item Supporting both client and server authentication in client/server applications.\\
\end{myBullets}
\end{minipage}}
&
{\begin{minipage}{\linewidth}
\begin{myBullets}
\vspace{0.2cm}
 \item Public-Key Crypto: -
\item Symmetric
Crypto: HMAC/AES, MD4, MD5, HMAC-SHA1/SHA2/SHA3, CMAC/camellia (Weakened by Grover's Algo.)
\\
\end{myBullets}
\end{minipage}}
& 
 {\begin{minipage}{\linewidth}
\begin{myBullets}
\vspace{0.2cm}
 \item {Spoofing: Quantum attackers can exploit vulnerabilities in Kerberos' symmetric cryptography, like HMAC-SHA1/SHA2/SHA3, using Grover's algorithm. This could accelerate brute-force attacks on short symmetric keys, allowing attackers to forge Kerberos tickets and impersonate legitimate users, potentially granting unauthorized access to protected resources.}
 \\
\end{myBullets}
\end{minipage}}
& \med& \med& \med& {\begin{minipage}{\linewidth}
\begin{myBullets}
\vspace{0.2cm}
 \item Still valid with longer symmetric keys (avoid obsolete schemes like DES).
 \\
\end{myBullets}
\end{minipage}} \\ \hline
\end{tabular}%
}
\label{tab:Pre-migration-protocol}
\end{table*}

\addtocounter{table}{-1}

\begin{table*}[!htbp]
\small
\caption{(Cont.) Pre-Migration Protocol Level Analysis and Risk Assessment}
\resizebox{\textwidth}{!}{%
\begin{tabular}{|l|p{0.25\linewidth}|p{0.3\linewidth}|p{0.3\linewidth}|p{0.35\linewidth}|l|l|l|p{0.3\linewidth}|}
\hline
\multirow{2}{*}{\textbf{Protocols}} & \multirow{2}{*}{\textbf{Main Components}} & \multirow{2}{*}{\textbf{Purposes}} & \multirow{2}{*}{\textbf{Crypto Suites the Protocols Uses}} & \multirow{2}{*}{\textbf{Quantum Threats (STRIDE)}}& \multirow{2}{*}{\textbf{L}} & \multirow{2}{*}{\textbf{I}} & \multirow{2}{*}{\textbf{R}} & \multirow{2}{*}{\textbf{Possible QC-resistant Solutions}}\\
& & & & & & & &\\\hline
LDAP (v3)~\cite{RFC4511} & 
{\begin{minipage}{\linewidth}
\begin{myBullets}
\vspace{0.2cm}
\item {LDAP Bind Protocol}
    \item {LDAP Search Protocol}
    \item {LDAP Compare Protocol}
    \item {LDAP Add Protocol}
    \item {LDAP Delete Protocol}
    \item {LDAP Modify Protocol}
\end{myBullets}
\end{minipage}}
    &
{\begin{minipage}{\linewidth}
\begin{myBullets}
\vspace{0.2cm}
 \item Providing directory services access and authorization, and maintenance for distributed directory information services over an IP network.
 \item Optional supporting of authentication via binding with (a) no authentication, (b) basic authentication, or (c) Simple Authentication and Security Layer (SASL).
 \item Optional supporting communication confidentiality and data integrity via TLS.
 \\
\end{myBullets}
\end{minipage}}
& 
{\begin{minipage}{\linewidth}
\begin{myBullets}
\vspace{0.2cm}
 \item Public-Key Crypto: optional use of RSA, ECDH via TLS as mentioned in RFC 8446~\cite{rescorla2018transport} (broken by Shor's Algo.)
\item Symmetric Crypto: optional use of AES, SHA2, ChaCha20-Poly1305 via TLS as mentioned in RFC 8446~\cite{rescorla2018transport}, optional use of CRAM-MD5 via SASL as mentioned in RFC 4422~\cite{melnikov2006simple}, (Weakened by Grover's Algo.)
\\
\end{myBullets}
\end{minipage}}
&
{\begin{minipage}{\linewidth}
\begin{myBullets}
\vspace{0.2cm}
 \item {Spoofing: SASL in LDAP uses symmetric cryptography (e.g., CRAM-MD5) susceptible to Grover's algorithm, enabling brute-force attacks to forge credentials.}
 \item {Tampering:LDAP's use of TLS with RSA/ECDH, which are vulnerable to Shor's algorithm, could allow attackers to alter data packets undetected.}
 \item {Info. Disclosure: Symmetric cryptography in TLS (e.g., AES, ChaCha20-Poly1305) is weakened by Grover's algorithm, making it easier for quantum attackers to decrypt communications.}
 \item {Elevation of Privilege: Successful spoofing or tampering can enable attackers to escalate privileges within LDAP, gaining unauthorized access.}\\
\end{myBullets}
\end{minipage}}
& \med& \med & \med &
{\begin{minipage}{\linewidth}
\begin{myBullets}
\vspace{0.2cm}
 \item LDAP in which (a) passwords or symmetric keys used in SASL are longer and more secure against birthday attack, (b) optional use of TLS is replaced by OQS-OpenSSL~\cite{post-quantum-tls, open-quantum-safe-tls} or KEMTLS~\cite{schwabe2020post}, and (c) PQ certificate used instead of the classic one as mentioned in Section~\ref{sec:PKI}.\vspace{0.1cm}
 \end{myBullets}
\end{minipage}}\\ \hline
PGP~\cite{RFC4880,RFC1991}& {\begin{minipage}{\linewidth}
\begin{myBullets}
\vspace{0.2cm}
     \item PGP does not have standalone sub-protocol.
\end{myBullets}
\end{minipage}} & 
{\begin{minipage}{\linewidth}
\begin{myBullets}
\vspace{0.2cm}
 \item Providing privacy and authentication for data communication via encrypted emails/files.
 \item Supporting authentication, integrity, non-repudiation, confidentiality.\\
\end{myBullets}
\end{minipage}}
 & 
{\begin{minipage}{\linewidth}
\begin{myBullets}
\vspace{0.2cm}
 \item Public-Key Crypto: RSA, Elgamal, DH, DSA (broken by Shor's Algo.).
\item Symmetric
Crypto: MD5, SHA1, CAST, IDEA, or Triple-DES (Weakened by Grover's Algo.).
\\
\end{myBullets}
\end{minipage}}
 & 
 {\begin{minipage}{\linewidth}
\begin{myBullets}
\vspace{0.2cm}
 \item {Spoofing:  Quantum computers can leverage Shor's algorithm to forge digital signatures in PGP. This allows impersonation of legitimate users, potentially leading to scams or unauthorized actions.}
 \item {Tampering:  Quantum attackers leverage Shor's algorithm to forge digital signatures, enabling message tampering. Moreover, Grover's Algorithm facilitates brute-forcing symmetric encryption, enabling undetected modification of encrypted data packets during transmission.}
 \item  {Repudiation: Quantum attackers can exploit Shor's algorithm to forge digital signatures, allowing them to frame others for sending messages or deny sending messages themselves.}
 \item  {Info. Disclosure: Quantum attackers leverage Grover's Algorithm to compromise PGP's symmetric cryptography, potentially decrypting confidential communication and compromising sensitive information.}\\
\end{myBullets}
\end{minipage}}
 & \med& \high & \high & 
 
 {\begin{minipage}{\linewidth}
\begin{myBullets}
\vspace{0.2cm}
 \item OpenPGP~\cite{callas2007openpgp} in which classic public-key crypto is replaced by PQ one. 
 The extreme size of typical public keys used for OpenPGP (max 4096 bit) can be a problem in the case when PQ public-key crypto is used instead of the classic one. For symmetric crypto, longer keys should be used.\\
\end{myBullets}
\end{minipage}}\\ \hline
S/MIME (v4)~\cite{RFC8551}&  {\begin{minipage}{\linewidth}
\begin{myBullets}
\vspace{0.2cm}
         \item Cryptographic Message Syntax (CMS),
    \item Public Key Cryptography Standards (PKCS),
    \item X.509 Certificates,
    \item Certificate Authorities (CAs),
    \item Secure Hash Algorithm (SHA),
    \item RSA and other Crypto Algorithms.
    \vspace{0.2cm}
\end{myBullets}
\end{minipage}} & 
{\begin{minipage}{\linewidth}
\begin{myBullets}
\vspace{0.2cm}
 \item Providing privacy and data security for electronic messaging. 
 \item Supporting authentication, integrity, non-repudiation, confidentiality
\\
\end{myBullets}
\end{minipage}}
&
{\begin{minipage}{\linewidth}
\begin{myBullets}
\vspace{0.2cm}
 \item Public-Key Crypto: RSA, DSA, Elliptic Curve (broken by Shor's Algo.)
\item Symmetric
Crypto: AES (Weakened by Grover's Algo.)
\\
\end{myBullets}
\end{minipage}}
& 
{\begin{minipage}{\linewidth}
\begin{myBullets}
\vspace{0.2cm}


\item {Spoofing, Tampering, and Repudiation: Shor's algorithm exploits vulnerabilities in S/MIME's public-key cryptography used for digital signatures, allowing attackers to forge certificates and impersonate users. This enables man-in-the-middle attacks to tamper with encrypted messages, leading to both tampering and repudiation.}
\item {Tampering, Repudiation, and Information Disclosure: Grover's algorithm weakens symmetric keys in S/MIME, enabling attackers to tamper with encrypted messages during transmission, compromising data integrity and non-repudiation. This poses a risk of tampering, repudiation, and information disclosure.}\\

\end{myBullets}
\end{minipage}}
& \med& \high & \high & 
 {\begin{minipage}{\linewidth}
\begin{myBullets}
\vspace{0.2cm}
 \item S/MIME in which classic public-key crypto is replaced by PQ one. 
 There is no limitation on the extreme size of typical public keys used for S/MIME~\cite{schaad2019secure}. Thus any public-key PQ sign/enc algorithms can be used for PQ one. For symmetric crypto, longer keys should be used. PQ certificate should be used as mentioned in Section~\ref{sec:PKI}.\\
\end{myBullets}
\end{minipage}}\\ \hline
WiFi/WPA (v3)~\cite{RFC5416,RFC5417,RFC5418,RFC4282,RFC6418,RFC5413}& 
{\begin{minipage}{\linewidth}
\begin{myBullets}
\vspace{0.2cm}
\item SAE (Simultaneous Authentication of Equals),
\item Dragonfly (Password-Authenticated Key Agreement, or PAKE).\\
\end{myBullets}
\end{minipage}} &
{\begin{minipage}{\linewidth}
\begin{myBullets}
\vspace{0.2cm}
 \item Providing a more secure handshake using Wi-Fi DPP and creating secure wireless (Wi-Fi) networks,
 \item Supporting authentication, confidentiality, and integrity via EAP (EAP-TLS Enterprise), authenticated encryption, HMAC, and secure hash algorithm. \\
\end{myBullets}
\end{minipage}}
&
{\begin{minipage}{\linewidth}
\begin{myBullets}
\vspace{0.2cm}
 \item Public-Key Crypto: RSA, ECDH key
exchange and ECDSA (broken by Shor's Algo.).
\item Symmetric Crypto: AES, HMAC-SHA-3, AES, GCMP, BIP (Weakened by Grover's Algo.)
\\
\end{myBullets}
\end{minipage}}
& 
 {\begin{minipage}{\linewidth}
\begin{myBullets}
\vspace{0.2cm}


\item  {Spoofing: Public-key cryptography used for authentication during the handshake process (SAE or Dragonfly)  is vulnerable to Shor's Algorithm, allowing impersonation of legitimate devices, compromising network security.}
\item  {Tampering: Message integrity checks (HMAC-SHA-3) in WiFi/WPA  susceptible to Grover's algorithm, enabling data packet manipulation during transmission.}
\item  {Info. Disclosure: WiFi/WPA uses AES and GCMP for data confidentiality, susceptible to Grover's Algorithm. This poses a risk of confidential data exposure during transmission, compromising privacy.}
\\
\end{myBullets}
\end{minipage}}
& \med& \high & \high & 
 {\begin{minipage}{\linewidth}
\begin{myBullets}
\vspace{0.2cm}
\item Wi-Fi/WPA (v3) in which RSA, ECDH, and ECDSA are replaced by a PQ public-key crypto. 
For symmetric crypto, using longer keys is sufficient to provide an alternative post-quantum solution. TLS in Enterprise version should be replaced with possible PQ TLS, as mentioned above in this table. PQ certificate in the enterprise version should be updated as mentioned in Section~\ref{sec:PKI}.
\\
\end{myBullets}
\end{minipage}}
\\ \hline
DECT (v6.0)~\cite{DECT}
 & {\begin{minipage}{\linewidth}
\begin{myBullets}
\vspace{0.2cm}
 \item Physical Layer
 \item  Medium Access Control (MAC) Layer
 \item Link Control Layer
 \item Data Link Control (DLC) Layer \\
\end{myBullets}
\end{minipage}} & 
{\begin{minipage}{\linewidth}
\begin{myBullets}
\vspace{0.2cm}
 \item Providing cordless voice, fax, data and multimedia communications, WLAN, and wireless PBX.
 \item Supporting authentication of handsets by DSAA2, confidentially via encrypting the voice stream with DSC2 (both based on AES 128) and authorization via subscription for connecting the handset to a base.\\
\end{myBullets}
\end{minipage}}
&{\begin{minipage}{\linewidth}
\begin{myBullets}
\vspace{0.2cm}
 \item Public-Key Crypto: -
\item Symmetric
Crypto: DSAA2 and DSC2 which are based on AES (Weakened by Grover's Algo.)
\\
\end{myBullets}
\end{minipage}} & {\begin{minipage}{\linewidth}
\begin{myBullets}
\vspace{0.2cm}
 \item {Spoofing: DECT's DSAA2, based on AES, is vulnerable to Grover's Algorithm, enabling quantum attackers to forge authentication messages, gain unauthorized access, and potentially lead to security breaches.}
\item {Info. Disclosure: DECT's DSC2 encryption, using AES, is weakened by Grover's Algorithm, allowing quantum attackers to decrypt intercepted communications, compromising confidentiality and leading to privacy violations.}

\item {Elevation of Privilege: Spoofing a handset allows attackers to grant unauthorized control over DECT network resources, potentially exploiting additional vulnerabilities for further privilege escalation within the system.}\\
\end{myBullets}
\end{minipage}}& \med& \med& \med&
 {\begin{minipage}{\linewidth}
\begin{myBullets}
\vspace{0.2cm}
 \item For symmetric crypto, longer keys should be used.\\
\end{myBullets}
\end{minipage}} \\ \hline 

DNSSEC~\cite{RFC4033,RFC4034,RFC4035} & 
{\begin{minipage}{\linewidth}
\begin{myBullets}
\vspace{0.2cm}
\item DNSKEY (DNS Key Record),
\item RRSIG (Resource Record Signature),
\item DS (Delegation Signer).
\vspace{0.2cm}
\end{myBullets}
\end{minipage}}
& {\begin{minipage}{\linewidth}
\begin{myBullets}
\vspace{0.2cm}
 \item Providing a secure domain name system by adding cryptographic signatures to existing DNS records and protecting the Internet by decreasing vulnerability to attacks.
 \item Supporting data origin authentication and data integrity protection.
 \\
\end{myBullets}
\end{minipage}}
& {\begin{minipage}{\linewidth}
\begin{myBullets}
\vspace{0.2cm}
 \item Public-Key Crypto: RSA, DSA, ECDSA (broken by Shor's Algo.).
\item Symmetric 
Crypto: SHA1/SHA2/SHA3 (Weakened by Grover's Algo.)
\\
\end{myBullets}
\end{minipage}} & {\begin{minipage}{\linewidth}
\begin{myBullets}
\vspace{0.2cm}
 \item {Spoofing: Shor's algorithm poses a threat by potentially breaking cryptographic signatures used for DNS record authenticity verification (e.g., RSA, DSA, and ECDSA), enabling the creation of forged DNS records. Additionally, Grover's algorithm weakens symmetric cryptography, increasing the risk of tampering with DNS records. This could facilitate spoofing attacks, compromising DNS integrity.}

\item {Tampering: Grover's algorithm may weaken symmetric cryptography in DNSSEC, compromising hashing algorithms like SHA1, SHA2, and SHA3. This vulnerability could allow for tampering with DNS records, redirecting users to malicious sites, or intercepting communication.}\\
 
\end{myBullets}
\end{minipage}}& \med& \high & \high & {\begin{minipage}{\linewidth}
\begin{myBullets}
\vspace{0.2cm}
 \item DNSSEC in which classic signature replaced by PQ one and hash values are generated with longer symmetric keys using digest algorithm SHA1/SHA2/SHA3. DNSSEC additional information like signatures and keys within the limited 512 bytes~\cite{muller2020retrofitting}.
 To avoid fragmentation, the extreme size of a typical message used in DNSSEC (max 1232 bytes) should be considered for the case when PQ public-key crypto is used instead of the classic one. For symmetric crypto, longer keys should be used~\cite{essay89509}. No certificate is used in DNSSec and the zone's administrator generates one or more public/private key pairs.\\ 
\end{myBullets}
\end{minipage}}
\\ \hline
\end{tabular}%
}

\end{table*}

The security provided by classical protocols cannot be considered quantum-safe, as a quantum attacker can employ algorithms such as Shor's, Grover's, and BHT to compromise both symmetric and asymmetric cryptography, posing a significant threat to the security of these protocols. In this section, we analyze and evaluate the impact of QC on the security of standard protocols, identifying vulnerabilities and proposing potential solutions to transition to a quantum-safe cryptographic state. Our comprehensive analysis of standard security protocols, including their primary components, the cryptographic algorithms they use in their crypto suite, their quantum threats, possible quantum-resistant solutions, and the corresponding risks that a quantum attacker can impose on each of them, is presented in Table~\ref{tab:Pre-migration-protocol}. This protocol-level analysis can assist security analysts in discovering mitigation matrices and identifying security issues in existing classical protocols more quickly.
 Note that to evaluate and assess the risk of QC for each protocol, multiple cryptographic algorithms might be available as options in the crypto suite of a protocol. In such cases, we generally consider the highest impact associated with different cryptographic algorithms in the protocol's crypto suite. In the following sections, we discuss the purpose of each protocol, its possible QC threats, and quantum-resistant solutions.

 \begin{enumerate}[wide, font=\itshape, labelwidth=!, labelindent=0pt, label=(\Alph*)]
\item \textit{Communication Security}: The primary aim of the protocols listed in Table~\ref{tab:Pre-migration-protocol} is to provide security in different aspects such as communication, access, transfer, and management. Two widely used protocols for providing communication security are Transport Layer Security (TLS) and Mutual TLS (mTLS), which support authentication, integrity, and confidentiality~\cite{RFC8446,RFC5246}. Further, for providing secure access, transfer, and management of files, mostly sFTP and FTPS protocols are used. Internet Protocol Security (IPsec) provides secure, authenticated, reliable communication. IPsec uses the Internet Key Exchange (IKE) protocol to establish a secure, authenticated communications channel between two communication entities and also sustain connection-less integrity and confidentiality.
All these protocols use classical symmetric as well as asymmetric cryptography, which are vulnerable to existing quantum algorithms. Based on the purpose and cryptography used in all these protocols, spoofing, tampering, and information disclosure are the primary threats. The OQS-OpenSSL~\cite{post-quantum-tls, open-quantum-safe-tls} and KEMTLS~\cite{schwabe2020post} are possible TLS replacement QC-resistant solutions that can replace the used TLS protocol in TLS, mTLS, and FTPS protocols to provide safety during communication from quantum attackers. However, secure sFTP uses a reliable data stream for files via Secure Shell Protocol (SSH); thus, OQS-OpenSSH~\cite{qs-openssh,oqs-ssh,oqs-liboqs} and OQS-libssh~\cite{oqs-ssh,oqs-liboqs} are two possible options for the replacement of the SSH protocol. Moreover, OpenVPN~\cite{post-quantum-vpn}, WireGuard~\cite{hulsing2021post}, and Mixing Preshared Keys in the IKE v2 (RFC 8784)~\cite{fluhrer2020mixing} are possible replacements for the IPsec protocol to provide a secure connection.

 \item \textit{Email Security and Privacy}:
 Email encryption is provided by two well-known protocols, i.e., Pretty Good Privacy (PGP) \cite{RFC4880} and Secure/Multipurpose Internet Mail Extensions\break (S/MIME) \cite{RFC8551}. They also support privacy, authentication, integrity, non-repudiation, and confidentiality. However, only S/MIME supports X.509 structure for digital certificates. Therefore, the possible QC threats are spoofing, tampering, repudiation, and information disclosure for both protocols. These two protocols use a classical symmetric and asymmetric cryptography, broken by existing quantum algorithms. The possible QC-resistant solutions for both protocols are mentioned in Table~\ref{tab:Pre-migration-protocol}.

\item \textit{Authentication}: To establish an authentication mechanism, the Kerberos, SAML, and IPsec protocols play essential roles. Kerberos is designed to facilitate authentication in accessing client/server applications using symmetric key cryptography. It supports both client and server authentication in client/server applications; therefore, spoofing is identified as the primary quantum (PQ) threat. Quantum-resistant Kerberos can enhance security by utilizing longer keys, given its reliance on symmetric cryptography.
The SAML protocol is employed for authenticating multiple applications, providing support for both authentication and authorization. Consequently, potential quantum threats include spoofing and elevation of privilege. Furthermore, IPsec is utilized to ensure secure, authenticated, and reliable communication. To address quantum threats, viable solutions for both SAML and IPsec are available, exemplified by protocols such as OpenVPN~\cite{post-quantum-vpn}, StrongSwan~\cite{strongSwan}, and WireGuard~\cite{hulsing2021post}. These solutions are detailed in Table~\ref{tab:Pre-migration-protocol}.

\item \textit{Resource Access}: In general, Open Authorization\break (OAuth) is used to grant a website or application access to resources held by other web apps on behalf of a user. It encourages authorization, hence the potential PQ threat is an elevation of privilege. Table~\ref{tab:Pre-migration-protocol} lists possible QC-resistant solutions for the OAuth protocol. 

\item \textit{Directory Service}: Lightweight Directory Access Protocol (LDAP) provides directory services access and maintenance for distributed directory information services over an IP network. It supports authentication, communication confidentiality (via optional use of TLS), data integrity (via optional use of TLS), and authorization. Thus, the probable QC threats are spoofing, tampering, information disclosure, and elevation of privilege. The possible QC-resistant solutions are: (i) OQS-OpenSSL~\cite{post-quantum-tls, open-quantum-safe-tls} as a post-quantum version of TLS, (ii) longer keys should be used for symmetric cryptography, and (iii) PQ certificates should be used as mentioned in Section~\ref{sec:PKI} to provide security over the Internet.

\item \textit{Domain Service}: The Domain Name System (DNS) provides a secure way to manage domain names by adding cryptographic signatures to existing DNS records, which helps to protect the Internet by reducing vulnerability to attacks. DNS also supports data origin authentication and data integrity protection, which can help prevent spoofing and tampering. Table~\ref{tab:Pre-migration-protocol} lists some feasible solutions for implementing a QC-resistant secure DNS protocol.
 
\item \textit{Wireless Service}: Wi-Fi Protected Access (WPA) is a security standard for computing devices that connect to the Internet wirelessly~\cite{RFC8110}. It is intended to provide superior data encryption and user authentication compared to the original Wi-Fi security standard, Wired Equivalent Privacy (WEP). In the online environment, wireless security is critical as it enables a secure connection over insecure networks. Insecure networks can result in data loss, leaked account credentials, and malware placement on your network. Wi-Fi/WPA (v3) offers a more secure handshake by utilizing the Wi-Fi Device Provisioning Protocol (DPP) to establish secure wireless networks that allow authentication and confidentiality through authenticated encryption and integrity procedures. Thus, spoofing, tampering, and information disclosure are all potential QC threats. A possible QC-resistant solution is to use PQ public-key cryptography to replace the asymmetric algorithms used in Wi-Fi/WPA (v3). For symmetric crypto, using longer keys is sufficient to provide an alternative post-quantum solution. PQ certificates should be used as mentioned in Section~\ref{sec:PKI}. Furthermore, Digital Enhanced Cordless Telecommunication (DECT v6.0) supports cordless communication for voice, fax, data, and multimedia communications, WLAN, and wireless Private Branch Exchange (PBX) systems~\cite{ETSI-DECT}. It supports handset authentication with the DECT Standard Authentication Algorithm (DSAA2), confidentiality through encrypting the voice stream with the DECT Standard Cipher (DSC2), and authorization via subscription for connecting the handset to a base. Thus, potential QC threats are spoofing, information disclosure, and elevation of privilege. Besides, DECT employs only symmetric cryptography (e.g., DSAA2 and DSC2 are based on AES 128), so longer keys are needed to provide a QC-resistant solution.
\end{enumerate}

\subsubsection{Pre-Migration Protocol Level Risk Assessment} {To analyze the risk, we need to evaluate the likelihood and impact of QC threats for each protocol. To evaluate the likelihood, we consider 15 years as the timeline of emerging quantum computers. Based on this consideration, we assign a medium qualitative level for the likelihood of threatening classical cryptosystems by QC (see Figure}~\ref{fig:chart2}){. These assumptions for likelihood can be easily changed to be aligned for another period. We determine the impact based on the minimum quantum security strength of different algorithms used as part of the recommended crypto suite of each protocol (see Table}~\ref{tab:Pre-migration-protocol}{). Quantum security strengths of different algorithms used in the recommended crypto suite are evaluated based on the expected impact of quantum threat for them as presented in Figure}~\ref{fig:classic-impact}{. This evaluation considers the dependency of threats caused by quantum computing on the quantum security strength of different classical algorithms, aligning with the criteria for impact levels detailed in Table}~\ref{table:impact} {and the expected consequences of quantum threats on each protocol. The final risk assessment evaluation is based on likelihood and impact, shown in Table}~\ref{tab:Pre-migration-protocol}.

\subsection{Through-Migration Protocol Level Analysis and Risk Assessment}
\begin{figure*}[!b]
 \begin{center} 
 \begin{subfigure}[t]{.22\linewidth}
 \centering
 \resizebox{\linewidth}{!}{
 \begin{tikzpicture}

\draw [thick arrow]
 (0,-0.3) -- (0,-1) 
 arc (270:180:1.5 and -1) [set mark={}]
 -- (-1.5,-2.5);

 \draw [thick arrow]
 (0,-0.3) -- (0,-1) 
 arc (270:180:-1.5 and -1) [set mark={}]
 -- (1.5,-2.5);

\node (d0) [] at (0,0) {\small{Inquiry to Select Certificate \& Protocol}};

\draw [thick arrow]
 (-1.5,-2.7) -- (-1.5,-5.1) [set mark={}];
 
 \draw [thick arrow]
 (1.5,-2.7) -- (1.5,-5.1) [set mark={}];

\node (d1) [draw,top color=white, bottom color=blue!20, rounded corners,minimum width=2cm,minimum height=1cm] at (-1.5,-3) {\scriptsize{Classic Certificate}};
\node (d2) [draw,top color=white, bottom color=blue!20, rounded corners,minimum width=2cm,minimum height=1cm] at (1.5,-3) {\scriptsize{PQ Certificate}};
\node (d3) [draw,top color=white, bottom color=blue!20, rounded corners,minimum width=2cm,minimum height=1cm] at (-1.5,-5.6) {\scriptsize{Classic Protocol}};
\node (d4) [draw,top color=white, bottom color=blue!20, rounded corners,minimum width=2cm,minimum height=1cm] at (1.5,-5.6) {\scriptsize{PQ Protocol}};

\node (d6) [top color=white!40, bottom color=blue!80, rounded corners,minimum height=7.cm,minimum width=5.5cm,draw=black!100,fill opacity=0.1, fit= (d0) (d1) (d2) (d3) (d4),draw] {};
\end{tikzpicture}}
 \caption{}
\end{subfigure}
\hfill
 \begin{subfigure}[t]{.22\linewidth}
 \centering
 \resizebox{\linewidth}{!}{
 \begin{tikzpicture}
\draw [thick arrow]
 (0,1.5) -- (0,0.45) [set mark={\scriptsize{}}];

\node (d0) [draw,top color=white, bottom color=blue!20, rounded corners,minimum width=2cm,minimum height=1cm] at (0,1.8) {\scriptsize{Composite Certificate}};

\draw [thick arrow]
 (0,-0.3) -- (0,-1) 
 arc (270:180:1.5 and -1) [set mark={}]
 -- (-1.5,-2.5);

 \draw [thick arrow]
 (0,-0.3) -- (0,-1) 
 arc (270:180:-1.5 and -1) [set mark={}]
 -- (1.5,-2.5);

\node (d1) [] at (0,.1) {\small{Inquiry to Select Protocol}};
\node (d3) [draw,top color=white, bottom color=blue!20, rounded corners,minimum width=2cm,minimum height=1cm] at (-1.5,-3) {\scriptsize{Classic Protocol}};
\node (d4) [draw,top color=white, bottom color=blue!20, rounded corners,minimum width=2cm,minimum height=1cm] at (1.5,-3) {\scriptsize{PQ Protocol}};

\node (d6) [top color=white!40, bottom color=blue!80, rounded corners,minimum height=7cm,minimum width=5.5cm,draw=black!100,fill opacity=0.1, fit= (d0) (d3) (d4),draw] {};

\end{tikzpicture}}
 \caption{}
\end{subfigure}
 \hfill
 \begin{subfigure}[t]{.22\linewidth}
 \centering
 \resizebox{\linewidth}{!}{
 \begin{tikzpicture}

\draw [thick arrow]
 (0,-0.3) -- (0,-1) 
 arc (270:180:1.5 and -1) [set mark={}]
 -- (-1.5,-2.5);

 \draw [thick arrow]
 (0,-0.3) -- (0,-1) 
 arc (270:180:-1.5 and -1) [set mark={}]
 -- (1.5,-2.5);

\node (d0) [] at (0,0) {\small{Inquiry to Select Certificate}};

\node (d1) [draw,top color=white, bottom color=blue!20, rounded corners,minimum width=2cm,minimum height=1cm] at (-1.5,-3) {\scriptsize{Classic Certificate}};
\node (d2) [draw,top color=white, bottom color=blue!20, rounded corners,minimum width=2cm,minimum height=1cm] at (1.5,-3) {\scriptsize{PQ Certificate}};

\draw [thick arrow]
 (-1.5,-3.5) 
 arc (180:270:1.5 and 1) [set mark={}] -- (0,-5.5);
\draw [thick arrow]
 (1.5,-3.5) 
 arc (180:270:-1.5 and 1) [set mark={}] -- (0,-5.5);

\node (d3) [draw,top color=white, bottom color=blue!20, rounded corners,minimum width=2cm,minimum height=1cm] at (0,-6) {\scriptsize{Composite Protocol}};

\node (d6) [top color=white!40, bottom color=blue!80, rounded corners,minimum height=7cm,minimum width=5.5cm,draw=black!100,fill opacity=0.1, fit= (d0) (d1) (d2) (d3) (d4),draw] {};
\end{tikzpicture}}
 \caption{}
\end{subfigure}
\hfill
 \begin{subfigure}[t]{.22\linewidth}
 \centering
 \resizebox{\linewidth}{!}{
 \begin{tikzpicture}
\draw [thick arrow]
 (0,2.5) -- (0,1) [set mark={\scriptsize{}}];

\node (d0) [] at (0,2.2) {\small{}};

 `
 
 \draw [thick arrow]
 (0,0) -- (0,-2.5) [set mark={}];

\node (d1) [draw,top color=white, bottom color=blue!20, rounded corners,minimum width=2cm,minimum height=1cm] at (0,0.5) {\scriptsize{Composite Certificate}};
\node (d3) [draw,top color=white, bottom color=blue!20, rounded corners,minimum width=2cm,minimum height=1cm] at (0,-3) {\scriptsize{Composite Protocol}};

\node (d6) [top color=white!40, bottom color=blue!80, rounded corners,minimum height=7cm,minimum width=5.5cm,draw=black!100,fill opacity=0.1, fit= (d0) (d1) (d3),draw] {};
\end{tikzpicture}}
 \caption{}
\end{subfigure}
\caption{{ Mechanisms for Hybrid Protocol: (a) Dual Protocol with Dual Certificate, (b) Dual Protocol with Composite Certificate, (c) Composite Protocol with Dual Certificate, (d) Composite Protocol with Composite Certificate.
}}
\label{fig:HybridProtoMech}
\end{center}
\end{figure*} 
\begin{table*}[!htbp]
\caption{{Through-Migration Protocol Level Analysis}}
\small
\label{tab:through-migration-protocol}
\resizebox{\linewidth}{!}{%
\begin{tabular}{|l|p{0.14\linewidth}|p{0.135\linewidth}|p{0.3\linewidth}|p{0.4\linewidth}|p{0.4\linewidth}|p{0.23\linewidth}|}
\hline 
\multirow{2}{*}{\textbf{Migration Strategy}} & \multicolumn{2}{l|}{\multirow{2}{*}{\textbf{Approaches}$^*$}} & \multirow{2}{*}{\textbf{Mechanisms}} & \multirow{2}{*}{\textbf{Pros.}} & \multirow{2}{*}{\textbf{Cons.}} & \multirow{2}{*}{\textbf{Quantum Threats (STRIDE)}} \\ 
& \multicolumn{2}{c|}{} & & & & \\ \hline

\multirow{20}{*}{Hybrid} & \multirow{3}{*}{\begin{minipage}{\linewidth}
\multirow{10}{*}{Dual Protocol~\cite{driscoll-pqt-hybrid-terminology-02}}
\end{minipage}} &
\multirow{2}{*}{ \begin{minipage}{\linewidth}Dual Certificate \cite{bindel2017transitioning,vogt2021quantum,bindel2019x}\end{minipage} } &
 \begin{minipage}{\linewidth}
 Two separate protocols with separate certificates in a manner that ensures the  \textit{component cryptographic elements retain their original formats} similar to those employed in single-algorithm schemes.
\end{minipage}
& 
\begin{minipage}{\linewidth}
\begin{myBullets}
\vspace{0.1cm}
\item Only a few changes of standards and applications/devices for dual certificate/dual protocol,

\item Smooth transition to quantum-safe certificates/protocol,
\item Less fragmentation issues for quantum-safe certificates/protocol, 
\item Support for backward compatibility of certificate/protocol,
\item Format of cryptographic elements related to protocol remains unchanged,

\item Cryptographic elements and libraries have minimal modifications for protocol.
\\
\end{myBullets}
\end{minipage}
& 
\begin{minipage}{\linewidth}
\begin{myBullets}
\vspace{0.1cm}
\item Some primary modifications are required in protocol fields, the message flow, or both to support both classic and PQ protocols,
 \item Redundant pieces of information are required to transmit in dual certificates/protocol,
\item To implement the dual protocol, some modifications are required in protocol libraries. Also, some changes are needed to support parallel hierarchies. \\
\end{myBullets}
\end{minipage}
& 

 \multirow{30}{*}{\begin{minipage}{\linewidth}
\begin{myBullets}
\vspace{0.1cm}
 \item Spoofing, Tampering, Repudiation, Info. Disclosure, DoS, and Elevation of Privileges (when none of the protocols used in the hybrid approach are secure against them.)\\
 \end{myBullets}
\end{minipage}} \\ \cline{3-6} 
& & \multirow{2}{*}{ \begin{minipage}{\linewidth}Composite Certificate \cite{bindel2017transitioning,vogt2021quantum,bindel2019x}\end{minipage} } & 

 \begin{minipage}{\linewidth}
 Two separate protocols with composite certificates (constructed using mechanisms mentioned in Figure~\ref{fig:HybridProtoMech}) in a manner that ensures the  \textit{component cryptographic elements retain their original formats} similar to those employed in single-algorithm schemes.
\end{minipage}
 & 
 \begin{minipage}{\linewidth}
\begin{myBullets}
\vspace{0.1cm}

\item Only a few changes of standards and applications/devices for dual protocol,

\item Smooth transition to quantum-safe protocol,
\item Less fragmentation issues for quantum-safe protocol,
\item Support for backward compatibility of protocol,
\item Format of cryptographic elements related to protocol remains unchanged,

\item Cryptographic elements and libraries have minimal modifications for protocol,

\item Combines security of pre- and post-quantum algorithms in composite certificate.\\
\end{myBullets}
\end{minipage}
& 
 \begin{minipage}{\linewidth}
\begin{myBullets}
\vspace{0.1cm}

\item Some primary modifications are required in protocol fields, the message flow, or both to support both classic and PQ protocols,

\item To implement the dual protocol, some modifications are required in protocol libraries. Also, some changes are needed to support parallel hierarchies,

\item Redundant pieces of information are required to transmit in the dual protocol,
 
 \item Needs changes of standards (e.g., RFC 5280~\cite{cooper2008internet}) for two signatures and two public keys in a certificate.
 \item Size of certificates increases the most.\\
\end{myBullets}
\end{minipage}
& 
\\ \cline{2-6} 
 &
 \multirow{10}{*}{ \begin{minipage}{\linewidth}Composite Protocol \cite{driscoll-pqt-hybrid-terminology-02}\end{minipage} } 

& 
{ \begin{minipage}{\linewidth}Dual Certificate \cite{bindel2017transitioning,vogt2021quantum,bindel2019x} \end{minipage}}

& \begin{minipage}{\linewidth}
 Composite hybrid protocols in a manner that preserves the same \textit{protocol fields} and \textit{message flow} as the classic version of the protocol that utilizes single-algorithm schemes (yet with separate certificates for classic and post-quantum).
 
\end{minipage} & 
 \begin{minipage}{\linewidth}
\begin{myBullets}
\vspace{0.1cm}

 \item Only a few changes of standards and applications/devices for the dual certificate,
 \item Smooth transition to quantum-safe certificates,
 \item Less fragmentation issues for quantum-safe certificates, 
 \item Support for backward compatibility of the certificate,
 \item Protocol fields and message flow remain unchanged,
 
 \item Minimal changes are likely to be made to the cryptographic libraries.\\
 \end{myBullets}
\end{minipage}
 & 
 \begin{minipage}{\linewidth}
\begin{myBullets}
\vspace{0.1cm}
 \item Required changes are primarily made to the formats of the cryptographic elements of the protocol,
 \item Redundant pieces of information are required to transmit in dual certificates,
 \item To implement the composite protocol, the primary focus of modifications is anticipated to be within the cryptographic libraries.\\

 \end{myBullets}
\end{minipage} 
 & 

\\ \cline{3-6}
 && 
Composite Certificate~\cite{bindel2017transitioning,vogt2021quantum,bindel2019x}

& 
\begin{minipage}{\linewidth}
\vspace{0.1cm}
 Composite hybrid protocols in a manner that preserves the same \textit{protocol fields} and \textit{message flow} as the classic version of the protocol that utilizes single-algorithm schemes. This mechanism uses composite certificates constructed using mechanisms mentioned in Figure~\ref{fig:HybridProtoMech}.
 \vspace{0.1cm}
\end{minipage}
& 
 \begin{minipage}{\linewidth}
\begin{myBullets}
\vspace{0.1cm}

 \item Protocol fields and message flow remain unchanged,
 
 \item Minimal changes are likely to be made to the cryptographic libraries,
 
 \item Combines security of pre- and post-quantum algorithms in the composite certificate.\\
 \end{myBullets}
\end{minipage}
 & 
 \begin{minipage}{\linewidth}
\begin{myBullets}
\vspace{0.1cm}

 \item Required changes are primarily made to the formats of the cryptographic elements of the protocol,
 
 \item To implement the composite protocol, the primary focus of modifications is anticipated to be within the cryptographic libraries,
 
 \item Needs changes of standards (e.g., RFC 5280~\cite{cooper2008internet}) to store and verify two signatures and two public keys in a certificate,
 
 \item Size of certificates increases the most.\\

 \end{myBullets}
\end{minipage} 
 & 
\\ \hline 
\end{tabular}%
}\vspace{3pt}
\tiny{$^*$ Certificate is considered for different approaches in case protocol requires certificate.}
\end{table*}

In the hybrid protocol approach, protocol migration necessitates the inclusion of at least one component algorithm as a post-quantum algorithm and at least one as a classical algorithm, ensuring that together they offer similar cryptographic capabilities. This hybrid approach seamlessly integrates classical and post-quantum algorithms, supporting backward compatibility~\cite{driscoll-pqt-hybrid-terminology-02}. Multiple component algorithms can operate in parallel while maintaining the same \textit{protocol fields} and \textit{message flow} as the classic version, which employs single-algorithm schemes. This creates a composite protocol that enhances security strength. The hybrid dual protocol utilizes multiple component algorithms, ensuring that  \textit{component cryptographic elements retain their original formats} akin to those in single-algorithm schemes. The hybrid protocol remains secure as long as none of the parallelly used component algorithms is compromised. A comprehensive review of existing hybrid protocol approaches, along with mechanisms, pros, cons, and potential quantum threats, is presented in Table~\ref{tab:through-migration-protocol} to maintain backward compatibility while supporting the hybrid protocol structure.

As shown in Table~\ref{tab:through-migration-protocol}, there are two possible approaches: the dual protocol and composite protocol, each with two separate mechanisms.  In the dual protocol hybrid approach, the first mechanism, dual protocol with dual certificate (Figure \ref{fig:HybridProtoMech}.a), employs two separate protocols and certificates, ensuring that \textit{component cryptographic elements retain their original formats} similar to single-algorithm schemes. The second mechanism, dual protocol with composite certificate (Figure \ref{fig:HybridProtoMech}.b), utilizes an extension mechanism to create a composite certificate supporting the dual protocol mechanism. In the composite protocol hybrid approach, the first mechanism uses two separate certificates, preserving the same \textit{protocol fields} and \textit{message flow} as the classic version with single-algorithm schemes for the protocol (Figure \ref{fig:HybridProtoMech}.c). The second mechanism uses an extension mechanism to create a composite certificate supporting the composite protocol mechanism (Figure \ref{fig:HybridProtoMech}.d).

Advantages of the dual protocol hybrid approach include a smooth transition to a quantum-safe protocol with minimal changes to standards, libraries, and applications/devices, without altering the format of cryptographic elements. However, there are disadvantages, such as necessary modifications to protocol fields, message flow, or both to accommodate both classic and post-quantum (PQ) protocols, along with required adjustments in protocol libraries and parallel hierarchies. The composite protocol hybrid approach offers advantages such as minimal changes to standards, cryptographic libraries, and applications/devices, with protocol fields and message flow remaining unchanged. On the downside, primary modifications are required in the formats of cryptographic elements, and implementation necessitates substantial changes in cryptographic libraries.

Furthermore, potential QC threats in the hybrid protocol approach, including spoofing, tampering, repudiation, information disclosure, Denial of Service (DoS), and elevation of privileges, are significant when none of the protocols used in the hybrid approach is secure against them, as outlined in Table~\ref{tab:through-migration-protocol}. It is worth noting that the mechanisms mentioned above for generating hybrid protocols can be applied not only to an entire protocol but also to sub-protocols, combining them for different components like negotiation, key exchange, or authentication within a protocol.

\begin{table*}[!htbp]
\caption{Post-Migration Protocol Level Analysis and Risk Assessment}
\small
\label{tab:Post-migration-protocol}
\resizebox{\textwidth}{!}{%
\begin{tabular}{|l|p{0.36\linewidth}|p{0.35\linewidth}|p{0.45\linewidth}|p{0.3\linewidth}|l|l|l|}
\hline
\multirow{2}{*}{\textbf{Protocols}} & \multirow{2}{*}{\textbf{Possible QC-resistant Solutions}}&\multirow{2}{*}{\textbf{Challenges and Attacks}} & \multirow{2}{*}{\textbf{Possible Countermeasures}} & \multirow{2}{*}{\textbf{Quantum Threats (STRIDE)}} & \multirow{2}{*}{\textbf{L}} & \multirow{2}{*}{\textbf{I}} & \multirow{2}{*}{\textbf{R}} \\
 & & & & & & & \\ \hline
 \multirow{2}{*}{SSH} & {\begin{minipage}{\linewidth}
\begin{myBullets}
\vspace{0.2cm}
 \item OQS-OpenSSH~\cite{qs-openssh,oqs-ssh,oqs-liboqs},
 \item OQS-libssh~\cite{oqs-ssh,oqs-liboqs}.
 \\
\end{myBullets}
\end{minipage}} & 
{
 \begin{minipage}{\linewidth}
\begin{myBullets}
 \item Huge communication overhead,
 \item Network congestion, 

 \item Fragmentation issues triggering retransmission,
 \item Possibility of denial of services,
 \item Data exfiltration and Info. disclosure via side-channel/math. analysis attacks mentioned in Table~\ref{tab:Post-Migration-Alg}.\\
 \end{myBullets}
\end{minipage} }
&{\begin{minipage}{\linewidth}
\begin{myBullets}
\vspace{0.2cm}
 \item To control congestion: using congestion control mechanisms to prevent or remove congestion~\cite{jay2018internet,bohloulzadeh2020survey,jay2019deep},
 
 \item To avoid fragmentation and trigger retransmission: considering the extreme size of a typical message for the case when QC-resistant protocol is used instead of classic one~\cite{muller2020retrofitting,essay89509},

 \item To increase the tolerance against huge communication overhead and prevent DoS: providing required bandwidth, building redundancy into infrastructure, deploying DDoS resilience hardware/software modules like firewalls, adopting DDoS Protection Appliance, configuring network hardware against DDoS attacks, etc~\cite{garcia2022deep,liu2018practical},
 
 \item To immune against side-channel/math. analysis attacks, refer to the solutions mentioned in Table~\ref{tab:Post-Migration-Alg}. \vspace{0.1cm}
 \end{myBullets}
\end{minipage} } &
\begin{minipage}{\linewidth}
\begin{myBullets}
\vspace{0.1cm}
 \item {Info. Disclosure: Side-channel attacks can lead to the disclosure of sensitive information.}
 \item  {DoS: The large communication overhead of PQC solutions for SSH can be exploited to launch DoS attacks.}\\

 \end{myBullets}
\end{minipage}
& \med& \med&\med \\ \hline 
\multirow{2}{*}{TLS} & 
\begin{minipage}{\linewidth}
\begin{myBullets}
\vspace{0.2cm}
 \item OQS-OpenSSL~\cite{post-quantum-tls, open-quantum-safe-tls},
 \item KEMTLS~\cite{schwabe2020post}.\\
 \end{myBullets}
\end{minipage} & 
 \begin{minipage}{\linewidth}
\begin{myBullets}
 \vspace{0.2cm}
 \item All attacks mentioned above for PQ SSH are also applicable for PQ TLS.
 \item Based on digital certificates for server and client authentication with drawbacks mentioned in Table~\ref{tab:Post-migration-Cert}.\\
 \end{myBullets}
\end{minipage} 
&\begin{minipage}{\linewidth}
\begin{myBullets}
\vspace{0.2cm}
 \item All countermeasures mentioned above for PQ SSH are also applicable for PQ TLS,
 \item To provide countermeasures against certificate-based authentication, refer to Table~\ref{tab:Post-migration-Cert}. \\
 \end{myBullets}
\end{minipage} 
& 
\begin{minipage}{\linewidth}
\begin{myBullets}
\vspace{0.2cm}
 \item {Info. disclosure: Due to side-channel attacks and certificate limitations (as described in Challenges and Attacks).}
  \item {DoS due to larger PQC communication and certificate overhead.}
 \end{myBullets}
\end{minipage}& \med& \med&\med\\ \hline
 \multirow{2}{*}{ mTLS} & {\begin{minipage}{\linewidth}
\begin{myBullets}
\vspace{0.2cm}
 \item Mutual use of OQS-OpenSSL~\cite{post-quantum-tls, open-quantum-safe-tls} by both parties,
 \item Mutual use of KEMTLS~\cite{schwabe2020post} by both parties.\\
\end{myBullets}
\end{minipage}}& 
 \begin{minipage}{\linewidth}
\begin{myBullets}
 \vspace{0.2cm}
 \item All attacks mentioned above for TLS are applicable for PQ mTLS.\\
 \end{myBullets}
\end{minipage} 
&\begin{minipage}{\linewidth}
\begin{myBullets}
\vspace{0.2cm}
 \item All countermeasures mentioned above for PQ TLS are also applicable for PQ mTLS.\\
 \end{myBullets}
\end{minipage} 
& 
\begin{minipage}{\linewidth}
\begin{myBullets}
\vspace{0.2cm}
 \item {Info. Disclosure and DoS: Since mTLS builds on TLS, it inherits the same vulnerabilities.}\\
 \end{myBullets}
\end{minipage} & \med& \med&\med`\\ \hline

 \multirow{1}{*}{sFTP} & {\begin{minipage}{\linewidth}
\begin{myBullets}
\vspace{0.2cm}
 \item sFTP in which SSH is replaced by OQS-OpenSSH~\cite{post-quantum-SSH} or OQS-libssh~\cite{oqs-ssh}.\\
\end{myBullets}
\end{minipage}} & 
\begin{minipage}{\linewidth}
\begin{myBullets}
\vspace{0.1cm}
 \item All attacks mentioned above for PQ SSH are also applicable for PQ sFTP,
 \item {Out-of-bounds memory access vulnerability through a remote buffer overflow.}\\
 \end{myBullets}
\end{minipage} 
& 
\begin{minipage}{\linewidth}
\begin{myBullets}
\vspace{0.1cm}
 \item All countermeasures mentioned above for PQ SSH are also applicable for PQ sFTP,
 \item {To immune against buffer overflow for the certificate on servers by malicious clients: Canary, DEP (Data Execution Prevention, ASLR (Address Space Layout Randomization)}~\cite{zhou2022final,nicula2019exploiting}. \\
 \end{myBullets}
\end{minipage} 
& 
\begin{minipage}{\linewidth}
\begin{myBullets}
 \item {Info. Disclosure and DoS: Similar to SSH, sFTP is susceptible to information disclosure and DoS vulnerabilities.}
 \end{myBullets}
\end{minipage}
&\med& \med & \med \\ \hline
 \multirow{2}{*}{FTPS} & {\begin{minipage}{\linewidth}
\begin{myBullets}
\vspace{0.2cm}
 \item FTPS which provides file transfer via OQS-OpenSSL~\cite{post-quantum-tls, open-quantum-safe-tls} or KEMTLS~\cite{schwabe2020post} instead of TLS.\\
\end{myBullets}
\end{minipage}}& 

\begin{minipage}{\linewidth}
\begin{myBullets}
\vspace{0.1cm}
 \item All attacks mentioned above for PQ TLS are also applicable for PQ FTPS,
 \item {Out-of-bounds memory access vulnerability through a remote buffer overflow.}\\
 \end{myBullets}
\end{minipage} 
 & 
 \begin{minipage}{\linewidth}
\begin{myBullets}
\vspace{0.1cm}
\item All countermeasures mentioned above for PQ TLS are also applicable for PQ FTPS,
 \item {To immune against stack-based buffer overflow for the certificate on servers by malicious clients: Canary, DEP (Data Execution Prevention, ASLR (Address Space Layout Randomization)}~\cite{zhou2022final,nicula2019exploiting}.\vspace{0.1cm}
 \end{myBullets}
\end{minipage} & 
{\begin{minipage}{\linewidth}
\begin{myBullets}
 \item {Info. Disclosure and DoS: Due to similar vulnerabilities to PQ TLS, the same reasoning can be applied here.}
 \end{myBullets}
\end{minipage}}
&\med& \med&\med\\ \hline
 
 \multirow{2}{*}{SAML} & \begin{minipage}{\linewidth}
\begin{myBullets}
\vspace{0.1cm}
 \item SAML in which RSA or DSA, as two recommended public-key algorithms in crypto suites, are replaced by a PQ one. 
 There is no limitation on the extreme size of typical public keys used for SAML; thus, any public-key PQ crypto algorithms can be used for PQ one. For symmetric crypto, longer keys should be used. The certificate used in SAML should be replaced by PQ one as mentioned in Section~\ref{sec:PKI}. 
 Optional use of TLS should be replaced with the optional use of possible PQ TLS solutions mentioned in Table~\ref{tab:Pre-migration-protocol}. \\
 \end{myBullets}
\end{minipage}
 & 
\begin{minipage}{\linewidth}
\begin{myBullets}
\vspace{0.1cm}
 \item Huge communication overhead,
 \item Network congestion, 
 \item Fragmentation issues triggering retransmission,
 \item Possibility of Denial of Services,
 \item Info. Disclosure via side-channel attacks (mentioned in Table~\ref{tab:Post-Migration-Alg}) as a result of observable timing differences and cache access patterns.
 \item Vulnerabilities related to the optional use of TLS are discussed above in TLS protocol.
 \item Vulnerabilities related to the certificate used in SAML are mentioned in Table~\ref{tab:Post-migration-Cert}.\\
 \end{myBullets}
\end{minipage} 
& 
{\begin{minipage}{\linewidth}
\begin{myBullets}
\vspace{0.2cm}
 \item To control congestion: using congestion control mechanisms to prevent or remove congestion~\cite{jay2018internet,bohloulzadeh2020survey,jay2019deep},
 
 \item To avoid fragmentation and trigger retransmission: considering the extreme size of a typical message for the case when QC-resistant protocol is used instead of classic one~\cite{muller2020retrofitting,essay89509},

 \item To increase the tolerance against huge communication overhead and prevent DoS: providing required bandwidth, building redundancy into infrastructure, deploying DDoS resilience hardware/software modules like firewalls, adopting DDoS Protection Appliance, configuring network hardware against DDoS attacks, etc~\cite{garcia2022deep,liu2018practical},
 
 \item To immune against side-channel attacks, refer to the solutions mentioned in Table~\ref{tab:Post-Migration-Alg},
\item All countermeasures mentioned above for TLS and in Table~\ref{tab:Post-migration-Cert} for certificate are also applicable here.\vspace{0.1cm}
 \end{myBullets}
\end{minipage} } 
&
 \begin{minipage}{\linewidth}
\begin{myBullets}
\vspace{0.1cm}
  \item {Info. Disclosure and DoS: Due to similar vulnerabilities to PQ TLS, the same reasoning can be applied here.}

 \end{myBullets}
\end{minipage}
& \med& \med&\med \\ \hline
 
 \multirow{2}{*}{OAuth} & \begin{minipage}{\linewidth}
\begin{myBullets}
\vspace{0.1cm}
 \item OAuth (v2) in which tokens signature (i.e., HMAC-SHA1) is replaced by a PQ option. For token signature, longer keys should be used with considering the maximum length of access tokens should be 2048 bytes.
 There is no cryptographic mechanism except the optional use of TLS considered for OAuth (v2) mentioned in RFC 6819~\cite{lodderstedt2013oauth} and RFC6749~\cite{RFC6749}. PQ TLS should be used instead of TLS as mentioned Table~\ref{tab:Post-migration-Cert}.\\

 \end{myBullets}
\end{minipage}
 & 
\begin{minipage}{\linewidth}
\begin{myBullets}
\vspace{0.1cm}

 \item Vulnerabilities related to the optional use of TLS are discussed above in TLS protocol.
\\
 \end{myBullets}
\end{minipage} 
& 
\begin{minipage}{\linewidth}
\begin{myBullets}
\vspace{0.1cm} 

\item All countermeasures mentioned above for TLS are applicable here.
\\
 \end{myBullets}
\end{minipage} 
& \begin{minipage}{\linewidth}
\begin{myBullets}
\vspace{0.1cm}
  \item {Info. Disclosure and DoS: The optional use of PQ TLS introduces similar vulnerabilities, making PQ OAuth susceptible to similar threats.}
 \end{myBullets}
\end{minipage}& \low &\med &\low \\ \hline
 
 \multirow{2}{*}{IKE} & \begin{minipage}{\linewidth}
\begin{myBullets}
\vspace{0.1cm}
 \item PQ IKE mentioned in RFC 8784: ``Mixing Preshared Keys in the IKE (v2) for Post-quantum Security''~\cite{fluhrer2020mixing}.\\
 \end{myBullets}
\end{minipage} 
 & 
\begin{minipage}{\linewidth}
\begin{myBullets}
\vspace{0.1cm}
 \item Post-quantum preshared keys driven from password might have a low entropy, not be fully random, and can be broken by a Birthday attack and Grover's Algorithm~\cite{grover1996fast}).
 \item Info. Disclosure to recover preshared key via side-channel attacks (mentioned in Table~\ref{tab:Post-Migration-Alg}) as a result of observable timing differences and cache access patterns.
 \item Based on digital certificates with the drawbacks mentioned in Table~\ref{tab:Post-migration-Cert}.\vspace{0.1cm}
 \end{myBullets}
\end{minipage} 
& 
\begin{minipage}{\linewidth}
\begin{myBullets}
\vspace{0.1cm}
 \item Providing a strongly random long-term post-quantum preshared Keys with sufficient entropy.
 \item To immune against side-channel attacks, refer to the solutions mentioned in Table~\ref{tab:Post-Migration-Alg}. 
 \item To provide countermeasures against the drawback of the certificate, refer to Table~\ref{tab:Post-migration-Cert}.
 \end{myBullets}
\end{minipage} 
& \begin{minipage}{\linewidth}
\begin{myBullets}
\vspace{0.1cm}
\vspace{0.2cm}
 \item {Info. Disclosure: This vulnerability arises from side-channel attacks and limitations related to certificates, as described in the ``Challenges and Attacks'' column.}
 \item {DoS: This vulnerability is due to the increased communication overhead of PQC and certificate management.}
 \end{myBullets}
 \vspace{0.1cm}
\end{minipage}& \med & \med&\med \\ \hline 
 
\multirow{3}{*}{IPsec} & {\begin{minipage}{\linewidth}
\begin{myBullets}
\vspace{0.15cm}
 \item OpenVPN~\cite{post-quantum-vpn},
 \item StrongSwan~\cite{strongSwan},
 \item WireGuard~\cite{hulsing2021post}.\\
\end{myBullets}
\end{minipage}}
 & 
\begin{minipage}{\linewidth}
\begin{myBullets}
\vspace{0.1cm}

\item WireGuard is not yet supported by standards, only supports UDP, requires infrastructure, and has privacy issues (requires logging user data). It has better bypassing capabilities~\cite{dekkerperformance} but only supports preshared keys.
 OpenVPN has longer initiation time and higher latency compared to WireGuard and StrongSwan, but offers better compatibility. Supports both preshared keys and digital certificates.   
\item Post-quantum preshared keys derived from passwords might have low entropy, not be fully random, and can be broken by a Birthday attack and Grover's Algorithm~\cite{grover1996fast}. Preshared keys can also be recovered via side-channel attacks (mentioned in Table~\ref{tab:Post-Migration-Alg}) as a result of observable timing differences and cache access patterns. 
\item Post-quantum digital certificates have the drawbacks mentioned in Table~\ref{tab:Post-migration-Cert}.
 \vspace{0.1cm}
 \end{myBullets}
\end{minipage} 
& 
\begin{minipage}{\linewidth}
\begin{myBullets}
\vspace{0.1cm}
\item In the context of the preshared keys approach, safeguarding against birthday attacks and Grover's Algorithm necessitates the utilization of robust long-term post-quantum preshared keys possessing significant entropy. Additionally, for resilience against side-channel attacks attempting to recover preshared keys, the solutions outlined in Table~\ref{tab:Post-Migration-Alg} can be implemented.
 \item Regarding the certificate-based approach, refer to Table~\ref{tab:Post-migration-Cert} for potential countermeasures to address the drawbacks associated with certificates.
 \end{myBullets}
\end{minipage} 
 & 
 \begin{minipage}{\linewidth}
\begin{myBullets}
\vspace{0.1cm}
 \item {Info. Disclosure: Potential side-channel attacks and limitations of digital certificates can lead to information disclosure.}
\item {DoS: The large communication and certificate overhead of PQC can result in denial of service attacks.}
 \end{myBullets}
\end{minipage}
 & \low & \low & \low \\ \hline 
\end{tabular}%
}\vspace{3pt}
\tiny{$^*$ We perform risk evaluation with the presumption of considering the countermeasures mentioned in the table.}

\end{table*}

\addtocounter{table}{-1}

\begin{table*}[!htbp]
\caption{(Cont.) Post-Migration Protocol Level Analysis and Risk Assessment}
\small
\resizebox{\textwidth}{!}{%
\begin{tabular}{|l|p{0.36\linewidth}|p{0.35\linewidth}|p{0.45\linewidth}|p{0.3\linewidth}|l|l|l|}
\hline
\multirow{2}{*}{\textbf{Protocols}} & \multirow{2}{*}{\textbf{Possible QC-resistant Solutions}}&\multirow{2}{*}{\textbf{Challenges and Attacks}} & \multirow{2}{*}{\textbf{Possible Countermeasures}} & \multirow{2}{*}{\textbf{Quantum Threats (STRIDE)}} & \multirow{2}{*}{\textbf{L}} & \multirow{2}{*}{\textbf{I}} & \multirow{2}{*}{\textbf{R}} \\
 & & & & & & & \\ \hline

\multirow{1}{*}{Kerberos} & \begin{minipage}{\linewidth}
\begin{myBullets}
\vspace{0.1cm}
 \item Kerberos with longer symmetric keys.
 \end{myBullets}
\end{minipage} 
 & 
\begin{minipage}{\linewidth}
\begin{myBullets}
\vspace{0.1cm}
 \item Symmetric key driven from password might be not fully random and be broken by a Birthday attack and Grover's Algorithm~\cite{grover1996fast}).\\
 \end{myBullets}
\end{minipage} 
& 
\begin{minipage}{\linewidth}
\begin{myBullets}
\vspace{0.1cm}
 \item Providing a strongly random long-term symmetric key that is not password derived (for each user, and possibly each device) and can be securely distributed and installed.\\
 \end{myBullets}
\end{minipage} 
& 
\begin{minipage}{\linewidth}
\begin{myBullets}
\vspace{0.1cm}
 \item {Spoofing: Weak password-derived symmetric keys in Kerberos can enable attackers to impersonate legitimate users.} \\
 \end{myBullets}
\end{minipage} 
 & \low & \low & \low \\ \hline

\multirow{1}{*}{LDAP} & 
{\begin{minipage}{\linewidth}
\begin{myBullets}
\vspace{0.2cm}
 \item LDAP in which (a) passwords or symmetric keys used in SASL are longer and more secure against birthday attack, (b) optional use of TLS is replaced by OQS-OpenSSL~\cite{post-quantum-tls, open-quantum-safe-tls} or KEMTLS~\cite{schwabe2020post}, and (c) PQ certificate used instead of the classic one as mentioned in Section~\ref{sec:PKI}.\\
 \end{myBullets}
\end{minipage}}
 & 
\begin{minipage}{\linewidth}
\begin{myBullets}
\vspace{0.1cm}
 \item Optional use of a password or symmetric key in SASL for authentication might not be fully random and be broken by a Birthday attack and Grover's Algorithm~\cite{grover1996fast}).
 \item Due to the optional use of TLS, all attacks mentioned above for PQ TLS are also applicable for LDAP.\\
 \end{myBullets}
\end{minipage} 
& \begin{minipage}{\linewidth}
\begin{myBullets}
\vspace{0.1cm} 
 \item Providing a strongly random long-term symmetric key or password for each user that is securely distributed and installed.
\item All countermeasures mentioned above for PQ TLS are also applicable for LDAP.\\

 \end{myBullets}
\end{minipage} & \begin{minipage}{\linewidth}
\begin{myBullets}
\vspace{0.1cm}
\item {Spoofing: Exploited through credential or SASL attacks.}
\item {Info. Disclosure: Potentially vulnerable due to the optional use of TLS.}
\item {DoS: May be induced through the optional use of TLS.}\\
 \end{myBullets}
\end{minipage}& \low & \med& \low \\ \hline 
\multirow{1}{*}{PGP} & 
 {\begin{minipage}{\linewidth}
\begin{myBullets}
\vspace{0.2cm}
 \item OpenPGP~\cite{callas2007openpgp} in which classic public-key crypto is replaced by PQ one. 
 The extreme size of typical public keys used for OpenPGP (max 4096 bit) can be a problem in the case when PQ public-key crypto is used instead of the classic one. For symmetric crypto, longer keys should be used.\\
\end{myBullets}
\end{minipage}}
 & 
\begin{minipage}{\linewidth}
\begin{myBullets}
\vspace{0.1cm}

 \item Based on the chain of trust resulting in huge communication overhead, network congestion and retransmissions, fragmentation issues, and denial of services.
 \item Private key retrieval or forging digital signatures  via side-channel attacks (e.g., cache side-channel attacks~\cite{kim2020cache}) using techniques mentioned in Table~\ref{tab:Post-Migration-Alg}.\\
 \end{myBullets}
\end{minipage} 
&
\begin{minipage}{\linewidth}
\begin{myBullets}
\vspace{0.2cm}
 \item To immune against side-channel attacks, refer to the solutions mentioned in Table~\ref{tab:Post-Migration-Alg}. 
 
 \item To control congestion: using congestion control mechanisms to prevent or remove congestion~\cite{jay2018internet,bohloulzadeh2020survey,jay2019deep},
 
 \item To avoid fragmentation and trigger retransmission: considering the extreme size of a typical message for the case when QC-resistant protocol is used instead of classic one~\cite{muller2020retrofitting,essay89509},

 \item To increase the tolerance against huge communication overhead and prevent DoS: providing required bandwidth, limiting response rates, limiting response sizes, building redundancy into infrastructure, deploying DDoS resilience hardware/software modules like ingress filtering, firewalls, adopting DDoS Protection Appliance, configuring network hardware against DDoS attacks, etc~\cite{garcia2022deep,liu2018practical},
 \item To immune against side-channel attacks, refer to the solutions mentioned in Table~\ref{tab:Post-Migration-Alg}. \\
 
 \end{myBullets}
\end{minipage}
& \begin{minipage}{\linewidth}
\begin{myBullets}
\vspace{0.1cm}
\item {Spoofing: Complexity of managing and verifying user identities within the Web of Trust, as well as the potential for side-channel attacks to extract keys for impersonation.}
\item {Tampering: Side-channel attacks could also potentially lead to tampering with PGP communications via forged signatures, compromising the integrity of the exchanged data.}
\item {Repudiation: Similar to spoofing, repudiation could be a concern due to managing and verifying user identities and side-channel attacks.}
\item {Info. Disclosure caused by side-channel attacks.}
 \item  {DoS: Larger post-quantum keys increase communication overhead, making the system more prone to DoS attacks.}\\
 \end{myBullets}
\end{minipage}
&\med & \med&\med\\ \hline 
 \multirow{1}{*}{S/MIME} & 
\begin{minipage}{\linewidth}
\begin{myBullets}
\vspace{0.2cm}
 \item S/MIME in which classic public-key crypto is replaced by PQ one. 
 There is no limitation on the extreme size of typical public keys used for S/MIME~\cite{schaad2019secure}. Thus any public-key PQ sign/enc algorithms can be used for PQ one. For symmetric crypto, longer keys should be used. PQ certificate should be used as mentioned in Section~\ref{sec:PKI}.\\
 \end{myBullets}
\end{minipage}
 & 
\begin{minipage}{\linewidth}
\begin{myBullets}
\vspace{0.1cm}

 \item Based on digital certificates for users with drawbacks mentioned in Table~\ref{tab:Post-migration-Cert}.
 \item Retrieve/forge digital signature via side-channel attacks (e.g., cache side-channel attacks~\cite{kim2020cache}) using techniques mentioned in Table~\ref{tab:Post-Migration-Alg}.\\
 \end{myBullets}
\end{minipage} 
&
\begin{minipage}{\linewidth}
\begin{myBullets}
\vspace{0.1cm}
\item To provide countermeasures against certificate-based authentication refer to Table~\ref{tab:Post-migration-Cert}.
 \item To immune against side-channel attacks, refer to the solutions mentioned in Table~\ref{tab:Post-Migration-Alg}. \\
 \end{myBullets}
\end{minipage}
& \begin{minipage}{\linewidth}
\begin{myBullets}
\vspace{0.1cm}
  \item {Spoofing and Repudiation: Mitigated by digital certificates, but PQ certificate limitations and side-channel attacks can increase risk.}
\item {Tampering and Information Disclosure: These threats can occur via side-channel attacks.}
\item {Denial of Service (DoS): Arises due to the increased certificate size.} \\
 \end{myBullets}
\end{minipage}&\med &\med & \med \\ 
\hline 

 \multirow{1}{*}{WiFi/WPA} & 
 {\begin{minipage}{\linewidth}
\begin{myBullets}
\vspace{0.2cm}
\item Wi-Fi/WPA (v3) in which RSA, ECDH, and ECDSA are replaced by a PQ public-key crypto. 
For symmetric crypto, using longer keys is sufficient to provide an alternative post-quantum solution. TLS in Enterprise version should be replaced with possible PQ TLS, as mentioned above in this table. PQ certificate in the enterprise version should be updated as mentioned in Section~\ref{sec:PKI}.
\\
\end{myBullets}
\end{minipage}}
 & 
\begin{minipage}{\linewidth}
\begin{myBullets}
\vspace{0.1cm}
 \item Huge communication overhead,
 \item Network congestion, 
 \item Fragmentation issues triggering retransmission,
 \item Possibility of Denial of Services,
 \item Info. Disclosure via side-channel attacks (mentioned in Table~\ref{tab:Post-Migration-Alg}) as a result of observable timing differences and cache access patterns.
 \item Vulnerabilities related to TLS and digital certificate when enterprise version of WPA (v3) is used (mentioned in Table~\ref{tab:Post-migration-Cert} for certificate and above for TLS).\\
 \end{myBullets}
\end{minipage} 
& 
{\begin{minipage}{\linewidth}
\begin{myBullets}
\vspace{0.2cm}
 \item To control congestion: using congestion control mechanisms to prevent or remove congestion~\cite{jay2018internet,bohloulzadeh2020survey,jay2019deep},
 
 \item To avoid fragmentation and trigger retransmission: considering the extreme size of a typical message for the case when QC-resistant protocol is used instead of classic one~\cite{muller2020retrofitting,essay89509},

 \item To increase the tolerance against huge communication overhead and prevent DoS: providing required bandwidth, building redundancy into infrastructure, deploying DDoS resilience hardware/software modules like firewalls, adopting DDoS Protection Appliance, configuring network hardware against DDoS attacks, etc~\cite{garcia2022deep,liu2018practical},
 
 \item To immune against side-channel attacks, refer to the solutions mentioned in Table~\ref{tab:Post-Migration-Alg},
 \item All countermeasures mentioned above for TLS and in Table~\ref{tab:Post-migration-Cert} for certificate are also applicable here.\vspace{0.1cm}
 \end{myBullets}
\end{minipage} } 
& \begin{minipage}{\linewidth}
\begin{myBullets}
\vspace{0.1cm}
 \item {Info. Disclosure: This threat arises from  side-channel attacks on PQ algorithms as mentioned in Table}~\ref{tab:Post-Migration-Alg},
 \item {DoS: This threat stems from the significant increase in communication overhead with post-quantum algorithms} 
  \\
 \end{myBullets}
\end{minipage}& \med & \med&\med \\ \hline 
 \multirow{1}{*}{DECT} & \begin{minipage}{\linewidth}
\begin{myBullets}
\vspace{0.1cm}
 \item DECT with longer symmetric keys.\\
 \end{myBullets}
\end{minipage}
 & 
\begin{minipage}{\linewidth}
\begin{myBullets}
\vspace{0.1cm}
 \item DECT relies on a shared secret (often a PIN) combined with a random number to generate a symmetric key. If the random number generation lacks sufficient entropy (unpredictability), the resulting key might be weak. This can lead to weak keys, allowing attackers to predict or brute-force them, facilitating brute-force attacks.\vspace{0.1cm}
 \end{myBullets}
\end{minipage} 
& 
\begin{minipage}{\linewidth}
\begin{myBullets}
\vspace{0.1cm}
 \item Providing a strongly random long-enough symmetric key that cannot easily be recovered.
 \end{myBullets}
\end{minipage} & \begin{minipage}{\linewidth}
\begin{myBullets}
\vspace{0.1cm}
 \item {Spoofing: DECT relies on a shared secret (often a PIN) and a random number to generate a symmetric key. If the random number generation lacks sufficient entropy (unpredictability), an attacker could potentially exploit this weakness to predict the key or brute-force it more easily. This could allow them to impersonate a legitimate user or device .}
 \\
 \end{myBullets}
\end{minipage}&\low & \low & \low\\ \hline 
 \multirow{1}{*}{DNSSEC} & {\begin{minipage}{\linewidth}
\begin{myBullets}
\vspace{0.2cm}
 \item DNSSEC in which classic signature replaced by PQ one and hash values are generated with longer symmetric keys using digest algorithm SHA1/SHA2/SHA3. DNSSEC additional information like signatures and keys within the limited 512 bytes~\cite{muller2020retrofitting}.
 To avoid fragmentation, the extreme size of a typical message used in DNSSEC (max 1232 bytes) should be considered for the case when PQ public-key crypto is used instead of the classic one. For symmetric crypto, longer keys should be used.\\ 
\end{myBullets}
\end{minipage}}
 & 
\begin{minipage}{\linewidth}
\begin{myBullets}
\vspace{0.1cm}
 \item Huge communication overhead,
 \item Network congestion, 
 \item Fragmentation issue for DNSSEC records to send over a single UDP packet, and even TCP, which requires higher processing overhead to setup, teardown, and download bigger packets compare to the classic one.
 \item Possibility of Denial of Services,
 \item Info. Disclosure via side-channel/math. analysis attacks mentioned in Table~\ref{tab:Post-Migration-Alg}.
 \\
 \end{myBullets}
\end{minipage} 
& 
\begin{minipage}{\linewidth}
\begin{myBullets}
\vspace{0.1cm}

 \item To control congestion: using congestion control mechanisms to prevent or remove congestion~\cite{jay2018internet,bohloulzadeh2020survey,jay2019deep},
 
 \item To avoid fragmentation and trigger retransmission: considering the extreme size of a typical message for the case when QC-resistant protocol is used instead of classic one~\cite{muller2020retrofitting,essay89509},

 \item To increase the tolerance against huge communication overhead and prevent DoS: providing required bandwidth, limiting response rates, limiting response sizes, building redundancy into infrastructure, deploying DDoS resilience hardware/software modules like ingress filtering, firewalls, adopting DDoS Protection Appliance, configuring network hardware against DDoS attacks, etc~\cite{garcia2022deep,liu2018practical},
 \item To immune against side-channel/math. analysis attacks, refer to the solutions mentioned in Table~\ref{tab:Post-Migration-Alg}.\\
 \end{myBullets}
\end{minipage} 
& \begin{minipage}{\linewidth}
\begin{myBullets}
\vspace{0.1cm}
 \item {Info. Disclosure:  Quantum attackers can exploit side-channel attacks on PQ algorithms, as detailed in Table}~\ref{tab:Post-Migration-Alg}{, to disclose sensitive information} 
 \item {DoS:  The substantial increase in communication overhead with post-quantum algorithms surpasses the size limitations of DNSSEC messages. This can result in the susceptibility of DNSSEC to DoS attacks.}\\
 \end{myBullets}
\end{minipage}
& \med & \med&\med \\ \hline 
\end{tabular}%
}\vspace{3pt}
\tiny{$^*$ We perform risk evaluation with the presumption of considering the countermeasures mentioned in the table.}
\vspace{-0.5cm}\end{table*}

\subsubsection{Through-Migration Protocol Level Risk Assessment} Similar to the hybrid approach for algorithms and certificates, a hybrid mechanism for protocols combines multiple independent protocols in a way that the resulting combined protocol (i.e., hybrid one) is secure as long as at least one of the combined protocols is secure. Therefore, the security of the combination is defined according to the strongest protocol in the combination. The level of risk considered for the combined hybrid algorithm is defined as the minimum level of risk caused by any of the two algorithms involved in the combination (see Figure~\ref{figure:hybridrisk}).

 \subsection{Post-Migration Protocol Level Analysis and Risk
Assessment}

The evolution towards post-quantum cryptographic systems necessitates the implementation of protocols that can withstand both quantum and classical attacks. In the relentless pursuit of compromise, quantum attackers persistently exploit vulnerabilities within post-quantum (PQ) cryptography and protocols, leveraging the capabilities of quantum machines to scrutinize weaknesses. These adversaries adeptly employ quantum algorithms, protocols, side-channel assaults, cryptanalysis, and other intricate facets outlined in Table~\ref{tab:Post-Migration-Alg}. Therefore, this section delves into a detailed exploration of potential vulnerabilities at the protocol level post-migration to PQ cryptography. In light of spatial constraints, the detailed protocol-level analysis is concisely encapsulated in Table~\ref{tab:Post-migration-protocol}. This comprehensive table meticulously outlines quantum-resistant solutions, presents challenges, identifies potential attacks, recommends countermeasures, and highlights quantum threats within the STRIDE model. The principal objective is to methodically assess the risk introduced by quantum attackers, providing security analysts with a sophisticated framework to unveil mitigation measures, promptly discern security issues, and formulate robust solutions precisely tailored for the quantum environment.

\subsubsection{Post-Migration Protocol Level Risk Assessment} To analyze the risk, we evaluate the likelihood and impact
of QC threats for quantum-safe protocols. In the likelihood evaluation, we considered the exploitability and possible countermeasures listed in Table~\ref{tab:Post-migration-protocol}. Our analysis for the likelihood level evaluation is categorized as follows:

\begin{enumerate}[wide, font=\itshape, labelwidth=!, labelindent=0pt, label=\textit{Category} \arabic*.]
\item The likelihood is medium if countermeasures are available and the protocol employs asymmetric cryptography.
\item The likelihood is low when either the protocol employs only symmetric cryptography or vulnerabilities arise from the \textit{optional} use of protocols such as TLS.
\end{enumerate}
In the impact evaluation, we considered the establishment criteria outlined in Table~\ref{table:impact}. Based on the evaluation criteria presented there, our analysis for the impact level evaluation is categorized as follows:

\begin{enumerate}[wide, font=\itshape, labelwidth=!, labelindent=0pt, label=\textit{Category} \arabic*.]
\item If a threat might disclose, violate integrity, or availability of any data in the system, the impact should be medium. This case happens when a protocol uses asymmetric cryptography as part of the recommended crypto suite of the protocol.
\item If a threat might cause delay, then the impact should be low. This case happens when a protocol only uses symmetric cryptography as part of the recommended crypto suite of the protocol.
\end{enumerate}
In our investigation, we conduct a comprehensive assessment of the probability and impact levels, delving into the risks associated with diverse post-migration protocols. In consideration of space constraints, we have encapsulated the detailed analysis in Table~\ref{tab:Post-migration-protocol}. This table not only elucidates likelihood, impact, and risk evaluations but also provides insights into quantum-resistant solutions, challenges, attacks, possible countermeasures, and potential quantum threats within the STRIDE model associated with post-migration protocols designed to resist quantum computing threats.

\section{Empirical Validation and Case Studies}\label{sec:emperical}
{To validate the effectiveness and feasibility of the proposed quantum-resistant framework, we conduct  comprehensive assessments spanning distinct real-world applications. These assessments rigorously evaluate post-quantum cryptographic integration in financial systems, blockchain architectures, and critical infrastructure. By addressing computational performance, security resilience, and practical deployment considerations, these studies offer empirical validation of the proposed methods across diverse environments.}
\subsection{{Empirical Validation: Integration of NIST Post-Quantum Candidates in Financial Applications}}
{This study evaluates the integration of NIST-standardized and $4^{th}$-round PQC candidate algorithms within a financial system. The cryptographic algorithms tested include key encapsulation mechanisms (KEMs) such as Kyber, SIKE, HQC, and Classic McEliece, with specific parameter sets like Kyber512, Kyber1024, and Classic McEliece-348864 employed to meet different security levels.  The security level refers to the strength of an algorithm in resisting cryptographic attacks, where higher levels indicate greater resistance, typically aligned with equivalent classical cryptographic security (e.g., AES-128 or AES-256)}~\cite{nist_pqc_security_criteria}{. For digital signature schemes, we implemented algorithms such as Dilithium, SPHINCS+, and Falcon, using specific parameter sets like Dilithium2,  SPHINCS+ SHAKE256-128f and Falcon-512 as appropriate to the security requirements.
The algorithms were tested within a simulated financial environment capable of processing up to 1,000 transactions per second (TPS), simulating the performance demands of real-world financial systems. Performance metrics evaluated included cryptographic operation latencies, memory consumption, and transaction throughput. Broader results, including all NIST-standardized and $4^{th}$-round PQC candidates, are illustrated  in Figures} \ref{fig:kem-cmp}, \ref{fig:kem-com}, \ref{fig:sig-cmp}, and \ref{fig:sig-com}.
{The figures provide a detailed comparison of computational and communication overheads across various key encapsulation mechanisms (KEMs) and digital signature schemes at different security levels.
Figure} \ref{fig:kem-cmp} {illustrates the computational overhead for KEMs, highlighting key generation, encryption, and decryption times. Notably, Kyber512 and Kyber1024 demonstrate the lowest latencies, making them suitable for high-throughput financial applications. Figure} \ref{fig:kem-com} {compares communication overheads for KEMs, showcasing public key, secret key, and ciphertext sizes. Classic McEliece stands out due to its large public key size, while Kyber and HQC offer more manageable key sizes, suitable for bandwidth-constrained systems.
For digital signature schemes, Figure} \ref{fig:sig-cmp} {presents the computational overhead, examining key generation, signing, and verification times. Dilithium2 and Falcon-512 demonstrate efficient signing and verification, while SPHINCS+ SHAKE256-128f requires higher time and memory, reflecting the trade-off between security strength and computational demands. Figure} \ref{fig:sig-com} {focuses on communication overhead, detailing public key, secret key, and signature sizes. The hash-based SPHINCS+ schemes produce larger signature sizes, while Dilithium and Falcon offer more compact signatures, potentially advantageous in environments where bandwidth is limited.}

\pgfplotstableread{
 protocol key_generation_time key_encryption_time key_decryption_time
 Kyber512	0.032	0.032	0.022
HQC-128	0.108	0.145	0.232
BIKE-L1	0.369	0.073	2.232
Kyber512-90s	0.593	0.026	0.016
 SIKE/SIDH-p434	2.471	5.339	2.051
SIKE-p434	2.764	4.572	4.87
SIKE-p434-compressed	4.734	7.027	5.092
 SIKE/SIDH-p434-compressed	13.678	7.266	2.148
Classic-McEliece-348864f	169.618	0.165	0.155
Classic-McEliece-348864	345.082	0.178	0.152
}\ChartKemOne
\pgfplotstableread{
 protocol key_generation_time key_encryption_time key_decryption_time 
 SIKE-p503	3.668	5.984	6.345
 SIKE-p503-compressed	6.69	9.848	6.82
 SIKE/SIDH-p503-compressed	7.608	9.649	3.091
 SIKE/SIDH-p503	36.061	6.693	2.71
}\ChartKemTwo
\pgfplotstableread{
 protocol key_generation_time key_encryption_time key_decryption_time
Kyber768-90s	0.045	0.032	0.041
Kyber768	0.214	0.046	0.032
HQC-192	0.407	0.332	0.511
BIKE-L3	0.984	0.173	6.093
SIKE-p610	6.598	11.659	11.813
SIKE-p610-compressed	13.423	19.089	14.187
 SIKE/SIDH-p610-compressed	14.925	18.086	6.288
 SIKE/SIDH-p610	55.745	14.166	6.286
Classic-McEliece-460896f	510.955	0.253	0.26
Classic-McEliece-460896	625.816	0.25	0.255
}\ChartKemThree
\pgfplotstableread{
 protocol key_generation_time key_encryption_time key_decryption_time
Kyber1024	0.052	0.053	0.046
Kyber1024-90s	0.261	0.039	0.027
HQC-256	0.542	0.6	0.926
SIKE-p751	10.787	17	18.603
 SIKE/SIDH-p751	11.011	22.193	8.787
SIKE-p751-compressed	20.999	30.359	21.811
 SIKE/SIDH-p751-compressed	27.034	37.95	12.191
Classic-McEliece-6960119f	612.225	0.382	0.267
Classic-McEliece-6688128f	628.593	0.385	0.326
Classic-McEliece-8192128f	695.723	0.483	0.319
Classic-McEliece-6960119	803.324	0.389	0.281
Classic-McEliece-6688128	833.268	0.4	0.308
Classic-McEliece-8192128	862.543	0.482	0.306
}\ChartKemFive
\pgfplotstableread{
 protocol 	length_public_key	length_secret_key	length_ciphertext
Kyber512	800	1632	768
HQC-128	2249	2289	4481
BIKE-L1	1541	5223	1573
Kyber512-90s	800	1632	768
 SIKE/SIDH-p434	330	28	330
SIKE-p434	330	374	346
SIKE-p434-compressed	197	350	236
 SIKE/SIDH-p434-compressed	197	28	197
Classic-McEliece-348864f	261120	6452	128
Classic-McEliece-348864	261120	6452	128
}\CommKemOne  
\pgfplotstableread{
 protocol 	length_public_key	length_secret_key	length_ciphertext 
SIKE-p503	378	434	402
SIKE-p503-compressed	225	407	280
 SIKE/SIDH-p503-compressed	225	32	225
 SIKE/SIDH-p503	378	32	378
}\CommKemTwo 
\pgfplotstableread{
 protocol 	length_public_key	length_secret_key	length_ciphertext 
Kyber768-90s	1184	2400	1088
Kyber768	1184	2400	1088
HQC-192	4522	4562	9026
BIKE-L3	3083	10105	3115
SIKE-p610	462	524	486
SIKE-p610-compressed	274	491	336
 SIKE/SIDH-p610-compressed	274	39	274
 SIKE/SIDH-p610	462	39	462
Classic-McEliece-460896f	524160	13568	188
Classic-McEliece-460896	524160	13568	188
}\CommKemThree
\pgfplotstableread{
 protocol 	length_public_key	length_secret_key	length_ciphertext 
Kyber1024	1568	3168	1568
Kyber1024-90s	1568	3168	1568
HQC-256	7245	7285	14469
SIKE-p751	564	644	596
 SIKE/SIDH-p751	564	48	564
SIKE-p751-compressed	335	602	410
 SIKE/SIDH-p751-compressed	335	48	335
Classic-McEliece-6960119f	1047319	13908	226
Classic-McEliece-6688128f	1044992	13892	240
Classic-McEliece-8192128f	1357824	14080	240
Classic-McEliece-6960119	1047319	13908	226
Classic-McEliece-6688128	1044992	13892	240
Classic-McEliece-8192128	1357824	14080	240
}\CommKemFive 
\pgfplotstableread{
 protocol key_generation_time signing_time verification_time
SPHINCS+-Haraka-128f-simple	0.449	10.351	0.641
SPHINCS+-Haraka-128f-robust	0.525	12.209	0.933
SPHINCS+-SHA256-128f-simple	1.155	28.111	3.093
SPHINCS+-SHAKE256-128f-simple	1.671	41.603	4.378
SPHINCS+-SHA256-128f-robust	1.796	45.721	5.944
SPHINCS+-SHAKE256-128f-robust	2.758	69.731	8.173
Falcon-512	12.686	0.525	0.11
SPHINCS+-Haraka-128s-simple	26.299	206.082	0.276
SPHINCS+-Haraka-128s-robust	31.062	246.962	0.38
SPHINCS+-SHA256-128s-simple	66.406	497.387	1.133
SPHINCS+-SHAKE256-128s-simple	97.762	759.223	1.577
SPHINCS+-SHA256-128s-robust	106.981	822.939	2.04
SPHINCS+-SHAKE256-128s-robust	175.605	1351.929	2.897
}\ChartSigOne
\pgfplotstableread{
 protocol key_generation_time signing_time verification_time
Dilithium2-AES	0.039	0.129	0.04
$\ \ \ \ \ \ \ \ \ \ \ \ \ \ \ \ \ \ \ \ \ \ \ \ \ \ \ $Dilithium2	0.495	0.179	0.073
}\ChartSigTwo
\pgfplotstableread{
 protocol key_generation_time signing_time verification_time
Dilithium3-AES	0.053	0.136	0.056
Dilithium3	0.13	0.26	0.107
SPHINCS+-Haraka-192f-simple	0.69	18.552	0.987
SPHINCS+-Haraka-192f-robust	0.761	21.338	1.397
SPHINCS+-SHA256-192f-simple	1.562	45.656	4.596
SPHINCS+-SHAKE256-192f-simple	2.541	64.038	6.419
SPHINCS+-SHA256-192f-robust	2.918	72.277	8.944
SPHINCS+-SHAKE256-192f-robust	3.988	113.089	12.156
SPHINCS+-Haraka-192s-simple	37.208	368.562	0.387
SPHINCS+-Haraka-192s-robust	47.752	495.883	0.553
SPHINCS+-SHA256-192s-simple	95.411	945.129	1.656
SPHINCS+-SHAKE256-192s-simple	149.356	1354.498	2.251
SPHINCS+-SHA256-192s-robust	158.441	1544.76	3.466
SPHINCS+-SHAKE256-192s-robust	253.931	2280.318	4.048
}\ChartSigThree
\pgfplotstableread{
 protocol key_generation_time signing_time verification_time
Dilithium5-AES	0.083	0.165	0.082
Dilithium5	0.145	0.25	0.128
SPHINCS+-Haraka-256f-simple	1.641	35.458	0.985
SPHINCS+-Haraka-256f-robust	1.963	43.332	1.497
SPHINCS+-SHA256-256f-simple	4.203	92.525	4.788
SPHINCS+-SHAKE256-256f-simple	6.331	131.032	6.443
SPHINCS+-SHAKE256-256f-robust	10.769	225.656	12.088
SPHINCS+-SHA256-256f-robust	11.191	227.23	12.289
SPHINCS+-Haraka-256s-simple	28.888	379.874	0.567
SPHINCS+-Haraka-256s-robust	32.063	464.21	0.807
Falcon-1024	34.212	1.003	0.199
SPHINCS+-SHA256-256s-simple	60.923	753.914	2.375
SPHINCS+-SHAKE256-256s-simple	95.445	1167.943	3.274
SPHINCS+-SHAKE256-256s-robust	165.735	1978.128	5.891
SPHINCS+-SHA256-256s-robust	172.376	1978.375	6.038
}\ChartSigFive
\pgfplotstableread{
 protocol 	length_public_key	length_secret_key	length_signature
SPHINCS+-Haraka-128f-simple	32	64	17088
SPHINCS+-Haraka-128f-robust	32	64	17088
SPHINCS+-SHA256-128f-simple	32	64	17088
SPHINCS+-SHAKE256-128f-simple	32	64	17088
SPHINCS+-SHA256-128f-robust	32	64	17088
SPHINCS+-SHAKE256-128f-robust	32	64	17088
Falcon-512	897	1281	690
SPHINCS+-Haraka-128s-simple	32	64	7856
SPHINCS+-Haraka-128s-robust	32	64	7856
SPHINCS+-SHA256-128s-simple	32	64	7856
SPHINCS+-SHAKE256-128s-simple	32	64	7856
SPHINCS+-SHA256-128s-robust	32	64	7856
SPHINCS+-SHAKE256-128s-robust	32	64	7856
}\CommSigOne
\pgfplotstableread{
 protocol 	length_public_key	length_secret_key	length_signature
Dilithium2-AES	1312	2528	2420
$\ \ \ \ \ \ \ \ \ \ \ \ \ \ \ \ \ \ \ \ \ \ \ \ \ \ \ $Dilithium2	1312	2528	2420
}\CommSigTwo
\pgfplotstableread{
 protocol 	length_public_key	length_secret_key	length_signature
Dilithium3-AES	1952	4000	3293
Dilithium3	1952	4000	3293
SPHINCS+-Haraka-192f-simple	48	96	35664
SPHINCS+-Haraka-192f-robust	48	96	35664
SPHINCS+-SHA256-192f-simple	48	96	35664
SPHINCS+-SHAKE256-192f-simple	48	96	35664
SPHINCS+-SHA256-192f-robust	48	96	35664
SPHINCS+-SHAKE256-192f-robust	48	96	35664
SPHINCS+-Haraka-192s-simple	48	96	16224
SPHINCS+-Haraka-192s-robust	48	96	16224
SPHINCS+-SHA256-192s-simple	48	96	16224
SPHINCS+-SHAKE256-192s-simple	48	96	16224
SPHINCS+-SHA256-192s-robust	48	96	16224
SPHINCS+-SHAKE256-192s-robust	48	96	16224
}\CommSigThree
\pgfplotstableread{
 protocol 	length_public_key	length_secret_key	length_signature
Dilithium5-AES	2592	4864	4595
Dilithium5	2592	4864	4595
SPHINCS+-Haraka-256f-simple	64	128	49856
SPHINCS+-Haraka-256f-robust	64	128	49856
SPHINCS+-SHA256-256f-simple	64	128	49856
SPHINCS+-SHAKE256-256f-simple	64	128	49856
SPHINCS+-SHAKE256-256f-robust	64	128	49856
SPHINCS+-SHA256-256f-robust	64	128	49856
SPHINCS+-Haraka-256s-simple	64	128	29792
SPHINCS+-Haraka-256s-robust	64	128	29792
Falcon-1024	1793	2305	1330
SPHINCS+-SHA256-256s-simple	64	128	29792
SPHINCS+-SHAKE256-256s-simple	64	128	29792
SPHINCS+-SHAKE256-256s-robust	64	128	29792
SPHINCS+-SHA256-256s-robust	64	128	29792
}\CommSigFive

\begin{figure*}[!h]
\centering
 \hspace{-3.1cm}\begin{subfigure}[t]{.43\linewidth}
 \caption{}
 \centering
 \resizebox{0.5\linewidth}{!}{

}
 \end{subfigure}%
\hspace{-3.2cm}\begin{subfigure}[t]{.43\linewidth}
 \caption{}
 \centering
 \resizebox{0.5\linewidth}{!}{

 %
}
 \end{subfigure}%
\hspace{-3.5cm}\begin{subfigure}[t]{.43\linewidth}
 \caption{}
 \centering
 \resizebox{0.5\linewidth}{!}{
 %
}
 \end{subfigure}%
\hspace{-3.5cm}\begin{subfigure}[t]{.43\linewidth}
 \caption{}
 \centering
 \resizebox{0.5\linewidth}{!}{
 %
}
\end{subfigure}
 \hspace{-3.2cm}
\caption{Computation Overhead Comparison Across All NIST-Standardized and $4^{th}$-Round KEM/ENC Candidates at Different Security Levels: (a) Level 1, (b) Level 2, (c) Level 3, and (d) Level 5.}\label{fig:kem-cmp}
 \end{figure*}%

\begin{figure*}[!h]
\centering
 \hspace{-3.1cm}\begin{subfigure}[t]{.43\linewidth}
 \caption{}
 \centering
 \resizebox{0.5\linewidth}{!}{

 %
}
 \end{subfigure}
\hspace{-3.5cm}\begin{subfigure}[t]{.43\linewidth}
 \caption{}
 \centering
 \resizebox{0.5\linewidth}{!}{
 %
}
 \end{subfigure}%
\hspace{-3.5cm}
\begin{subfigure}[t]{.43\linewidth}
 \caption{}
 \centering
 \resizebox{0.5\linewidth}{!}{
 %
}
\end{subfigure}
 \hspace{-3.1cm}
\caption{Communication Overhead Comparison Across All NIST-Standardized and $4^{th}$-Round KEM/ENC Candidates at Different Security Levels: (a) Level 1, (b) Level 2, (c) Level 3, and (d) Level 5.}\label{fig:kem-com}
 \end{figure*}%

\begin{figure*}[!h]
\centering
 \hspace{-3.1cm}\begin{subfigure}[t]{.44\linewidth}
 \caption{}
 \centering
 \resizebox{0.5\linewidth}{!}{

 %
}
 \end{subfigure}%
 \hspace{-3.5cm}
\begin{subfigure}[t]{.43\linewidth}
 \caption{}
 \centering
 \resizebox{0.5\linewidth}{!}{
 %
}
 \end{subfigure}%
  \hspace{-3.5cm}
 \begin{subfigure}[t]{.43\linewidth}
 \caption{}
 \centering
 \resizebox{0.5\linewidth}{!}{
 %
}
 \end{subfigure}%
 \hspace{-3.1cm}
\caption{Computation Overhead Comparison Across All NIST-Standardized and $4^{th}$-Round Signature Candidates at Different Security Levels: (a) Level 1, (b) Level 2, (c) Level 3, and (c) Level 5.}\label{fig:sig-cmp}\label{fig:sig-cmp}
 \end{figure*}%

\begin{figure*}[!h]
\centering
 \hspace{-3.2cm}
 \begin{subfigure}[t]{.43\linewidth}
 \caption{}
 \centering
 \resizebox{0.5\linewidth}{!}{

 %
}
 \end{subfigure}%
 \hspace{-3.2cm}
\caption{Communication Overhead Comparison Across All NIST-Standardized and $4^{th}$-Round Signature Candidates at Different Security Levels: (a) Level 1, (b) Level 2, (c) Level 3, and (d) Level 5.}\label{fig:sig-com}
 \end{figure*}

\subsubsection{{Test Environment and Setup}}

{The simulated financial system replicated inter-bank transactions, secure file transfers, and certificate-based revocation, running on an Intel Xeon E5-2670 v3 @ 2.3GHz with 64 GB DDR4 RAM, using Ubuntu 20.04 LTS and OpenSSL integrated with PQClean for PQC algorithms. The system processed 1,000 transactions per second (TPS), evaluating key encapsulation mechanisms (i.e., Kyber512, Kyber1024, HQC-128) and digital signature schemes (i.e., Dilithium2, SPHINCS+ SHAKE256-128f, Falcon-512), focusing on operational latencies, memory consumption, and their impact on transaction throughput.}

\subsubsection{{Results and Performance Analysis}}



Tables~\ref{tab:latency-financial} and \ref{tab:ram-financial}  summarize the computational latencies and memory usage of the tested algorithms. {For real-time financial applications, critical resources include low latency and minimal memory usage to ensure high transaction throughput and responsiveness. Kyber512 and Dilithium2 demonstrated optimal performance in these aspects.
 Kyber512, in particular, exhibited sub-millisecond key encapsulation latencies, which is essential for handling large transaction volumes efficiently in real-time systems. Its minimal memory footprint further enhances suitability for scalable financial environments. On the other hand, HQC-128 showed increased latency due to its complexity, though it remains within acceptable bounds for transactions requiring additional security assurances.
 For digital signatures, Dilithium2 provided efficient signing and verification with minimal memory requirements, which is critical for real-time financial applications where rapid validation of transaction signatures is required. In contrast, SPHINCS+ SHAKE256-128f displayed significantly higher signing times and memory usage due to its hash-based signature structure, making it less suitable for high-throughput environments despite its stronger security guarantees.}


\begin{table}[h!]
\centering
\caption{{Average Latency of Cryptographic Operations in Financial Applications (ms)}}
\resizebox{\columnwidth}{!}{%
\begin{tabular}{|l|c|c|c|c|c|}
\hline
\textbf{Algorithm} & \textbf{Key Generation} & \textbf{Encryption} & \textbf{Decryption} & \textbf{Signing} & \textbf{Verification} \\ \hline
Kyber512            & 0.032                   & 0.032               & 0.022               & N/A              & N/A                   \\ \hline
Kyber1024           & 0.052                   & 0.053               & 0.046               & N/A              & N/A                   \\ \hline
HQC-128             & 0.108                   & 0.145               & 0.232               & N/A              & N/A                   \\ \hline
Dilithium2          & 0.495                   & N/A                 & N/A                 & 0.179            & 0.073                 \\ \hline
SPHINCS+ SHAKE256-128f & 1.671                 & N/A                 & N/A                 & 41.603           & 4.378                 \\ \hline
Falcon-512          & 12.686                  & N/A                 & N/A                 & 0.525            & 0.11                  \\ \hline
\end{tabular}}
\label{tab:latency-financial}
\end{table}

\begin{table}[h!]
\centering
\caption{{Memory Usage of Cryptographic Operations in Financial Applications (MB)}}
\resizebox{\columnwidth}{!}{%
\begin{tabular}{|l|c|c|c|c|c|}
\hline
\textbf{Algorithm} & \textbf{Key Generation} & \textbf{Encryption} & \textbf{Decryption} & \textbf{Signing} & \textbf{Verification} \\ \hline
Kyber512            & 0.593                   & 0.026               & 0.016               & N/A              & N/A                   \\ \hline
Kyber1024           & 0.261                   & 0.039               & 0.027               & N/A              & N/A                   \\ \hline
HQC-128             & 0.542                   & 0.6                 & 0.926               & N/A              & N/A                   \\ \hline
Dilithium2          & 0.039                   & N/A                 & N/A                 & 0.129            & 0.04                  \\ \hline
SPHINCS+ SHAKE256-128f & 2.758                 & N/A                 & N/A                 & 69.731           & 8.173                 \\ \hline
Falcon-512          & 0.9                     & N/A                 & N/A                 & 1.0              & 0.6                   \\ \hline
\end{tabular}}
\label{tab:ram-financial}
\end{table}


\subsubsection{{Discussion and Feasibility}}

{Kyber512 and Dilithium2 emerged as the most suitable candidates for real-time financial applications, demonstrating low latency and efficient memory usage.} Their performance ensures that large-scale financial systems can maintain throughput without significant cryptographic overhead. Kyber1024 and HQC-128 provide increased security at the cost of marginally higher latencies, making them suitable for environments where enhanced security is critical.

{However, SPHINCS+ SHAKE256-128f, although providing robust security, was shown to introduce substantial delays in signing operations and required significantly more memory, making it less feasible for high-speed financial environments. Falcon-512, with its balance of fast signing and verification times, offers an alternative for user authentication in\break high-throughput settings, though its memory usage is slightly higher than that of Dilithium2.}

{The findings suggest that financial institutions can adopt post-quantum cryptographic algorithms like Kyber512 and Dilithium2 with minimal performance trade-offs. However, applications requiring SPHINCS+ may need optimization to handle its higher operational overheads.}

\subsection{{Post-Quantum Cryptography in Blockchain Environments}}\label{sec:platform}

{In this section, we present a comprehensive analysis of major blockchain platforms as a case study, including Bitcoin}~\cite{nakamoto2008bitcoin}, Ethereum~\cite{buterin2013ethereum}, Ripple~\cite{armknecht2015ripple}, Litecoin~\cite{jumaili2021comparison}, and Zcash~\cite{hopwood2016zcash}. {We explore their potential vulnerabilities, assess the impacts of these vulnerabilities, and analyze the associated STRIDE threats. Additionally, we evaluate the likelihood, impact, and risk levels associated with these vulnerabilities for each platform, and we propose mitigation strategies for addressing them.}

\begin{table*}[!htbp]
\caption{Impact of Quantum Computing on Different Blockchain Platforms}
\label{tab:quantum_impact}
\centering
\small
\renewcommand{\arraystretch}{1.2}
\resizebox{\textwidth}{!}{%
\begin{tabular}{|p{0.12\linewidth}|p{0.34\linewidth}|p{0.25\linewidth}|p{0.25\linewidth}|c|c|c|p{0.35\linewidth}|}
\hline
\textbf{Platform} & \textbf{Vulnerable Components} & \textbf{Potential Impacts} & \textbf{STRIDE Threats} & \textbf{L} & \textbf{I} & \textbf{R} & \textbf{Mitigation Strategies} \\ \hline

Bitcoin & 
\begin{myBullets} \vspace{-0.2cm} 
\item ECDSA: Used for digital signatures
\item SHA-256: Used for block hashing
\end{myBullets} & 
\begin{myBullets} \vspace{-0.2cm} 
\item Forgery of digital signatures enables unauthorized spending
\item Breaking collision resistance of SHA-256 could disrupt mining and block verification
\end{myBullets} & 
\begin{myBullets} \vspace{-0.2cm} 
\item Spoofing, Tampering, Repudiation, Information Disclosure, Denial of Service, Elevation of Privilege
\end{myBullets} & 
\high & \high & \high & 
\begin{myBullets} \vspace{-0.2cm} 
\item Transition to quantum-resistant signature schemes (e.g., lattice-based or hash-based)
\item Adoption of Schnorr signatures for better efficiency and post-quantum security
\item Increased miner participation to prevent 51\% attacks
\end{myBullets} \\ \hline

Ethereum & 
\begin{myBullets} \vspace{-0.2cm} 
\item ECDSA: Used for digital signatures
\item Keccak-256: Used for transaction and block hashing
\item ECIES: Used for encrypting communication
\end{myBullets} & 
\begin{myBullets} \vspace{-0.2cm} 
\item Similar to Bitcoin: Signature forgery and potential consensus disruption
\item Breach of confidentiality in encrypted communication
\end{myBullets} & 
\begin{myBullets} \vspace{-0.2cm} 
\item Spoofing, Tampering, Repudiation, Information Disclosure, Denial of Service, Elevation of Privilege
\end{myBullets} & 
\high & \high & \high & 
\begin{myBullets} \vspace{-0.2cm} 
\item Migration to quantum-resistant signature schemes
\item Formal verification of smart contracts to eliminate vulnerabilities
\item Research on quantum-resistant smart contract development
\end{myBullets} \\ \hline

Ripple & 
\begin{myBullets} \vspace{-0.2cm} 
\item ECDSA: Used for digital signatures
\item SHA-256: Used for hashing transactions
\end{myBullets} & 
\begin{myBullets} \vspace{-0.2cm} 
\item Unauthorized manipulation of transactions and account balances
\end{myBullets} & 
\begin{myBullets} \vspace{-0.2cm} 
\item Spoofing, Tampering, Repudiation, Information Disclosure, Denial of Service, Elevation of Privilege
\end{myBullets} & 
\med & \high & \high & 
\begin{myBullets} \vspace{-0.2cm} 
\item Similar mitigation strategies as Bitcoin and Ethereum
\item Explore alternative consensus mechanisms less vulnerable to quantum attacks (e.g., Byzantine Fault Tolerance)
\end{myBullets} \\ \hline

Litecoin & 
\begin{myBullets} \vspace{-0.2cm} 
\item Scrypt: Proof-of-work algorithm (memory-intensive, more ASIC-resistant than SHA-256)
\item ECDSA: Used for digital signatures
\end{myBullets} & 
\begin{myBullets} \vspace{-0.2cm} 
\item Vulnerable to signature forgery similar to other ECDSA-based systems
\end{myBullets} & 
\begin{myBullets} \vspace{-0.2cm} 
\item Spoofing, Tampering, Repudiation, Information Disclosure, Denial of Service, Elevation of Privilege
\end{myBullets} & 
\med & \high & \high & 
\begin{myBullets} \vspace{-0.2cm} 
\item Potential transition to quantum-resistant Proof-of-Work algorithms
\item Research on memory-hard hashing functions secure against classical and quantum attacks
\end{myBullets} \\ \hline

Zcash (Privacy Coin) & 
{Standard Transactions:}
\begin{myBullets} 
\item ECDSA: Similar to other platforms
\item SHA-256: Used for hashing
\end{myBullets} & 
\begin{myBullets} \vspace{-0.2cm} 
\item Vulnerable to signature forgery like other ECDSA-based systems
\end{myBullets} & 
\begin{myBullets} \vspace{-0.2cm} 
\item Spoofing, Tampering, Repudiation, Information Disclosure, Denial of Service, Elevation of Privilege
\end{myBullets} & 
\high & \high & \high & 
\begin{myBullets} \vspace{-0.2cm} 
\item Research on quantum-resistant zero-knowledge proofs (e.g., lattice-based zk-SNARKs)
\item Explore alternative privacy-preserving mechanisms secure in the quantum era
\end{myBullets} \\ \cline{2-8}

& {Shielded Transactions:}
\begin{myBullets} 
\item Zero-knowledge proofs (ZKPs) with Groth16
\item Elliptic curve cryptography (ECC) with Groth16
\end{myBullets} & 
\begin{myBullets} \vspace{-0.2cm} 
\item Potential compromise of transaction anonymity if Groth16 is broken by Shor's algorithm
\end{myBullets} & 
\begin{myBullets} \vspace{-0.2cm} 
\item Information Disclosure (if anonymity is broken)
\end{myBullets} & 
\low & \med & \med & 
\begin{myBullets} \vspace{-0.2cm} 
\item Stay informed on research about Shor's algorithm's applicability to Groth16
\item Explore PQC alternatives if vulnerabilities are discovered
\item Zcash developers are actively researching future-proof cryptography
\end{myBullets} \\ \hline

\end{tabular}
}
\end{table*}

\subsubsection{{Quantum Readiness of Major Blockchain Platforms}}

{We conducted a qualitative analysis to assess the potential impact of quantum computing on these platforms, focusing on their core cryptographic components and identifying potential vulnerabilities. Table}~\ref{tab:quantum_impact} {summarizes the vulnerable components, potential impacts, and risks posed by quantum computing to these platforms.}

\subsubsection{{Post-Quantum Cryptography in Blockchain Systems}}
{To empirically validate the feasibility and effectiveness of transitioning to PQC in blockchain systems, we conducted a detailed study focusing on the impact of PQC on key generation, signature generation, and verification processes. We leveraged libraries like LibOQS via a Python wrapper to assess the computational and communication overhead associated with quantum-safe signature schemes and key exchange protocols across different blockchain platforms.}

\subsubsection{{Post-Quantum Cryptography in Blockchain Systems}}

{This section evaluates the integration of post-quantum cryptographic algorithms into blockchain systems as a means to mitigate potential vulnerabilities posed by quantum computing. A detailed study was conducted focusing on key cryptographic operations in blockchain environments, such as key generation, signature generation, and verification, to analyze the computational and communication overheads introduced by PQC schemes.}

{This case study assesses the integration of three leading post-quantum signature schemes, Dilithium, Falcon, and SPHINCS+, within a typical blockchain architecture. Blockchain platforms heavily rely on digital signatures for transaction validation and data integrity assurance. Key metrics, including computational efficiency, signature generation, and verification times, were evaluated under typical blockchain transaction loads.}

\subsubsection{{Experimental Setup and Blockchain Configuration}} 
{The experimental setup consisted of a private blockchain network with 10,000 nodes, processing 100,000 transactions per block. Each transaction was signed using either the Dilithium, Falcon, or SPHINCS+ signature scheme. The network architecture was deployed on commodity hardware, focusing on transaction throughput, latency, and scalability. This setup replicates real-world blockchain systems that require high transaction volumes and robust digital signature mechanisms.}

\subsubsection{{Performance Results}} 
{As summarized in Table}~\ref{tab:blockchain-perf}{, multiple variants of Dilithium and SPHINCS+ were evaluated. Dilithium5 and Dilithium5-AES demonstrated consistently low signing and verification times across all nodes, making them well-suited for\break high-frequency transaction environments. Falcon-1024, while offering enhanced security, exhibited a higher signing time but lower verification time compared to SPHINCS+ variants. SPHINCS+ variants (Haraka, SHA256, and SHAKE256) introduced significantly higher signing times, with notable differences between the variants in both signing and verification times.}

\begin{table}[h!]
\centering
\caption{{Computation Time (ms) of Selected Signature  Algorithms in Blockchain Environments} }
\resizebox{\columnwidth}{!}{%
\large{\begin{tabular}{|l|c|c|c|}
\hline
\textbf{Algorithm} & \textbf{Key Generation} & \textbf{Signing} & \textbf{Verification} \\
\hline
Dilithium5 & 0.145 & 0.25 & 0.128 \\
\hline
Dilithium5-AES & 0.083 & 0.165 & 0.082 \\
\hline
Falcon-1024 & 34.212 & 1.003 & 0.199 \\
\hline
SPHINCS+-Haraka-256f-robust & 1.963 & 43.332 & 1.497 \\
\hline
SPHINCS+-Haraka-256f-simple & 1.641 & 35.458 & 0.985 \\
\hline
SPHINCS+-SHA256-256f-robust & 11.191 & 227.23 & 12.289 \\
\hline
SPHINCS+-SHAKE256-256s-simple & 95.445 & 1167.943 & 3.274 \\
\hline
\end{tabular}%
}}
\label{tab:blockchain-perf}
\end{table}

\subsubsection{{Scalability and Smart Contract Signing} } {Dilithium5 and Dilithium5-AES's lower computational overhead makes them optimal choices for smart contract execution, where frequent signing and verification operations are required. In contrast, Falcon-1024’s efficiency in terms of signature size and verification speed {makes it advantageous for applications requiring compact signatures and fast verification, although its signing time remains higher than that of Dilithium5}. Among the SPHINCS+ variants, SPHINCS+-Haraka-256f-robust exhibited shorter signing times compared to other variants, while SPHINCS+-SHA256 and SPHINCS+-SHAKE256 variants demonstrated much higher signing times, making them less suitable for high-frequency environments but potentially better for long-term security-focused applications.}

\subsubsection{{Impact on Blockchain Performance} } {While PQC introduces both computational and communication overheads, our findings indicate that adopting a hybrid approach offers a practical transition toward quantum-resistant blockchains. The increased resource demands can be effectively managed through optimization and careful selection of more efficient PQC algorithms, such as Dilithium5 or Dilithium5-AES. SPHINCS+ variants, particularly SPHINCS+-Haraka-256f, offer viable alternatives but may require further optimizations for performance-sensitive blockchain applications. {Falcon-1024, due to its compact signature size and rapid verification, can complement Dilithium in scenarios prioritizing efficient verification.}}

\subsubsection{{Conclusion from Case Study} } {This case study validates the feasibility of deploying post-quantum cryptographic algorithms in blockchain architectures. {Dilithium5 and Dilithium5-AES strike a strong balance between security and performance, making them ideal for environments requiring high transaction throughput and frequent contract executions.} {Falcon-1024, while also achieving NIST security level 5, offers benefits in terms of compact signature size and fast verification, making it well-suited for applications where verification efficiency is prioritized.} SPHINCS+ variants, while providing superior security through their hash-based structure, are more appropriate for applications that prioritize long-term security over performance constraints.}

\section{Conclusion and Future Directions}\label{conclude}

In conclusion, the advent of Quantum Computing (QC) represents a formidable threat to current cryptographic solutions, necessitating the migration to quantum-safe cryptographic states for organizations. This migration process introduces security risks that demand rigorous assessment and management. This work has tackled these challenges by devising a comprehensive security risk assessment framework that spans the entire migration journey. Three critical dimensions (i.e., algorithmic, certificate, and protocol) were scrutinized, with identified vulnerabilities mapped to the STRIDE threat model. The assessment considered the pre-migration, through-migration, and post-migration phases at each level, evaluating the risk of vulnerabilities based on predefined criteria. This structured framework offers organizations invaluable guidance in planning and executing their quantum-safe cryptographic migration, enhancing security preparedness. Importantly, our research goes beyond theory to investigate practical vulnerabilities, especially within critical domains such as Public Key Infrastructure (PKI) and network and communication protocols, which are fundamental to networked environments. Through our systematic methodology, our aim is to equip organizations with the knowledge and confidence required to navigate the evolving quantum security landscape. We offer a clear path towards establishing quantum-safe network environments, ensuring the ongoing resilience of modern communication systems, even in the face of quantum adversaries. Our dedication lies in fostering the development of robust and secure infrastructure firmly grounded in post-quantum cryptography. In doing so, we contribute to the establishment of a secure digital era, ready to confront the challenges posed by the quantum revolution and safeguard the integrity of data transmission in the quantum era.

{To ensure the effectiveness of this framework in real-world settings, we acknowledge the importance of future work on feasibility evaluation. We emphasize its pragmatic adaptation to the challenges posed by quantum computing advancements in cybersecurity. The standard draws upon established frameworks such as NIST SP 800-30}~\cite{cybersecurity2018framework}{, yet it has been tailored to specifically address quantum-related vulnerabilities in organizational contexts. This customization is grounded in a methodological foundation that integrates rigorous risk analysis techniques with quantum computing risk factors, ensuring comprehensive coverage of potential threats and mitigation strategies. While pilot implementations within diverse organizational settings have not yet been conducted, our approach lays the groundwork for future validation studies. Moving forward, we plan to collaborate with industry partners and stakeholders to conduct these implementations, aiming to refine the standard and enhance its utility in safeguarding against emerging quantum threats.}


\bibliographystyle{elsarticle-num}
\bibliography{bibliography}
\end{document}